# GASDYNAMIC FLOW CONTROL BY SUPERFAST LOCAL HEATING IN A STRONGLY NONEQUILIBRIUM PULSE PLASMA

*Andrey Yu. Starikovsky, Princeton University, USA*

*Nickolay L. Alexandrov, MIPT, Russia*

Paper presents a review of modern works on the gasdynamic flow control using a heat release in highly nonequilibrium pulsed plasma. The major attention is paid to the effects based on ultrafast (on nanosecond time scale at atmospheric pressure) local heating of the gas, since at present the main successes in high-speed flow control using gas discharges are associated with namely this thermal effect. The physical mechanisms that control the interaction of the discharges with gas flows are considered in detail. The first part of the review briefly describes the most popular approaches in plasma aerodynamics for the pulsed localized energy deposition: nanosecond surface barrier discharges, pulsed spark discharges, and nanosecond/femtosecond laser discharges. The mechanisms of ultrafast heating of air at high electric fields realized in these discharges, as well as during the decay of a discharge plasma, are analyzed. The second part of the review contains numerous examples of plasma control of gasdynamic flows. The control of the shock waves configuration in front of a supersonic object and control of its trajectory; the control of quasistationary separated flows and layers; the control of a laminar-turbulent transition; the control of static and dynamic separation of the boundary layer at large angles of attack; as well as the operation of plasma actuators in various weather conditions; and the use of plasma as an anti-icing tool for flying objects are analyzed.

# УПРАВЛЕНИЕ ГАЗОДИНАМИЧЕСКИМИ ПОТОКАМИ С ПОМОЩЬЮ СВЕРХБЫСТРОГО ЛОКАЛЬНОГО НАГРЕВА В СИЛЬНОНЕРАВНОВЕСНОЙ ИМПУЛЬСНОЙ ПЛАЗМЕ

*А.Ю. Стариковский, Принстонский университет, США*

*Н.Л. Александров, МФТИ, Россия*

Проведен обзор современных работ по управлению газодинамическими потоками с помощью сильнонеравновесной импульсной плазмы. Основное внимание уделено эффектам, в основе которых лежит сверхбыстрый (на наносекундных временах для атмосферного давления) локальный нагрев газа, поскольку в настоящее время главные успехи в управлении скоростными потоками с помощью газовых разрядов связаны именно с таким тепловым воздействием. Большое внимание уделено физическим механизмам, которые контролируют взаимодействие разряда с газовыми течениями. В первой части обзора кратко описаны наиболее популярные в плазменной аэродинамике подходы при организации импульсного энерговклада: наносекундные поверхностные барьерные разряды, импульсные искровые разряды и наносекундные оптические разряды. Отдельно проанализированы механизмы сверхбыстрого нагрева воздуха при высоких электрических полях, реализуемых в этих разрядах, а также при распаде разрядной плазмы. Во второй части обзора приведены многочисленные примеры плазменного управления газодинамическими потоками. Рассмотрены управление конфигурацией ударных волн перед сверхзвуковым объектом, управление его траекторией, управление квазистационарными отрывными течениями и слоями, управление ламинарно-турбулентным переходом, управление статическим и динамическим отрывом пограничного слоя на больших углах атаки, а также вопросы работы плазменных актуаторов в различных погодных условиях и использование плазмы для борьбы с обледенением летящего объекта.

# Оглавление

# Обозначения

| | |
|---|---|
| $x/c$ | Безразмерная координата по хорде крыла |
| $\bar{U}^*$ | Безразмерная усреднённая местная скорость, $\bar{U}/U_\infty$ |
| $F_c^+$ | Безразмерная частота возбуждающей силы, $F_f c/U_\infty$ |
| $T^*$ | Безразмерное время конвекции, $t\, U_\infty/c$ |
| $C_P$ | Коэффициент давления |
| $C_L$ | Коэффициент подъёмной силы |
| $C_M$ | Коэффициент момента силы |
| $C_D$ | Коэффициент силы сопротивления |
| $U$ | Мгновенная местная скорость |
| $V$ | Напряжение разряда |
| $n$ | Плотность газа |
| $n_i$ | Плотность ионов |
| $n_e$ | Плотность электронов |
| $\gamma$ | Показатель адиабаты, $C_p/C_v$ |
| $E_B$ | Пороговое пробойное электрическое поле |
| $E/n$ | Приведенная напряженность электрического поля |
| $K$ | Приведенная частота осцилляций крыла, $\pi f c/U_\infty$ |
| $U_S$ | Скорость звука |
| $U_\infty$ | Скорость свободного потока |
| $\nu_c$ | Транспортная частота столкновений электронов с другими частицами |
| $A$ | Угол атаки |
| $\bar{U}$ | Усреднённая местная скорость |
| $c$ | Хорда крыла |
| $F$ | Частота |
| $F_f$ | Частота возбуждающей силы |
| $\omega$ | Частота лазерного излучения |
| $f$ | Частота осцилляций крыла |
| $\nu_\varepsilon$ | Частота релаксации энергии электронов в столкновениях с другими частицами |
| $\Omega$ | Частота электромагнитного поля |
| $M$ | Число Маха, $U_\infty/U_S$ |
| $Re_c$ | Число Рейнольдса по хорде, $U_\infty c/\nu$ |
| $St$ | Число Струхаля, $F c/U_\infty$ |
| $E$ | Электрическое поле |
| $\omega_{pe}$ | Электронная плазменная частота |
| ns-SDBD | Поверхностный диэлектрический барьерный разряд с импульсным наносекундным питанием |
| AC-SDBD | Поверхностный диэлектрический барьерный разряд с синусоидальным питанием |
| $T_e$ | Эффективная температура электронов |
| $E_{eff}$ | Эффективное электрическое поле |

# 1. Введение

Активное управление потоком с помощью низкотемпературной плазмы в настоящее время представляет собой бурно развивающееся направление аэродинамики [1-3]. Одним из главных практических достоинств воздействия плазмы на поток является его быстродействие. Это воздействие может быть эффективным в широком диапазоне частот и газодинамических течений, начиная от стационарных потоков и кончая отрывными и турбулентными течениями [4]. К двум другим преимуществам плазменных систем относятся их малый вес и размеры. При учете их относительно малого энергопотребления все это позволяет разработать принципиально новые системы управления полетом на больших скоростях.

Использование плазменных систем в аэродинамике оказывается перспективным при управлении переходом от ламинарного течения к турбулентному в пограничном слое, управлении отрывом потока или его присоединением к твердой поверхности и изменении силы сопротивления и подъемной силы крыла. Также с помощью плазмы можно заметно продвинуться в таких вопросах, как воздействие на шумовые характеристики и колебания потока и управление структурой ударных волн и их взаимодействием в пограничном слое. Устройства, используемые для управления газодинамическим потоком с помощью плазмы, получили название плазменных актуаторов.

Первые предложения по использованию плазменных методов для управления воздушным потоком были сделаны достаточно давно. Одна из пионерских попыток воздействия на поток с помощью неравновесной плазмы была выполнена более 50 лет назад [5, 6]. Авторы этой работы исследовали управление отрывом потока посредством поверхностного барьерного диэлектрического разряда при наложении синусоидального высоковольтного импульса в диапазоне частот 50-570 Гц. Уменьшение силы сопротивления достигало 30%, а увеличение подъемной силы – 40% при обтекании профиля в диапазоне скоростей от 8.75 до 20.4 м/с. Был сделан вывод о том, что актуатор влиял на поток посредством ионного ветра, когда индуцированная им скорость составляла 20-25% от скорости свободного потока.

Работа [7] была одним из первых исследований влияния плазменных эффектов на распространение ударной волны в газе. Авторы наблюдали увеличение скорости ударной волны при ее распространении в газовом разряде с одновременным уменьшением амплитуды волны. Полученный в эксперименте рост скорости ударной волны сравнивался с результатами расчетов с учетом выделенной в разряде энергии. Полученное различие (1200-1300 м/с вместо 900 м/с) было приписано возможному нагреву газа за счет тушения колебательно- и электронно-возбужденных молекул и влиянию образования двойного слоя во фронте ударной волны.

Для модификации и управления потоком около сверхзвукового объекта было предложено несколько гипотетических схем. Эти схемы использовали новые подходы для генерации плазмы, магнитогидродинамического (МГД) управления потоком и генерации энергии, противоток горячего газа и другие тепловые эффекты [1].

Дальнейший рост интереса к управлению потоком посредством плазмы относится к 1993 году и связан с работами Рота с соавторами, которые снова применили поверхностный разряд для контроля отрыва потока в пограничном слое. В работах [8-12] продемонстрирована возможность подавлять отрыв потока на малых скоростях с помощью диэлектрического поверхностного ВЧ разряда.

Еще одно интересное экспериментальное исследование было представлено в [13]. Были проведены баллистические эксперименты для измерения силы сопротивления при движении сферы диаметром 15 мм со скоростью от 200 до 1350 м/с в неравновесной плазме, создаваемой в воздухе ВЧ разрядом. В работе зафиксирован аномальный отход ударной волны от сферы внутри зоны разряда на расстояния, существенно превышающие размеры тепловых неоднородностей на входе в плазму. Измерения коэффициента лобового сопротивления свидетельствуют о существенном снижении сопротивления на дозвуковых скоростях и о небольшом его повышении на сверхзвуковых скоростях. Уменьшение этого коэффициента при дозвуковых скоростях не соответствовало его ожидаемому снижению при получаемом в эксперименте повышении температуры и уменьшении числа Рейнольдса. Авторы показали, что наблюдаемые эффекты вызваны увеличенной скоростью потока с возмущениями, распространяющимися по всему полю скоростей в присутствии плазмы.

Возможности управления потоком с помощью плазмы можно разбить на две группы: 1) управление основным потоком, в том числе конфигурацией ударной волны в сверхзвуковых и гиперзвуковых режимах и 2) управление пограничным слоем. Управление основным потоком включает в себя воздействие на форму ударной волны, аэродинамические нарушения, уменьшение сопротивления, снижение тепловыделения, изменение направления течения, его ускорение и замедление, МГД генерация энергии. Управление пограничным слоем подразделяется на контроль перехода от ламинарного течения к турбулентному, воздействие на отрыв течения в пограничном слое, контроль подъемной силы и силы сопротивления, а также контроль акустических возмущений и увеличение эффективности смешения.

Имеется три различных физических механизма воздействия плазмы на аэродинамику. Это а) нагрев газа, б) электростатическая передача импульса от заряженных частиц к частицам газа и в) магнитогидродинамические эффекты, в том числе МГД ускорение потока и генерация электрической энергии на борту за счет кинетической энергии потока. Иногда необходимо учитывать диссоциации газа и изменение его среднего молекулярного веса. Однако значительная диссоциация или ионизация газа требуют большого количества энергии. Поэтому в задачах авиационно-космической промышленности в целом стараются сохранить возбуждение газа на минимальном уровне, что снижает роль эффектов, связанных с изменением состава газа в разряде.

Повышенный интерес к плазменной аэродинамике привел к растущему потоку публикаций в этой области. За последнее время появилось много впечатляющих демонстраций воздействия плазмы на газовые потоки. Часть из них нашла отражение в тематических обзорах, опубликованных в последнее время. Например, возможности плазменного управления сверхзвуковыми потоками (в том числе – затуханием ударных волн и изменением их формы) обсуждались в [1]. Преимущества плазменных актуаторов по сравнению с жидкостными и механическими актуаторами в отношении управления высокоскоростными потоками продемонстрированы в [14, 15]. Обзор работ по воздействию разрядной плазмы на газодинамические потоки с помощью электродинамических сил (ионный ветер) дан в [2]. Уменьшение сопротивления тела при больших скоростях потока посредством энерговклада рассматривалось в [16, 17]. Обзор работ по управлению отрывом потока в пограничном слое с помощью неравновесной плазмы дан в [18-22]. Влияние импульсного нагрева на взаимодействие ударных волн с пограничным слоем изучено в [23] для различных способов энерговклада. Обзор методов возбуждение газодинамических неустойчивостей с помощью локального импульсного нагрева представлен в [24, 25]. Кинетические процессы, определяющие взаимодействие наносекундного диэлектрического и квазистационарного поверхностных разрядов с газодинамическими течениями, рассмотрены в [26, 27]. Обзоры [28, 29] включают в себя много новых результатов на основе различных механизмов (теплового, электростатического и магнитогидродинамического), а также обсуждение их приложений для управления потоком в дозвуковых и сверхзвуковых режимах. Однако со времени появления этих обзоров прошло достаточно много времени. Появились новые важные результаты, уточняющие механизмы и эффект воздействия плазменных актуаторов на поток. Поэтому в данном обзоре лишь кратко упомянуты наиболее важные результаты плазменной аэродинамики, относящиеся к периоду до 2010-го года, и более подробно представлены новые работы и данные. Рассмотрение ограничено воздействием сильнонеравновесной импульсной плазмы на газодинамические потоки, в основе которого лежит сверхбыстрый (на наносекундных временах для атмосферного давления) локальный нагрев газа. Основные успехи в управлении скоростными потоками плазмой связаны в настоящее время именно с этим эффектом. Большое внимание уделено физическим механизмам, которые способствуют или препятствуют управлению газовыми потоками с помощью плазмы.

Первые разделы обзора посвящены наиболее популярным в плазменной аэродинамике подходам при организации импульсного энерговклада: наносекундным поверхностным барьерным разрядам, импульсным искровым разрядам и наносекундным оптическим разрядам. Отдельно рассматриваются механизмы сверхбыстрого нагрева воздуха при высоких электрических полях, реализуемых в этих разрядах, а также процессы, определяющие распад сильнонеравновесной разрядной плазмы, в котором происходит значительный нагрев воздуха. Во второй части обзора демонстрируются примеры плазменного управления газодинамическими потоками за счет импульсного нагрева воздуха. При этом рассматриваются управление конфигурацией ударных волн перед сверхзвуковым объектом, управление его траекторией, управление квазистационарными отрывными течениями слоями, управление отрывом пограничного слоя на больших углах атаки, управление динамическим отрывом потока, а также использование плазмы для борьбы с обледенением летящего объекта.

## 2. Наносекундные поверхностные барьерные разряды

Наносекундные поверхностные разряды широко применялись во второй половине прошлого века для создания предионизации в электроразрядных эксимерных лазерах [30]. При этом рабочей средой обычно являлись инертные газы или их смеси с галогенсодержащими соединениями. Пионерские эксперименты по исследованию возможности использования наносекундных поверхностных диэлектрических барьерных разрядов (surface dielectric barrier discharge (SDBD)) для управления потоками в воздухе атмосферного давления относятся уже к началу текущего столетия [31-33].

На рис. 1 приведена типичная схема электродов и разрядного промежутка, используемых для плазменных актуаторов на основе SDBD. Здесь два плоских электрода разделены диэлектрическим слоем, что позволяет избежать тока проводимости с одного электрода на другой. Высоковольтный электрод находится сверху, а второй электрод обычно заземлен. Разряд возникает на краю высоковольтного электрода после подачи на него импульса напряжения.

На рис. 2 с помощью ICCD фотографий демонстрируется развитие SDBD разряда обеих полярностей в воздухе атмосферного давления под действием высоковольтных импульсов амплитудой 28 кВ на высоковольтном электроде [32, 33, 35]. Полуширина импульса составляла 23 нс, а время нарастания и время спада импульса были 8 и 15 нс, соответственно. Частота повторения импульсов высокого напряжения была равна 1 кГц. Низковольтный электрод был покрыт пленкой PVC толщиной 0.4 мм. Диэлектрическая проницаемость пленки составляла $\varepsilon \approx 2.7$. Изображения развития разряда были получены с наносекундным временным разрешением с помощью ICCD камеры, сфокусированной на верхнюю плоскость внешнего электрода и диэлектрического слоя.

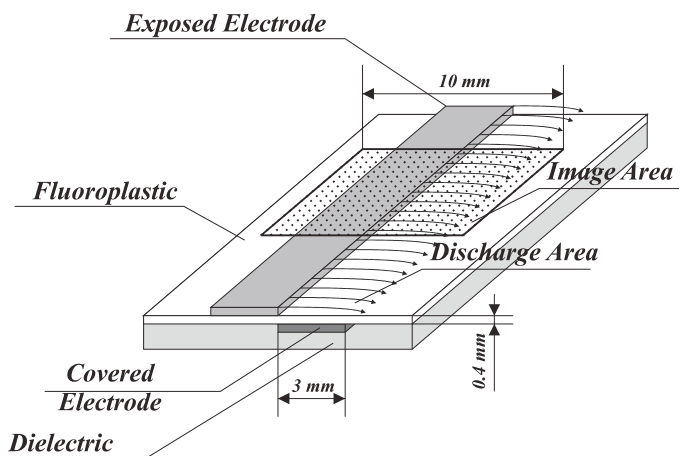

Рисунок 1. Схема разрядного промежутка SDBD [34].

Разряд развивался от края высоковольтного электрода над покрытым диэлектрическим слоем низковольтным электродом. Распространение катодонаправленного (положительного) разряда разделялось на четыре этапа. На первом этапе разряд развивался над заземленным электродом (1, 3 и 4 нс, рис.2,а). Скорость распространения разряда на этой стадии была приблизительно 1 мм/нс.

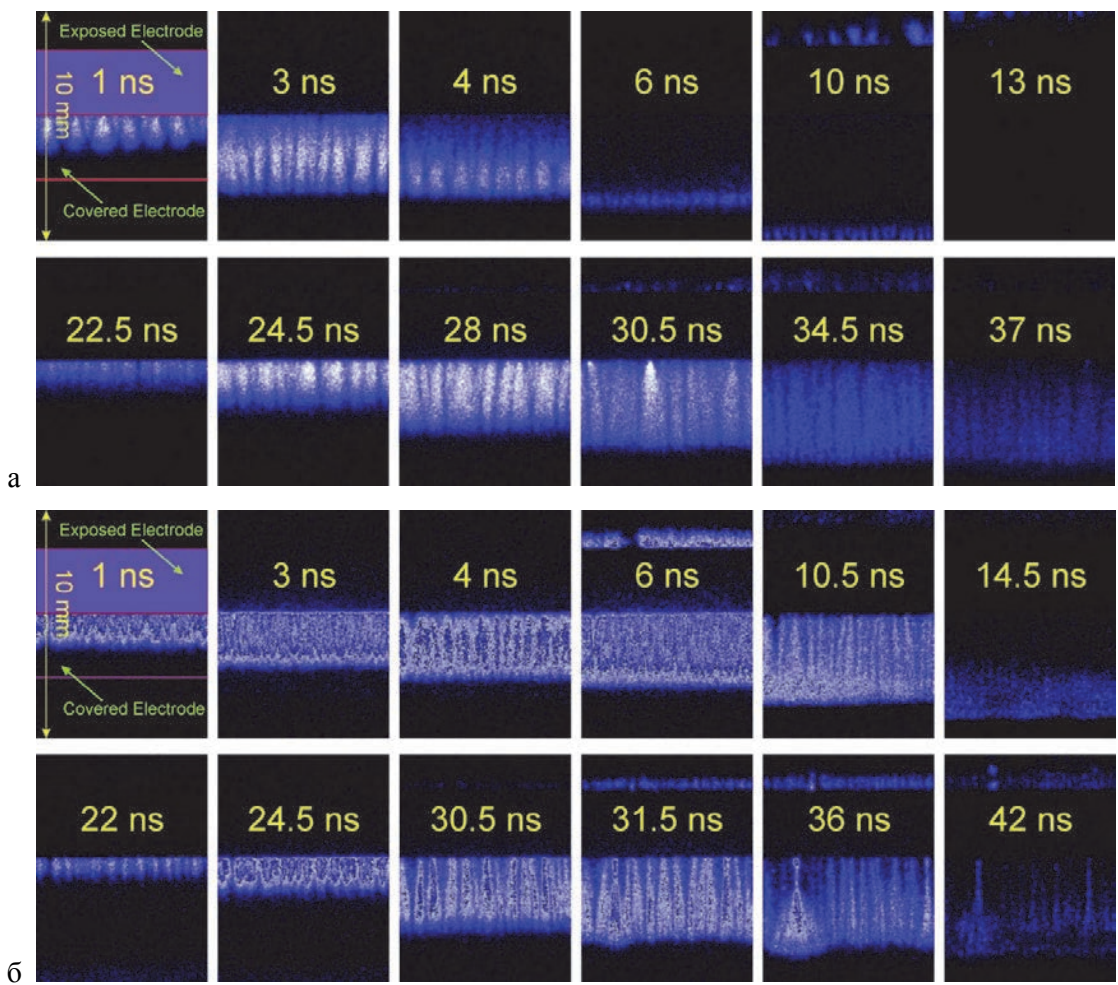

Рисунок 2. Изображения развития поверхностного наносекундного барьерного разряда с наносекундным временным разрешением [36]. Выдержка 0.5 нс. (а) Катодонаправленный разряд. (б) Анодонаправленный разряд.

Здесь можно наблюдать излучение всего канала стримера, а не только переднего фронта волны ионизации. Когда длина стримеров превышала длину нижнего электрода, их скорость становилась ниже (0.3 мм/нс). Этот этап занимал 5 нс (6-я и 10-я нс, рис. 2,а). После этого этапа начиналась "темная" фаза разряда, когда излучение разряда не наблюдалось (с 13 по 20 нс, рис. 2,а). Во время распространения стримеров происходила зарядка поверхности диэлектрика. Таким образом, когда задний фронт импульса высокого напряжения достигал электрода, потенциал электрода становился меньше, чем потенциал поверхности диэлектрика в разрядном промежутке. Это приводило к началу волны обратного разряда, которая способствовала удалению заряда с поверхности (22 и 37 нс, рис. 2,а). Вторая вспышка аналогична первой с той лишь разницей, что разряд не выходил за пределы низковольтного электрода, потому что вне этой области не было достаточного поверхностного заряда (34 и 37 нс, рис. 2,а).

Анодонаправленный (отрицательный) разряд (рис. 2,б) развивался почти так же, как и катодонаправленный. Основное отличие – у анодонаправленного разряда была ниже скорость распространения. Это приводило к сокращению "темной" фазы разряда (с 14 по 20 нс, рис. 2,б). Когда длина стримеров превышала длину нижнего электрода, структура разряда отличалась от случая катоднонаправленного разряда (ср. рис. 2,а, 6-я нс и рис. 2,б, 14-я нс). Здесь распределение разрядов по поверхности диэлектрика было более однородным, и стримеры были менее выраженными.

В настоящее время предложены схемы SDBD, в которых исключается развитие волны обратного пробоя [37-41], что может быть важно для более эффективного управления потоком газа. Подавление обратной волны пробоя в наносекундном SDBD осуществлялось при использовании в электродных схемах «диодной» поверхности, проводящей ток только в одном направлении.

Из приведенных фотографий следует, что разряд, развивающийся вдоль поверхности диэлектрика, имеет две фазы развития: прямую и обратную волны, связанные с передним и задним фронтом импульса напряжения [32, 33, 35]. Хотя есть определенное сходство представленных на рис. 2 плазменных филаментов с хорошо изученным стримерным разрядом в объеме [42, 43], но между этими разрядами есть и принципиальное отличие. В стримерном разряде излучает только его «головка», где и находится область сильного электрического поля, а плазменный канал со слабым электрическим полем не светится. Иная ситуация наблюдается в SDBD любой полярности, где светится не только фронт ионизации, но и каналы. Последнее косвенным образом указывает на сильное электрическое поле и в каналах таких разрядов далеко за передним фронтом ионизации.

Сильное влияние SDBD на газодинамический поток связано, прежде всего, с быстрым нагревом, который определяется такими характеристиками, как энерговклад в разряд и эффективность и быстрота передачи энергии от носителей электрического тока – электронов – в тепло. Кинетика нагрева газа связана с тем, в какие степени свободы молекул передается энергия электронов. За это в слабоионизованной неравновесной плазме отвечает приведенное электрическое поле $E/n$ ($n$ – концентрация нейтральных частиц), от которого зависят средняя энергия электронов в разряде и все другие электронные характеристики [42].

Параметр $E/n$ измерялся в SDBD различной полярности методом эмиссионной спектроскопии по отношению интенсивности излучения первой отрицательной системы ионов $N_2^+$ (391.7 нм) и второй положительной системы молекул $N_2$ (337.1 нм) [32, 33, 44-46]. На рис. 3 приведены результаты таких измерений [46] в случае воздуха атмосферного давления для одиночных импульсов напряжения длительностью 20 нс и временем роста 0.5 нс. Поскольку излучение собиралось сверху над разрядом (рис. 1), то полученное электрическое поле было усреднено по толщине плазменного слоя. Здесь максимум поля достигался в первые 1-2 нс; далее поле снижалось и могло иметь второй максимум меньшей амплитуды. Электрическое поле в разряде положительной полярности было больше аналогичной величины в случае отрицательной полярности. Полученные максимальные значения $E/n$ хорошо согласуются с другими аналогичными измерениями [32, 33, 44, 45]. Параметр $E/n$ в SDBD почти не менялся при уменьшении давления от атмосферного до 200 Торр [32, 33].

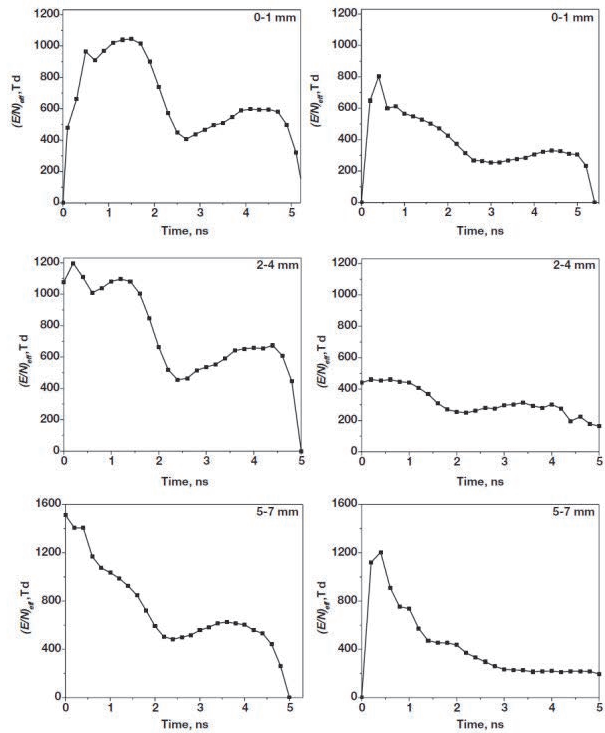

Рисунок 3. Эволюция во времени электрического поля на разных расстояниях от высоковольтного электрода в воздухе атмосферного давления при напряжении +24 кВ (левый столбец) и -24 кВ (правый столбец) [46].

К недостатку этого метода следует отнести ошибки измерения из-за сильной неоднородности плазмы SDBD. Здесь обычно излучение собирается над поверхностью диэлектрика со всего приповерхностного слоя, в котором создается плазма. Этот слой сильно неоднородный – параметры плазмы, включая электрическое поле, здесь могут меняться на расстояниях порядка $10^{-2}$ мм [47]. В результате излучение первой отрицательной системы ионов $N_2^+$ и второй положительной системы молекул $N_2$ могут приходить из разных частей слоя. В этом случае приведенное электрическое поле, восстановленное по отношению интенсивностей этих линий, может быть завышено относительно реальных значений поля в плазме [46, 47].

Для определения электрического поля в наносекундном SDBD в последнее время стали применяться подходы с использованием лазерных методов. Измерения электрического поля основывались на четырехволновом смешении [48] и генерации второй гармоники излучения пикосекундных Nd:YAG [49, 50] или фемтосекундных Ti-sapphire [51] лазеров в SDBD плазме воздуха. Эти подходы позволили следить за динамикой изменения компонентов электрического поля в различных точках разрядного промежутка. При этом электрическое поле усреднялось по ходу лазерного луча, который обычно был направлен вдоль поверхности диэлектрика перпендикулярно направлению развития разряда. В качестве примера на рис. 4 приведена динамика изменения компонентов электрического поля, измеренных в плазме SDBD на высоте 100 мкм над диэлектриком [51]. Разряд зажигался в воздухе при давлении 345 Торр под действием высоковольтных импульсов с амплитудой 20 кВ и полушириной импульса 25 нс. Частота следования импульсов составляла 20 Гц. На рисунке интервал времени 0-5 нс соответствует ситуации, когда волна ионизации еще не дошла до точки наблюдения, в которой фиксируется начальное электрическое поле, создаваемое остаточным поверхностным зарядом на диэлектрике после предыдущего импульса и потенциалом высоковольтного электрода. При прохождении волны ионизации (5-8 нс) поле растет, и в максимуме $E/n$ достигает 700 Тд (1 Тд = $10^{-17}$ В×см$^2$). После ухода ионизационной волны в точке наблюдения создается плазма, а поле во время импульса напряжения поддерживается на уровне немного ниже пробойного. При снижении напряжения на электроде к моменту времени $t$ = 22-26 нс электрическое поле также уменьшается. Оно начинает снова расти при $t$ > 30 нс, когда импульс совсем прекращается. Этот рост связан с созданием обратного электрического поля от накопленного в промежутке заряда. (Данная техника не позволяет измерять направление электрического поля.)

Таким образом, современные методы измерения электрического поля в плазме SDBD позволяют воспроизвести основные качественные закономерности развития разряда, ранее изучавшиеся с помощью скоростной съемки (см. рис. 2) и дают максимальные значения электрических полей, которые качественно согласуются с прежними измерениями методами эмиссионной спектроскопии (см. рис. 3). Отметим, что измерения поля лазерными методами до сих пор вызывают много вопросов, в частности – из-за особенностей калибровки таких измерений, предполагающих значительную экстраполяцию калибровочных зависимостей, полученных в допробойных полях, в режимы, когда поле превышает пробойное в десять и более раз. Другой сложный вопрос – в осреднении получаемых значений по релеевской области пучка. Поэтому данные измерения нужно рассматривать лишь как качественное подтверждение ранее полученных результатов.

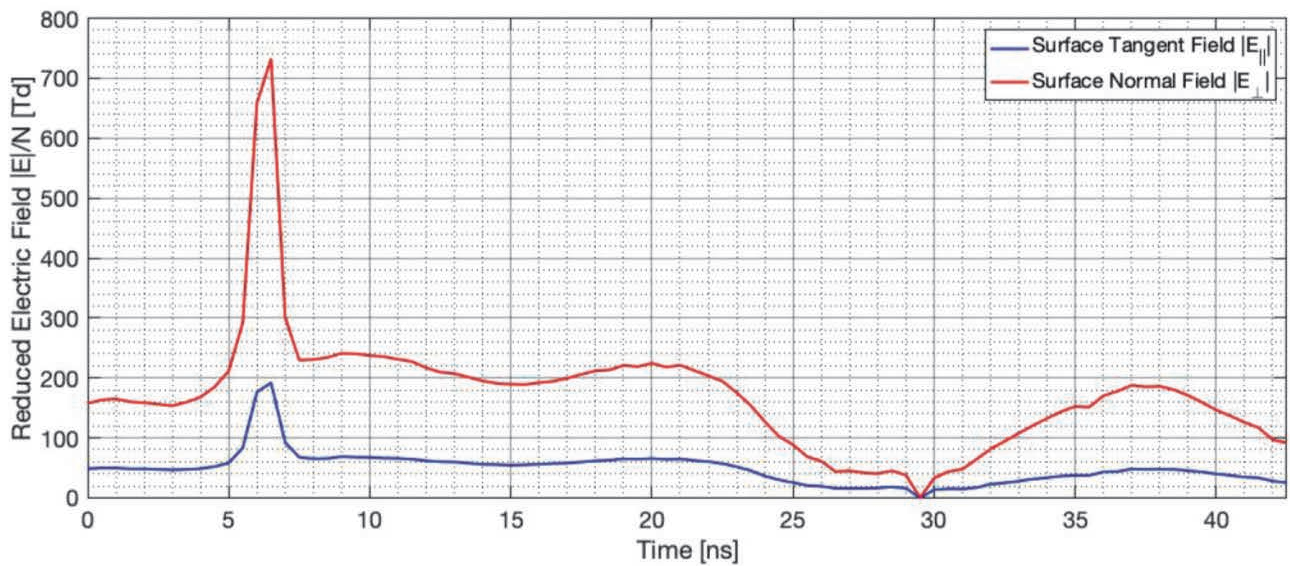

Рисунок 4. Эволюция во времени продольного и поперечного электрического поля для ns-SDBD в воздухе при давлении 345 Торр [51]. Измерения сделаны в точке, отстоящей от высоковольтного электрода на расстоянии ~ 2 мм и располагающейся на высоте 100 мкм над поверхностью диэлектрика.

Наиболее важной характеристикой быстрого нагрева газа в ns-SDBD является динамика изменения температуры газа во время разряда и в его послесвечении. Поступательная температура молекул в ns-SDBD и его послесвечении в воздухе атмосферного давления измерялась методом эмиссионной спектроскопии по излучению второй положительной системы азота (337.1 нм) [44]. Для измерения температуры в этой работе использовался дополнительный высоковольтный импульс напряжения, который "подсвечивал" распадающуюся плазму через фиксированный промежуток времени. На рис. 5 показана зависимость температуры газа от напряжения во время разрядной фазы и через 1 мкс после нее для разрядов различных полярности. В ходе эксперимента диафрагма шириной 1 мм устанавливалась таким образом, что регистрировалось излучение разряда из области шириной 1 мм, прилегающей к верхнему электроду. В случае отрицательного разряда температура оказалась несколько выше как во время разрядной фазы, так и в послесвечении. С повышением напряжения температура росла в случае обеих полярностей разряда. Из данных, представленных на рис. 5, следует, что воздух атмосферного давления можно заметно (на 150 К) нагреть уже во время разрядной стадии на временах порядка 10 нс, и что этот нагрев значительно усиливается и достигает сотен градусов через 1 мкс после импульса. Прямым доказательством быстрого нагрева в плазме SDBD также служит образование ударной волны. Распространение этой волны от области разряда фиксировалось во многих экспериментах (см., например, [32, 33, 52]).

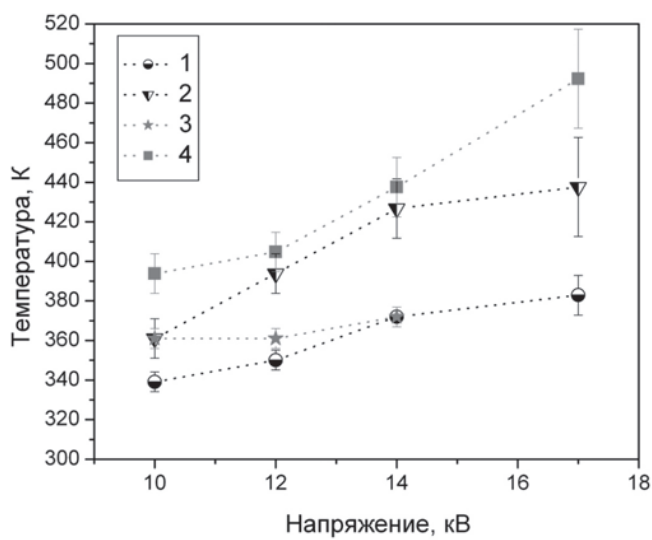

Рисунок 5. Зависимость температуры от амплитуды напряжения во время разряда (1,3) и в послесвечении разряда через 1 мкс (2, 4) для положительной (1, 2) и отрицательной (3, 4) полярностей [44].

Систематические измерения характеристик ns-SDBD, включая предельную длину и скорость распространения волны ионизации, а также толщину плазменного слоя, были выполнены при фотографировании разряда с короткой выдержкой с помощью ICCD камеры в воздухе при различных давлениях и импульсе напряжения продолжительностью 20 нс [53]. Длина распространения разряда уменьшалась от 50 до 5 мм при увеличении давления от 0.1 до 1 атм. При этом толщина плазменного слоя снижалась от 1.5 до 0.2 мм. Таким образом, увеличение давления воздуха приводит к пропорциональному снижению размеров плазменного слоя.

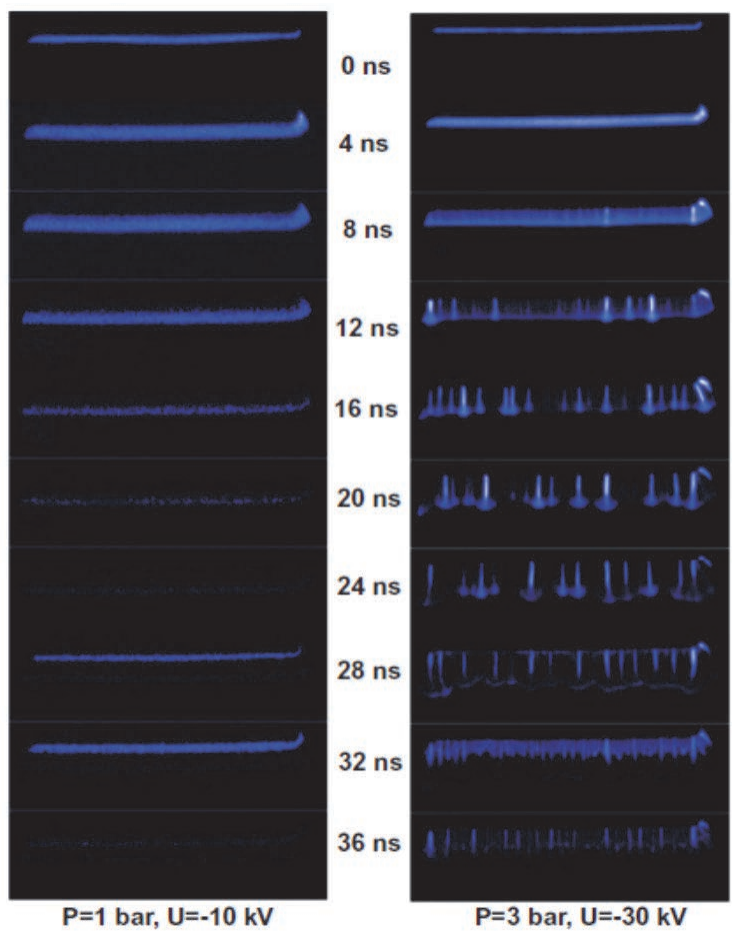

Рисунок 6. Изображения развития поверхностного наносекундного барьерного разряда в плоской геометрии с наносекундным временным разрешением [58]. Снимки при различных давлениях воздуха и амплитудах напряжения. Выдержка 0.5 нс, длительность импульса на полувысоте 25 нс.

Характеристики SDBD зависят не только от давления газа [54, 55] и параметров высоковольтного импульса, но и от свойств диэлектрического слоя, разделяющего электроды. В [50, 56] экспериментально исследовалось как вариация толщины диэлектрика и его электрических параметров влияет на характеристики разряда в воздухе атмосферного давления, прежде всего на предельную длину развития разряда. Оказалось, что рост диэлектрической проницаемости $\varepsilon$ от 2 до 35 приводит к существенному снижению длины развития разряда, в то время как изменение толщины диэлектрического слоя от 0.8 до 35 мм почти не сказывалось на длине разряда [56]. Кроме того, свойства импульсно-периодического разряда зависят от того, как быстро происходит релаксация заряда на поверхности диэлектрика. Из-за уменьшения эффекта остаточного заряда в случае использования диэлектрика с малым временем его релаксации (и при уменьшении частоты следования импульсов) волны ионизации во время разряда были более явно выраженными и характеризовались более высокими значениями максимального электрического поля и разрядного тока [50].

Представленная на рис. 2 форма развития наносекундного SDBD получила название квазиоднородной, в противовес форме с явно выраженными плазменными филаментами, наблюдаемой при повышенном энерговкладе и/или давлении [57, 58]. Эта форма разряда присуща не только импульсам наносекундной длительности и ранее встречалась при наложении высокочастотного синусоидального напряжения [47]. Воздействие SDBD на газодинамический поток зависит от режима, в котором разряд развивается. Поэтому переход от квазиоднородной формы разряда к форме с филаментами важен в практическом применении. Согласно наблюдениям [58], филаменты образовывались через несколько наносекунд после начала развития SDBD и только при отрицательной полярности напряжения (рис. 6). На рис. 7 приведена граница между областями по давлению воздуха и амплитуде напряжения между областью, где филаменты образуются, и областью, где реализуется квазиоднородная форма разряда. Подробный обзор экспериментальных работ по исследованию этого явления в наносекундном и других диапазонах SDBD приведен в [47].

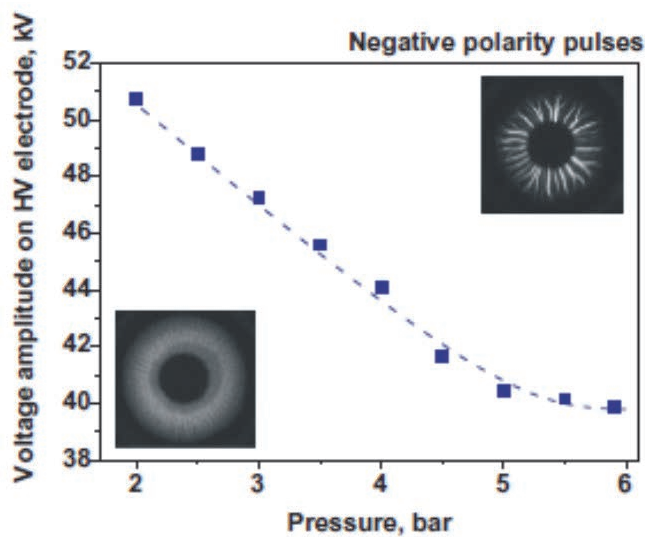

Рисунок 7. Области квазиоднородного разряда (под кривой) и разряда с филаментами (над кривой) для SDBD в воздухе при отрицательной полярности приложенного напряжения [58].

В [59] с помощью оптической эмиссионной спектроскопии исследовались свойства плазмы в филаментах SDBD. Было показано, что интенсивность излучения в плазменных филаментах воздуха превышает интенсивность излучения SDBD примерно в 50 раз. Это излучение, образуя сплошной спектр, преобладает в ультрафиолетовом диапазоне и заметно снижается в видимом и инфракрасном диапазонах. Дальнейшие экспериментальные исследования филаментов в SDBD показали, что они характеризуются более высоким (по сравнению с плазмой разряда без филаментов) значениями энерговклада и плотности и средней энергии электронов [60]. Филаментизация разряда приводит к увеличению энергии, вкладываемой в один канал (канал филаменты или стримерообразного плазменного образования, на переходном этапе), на порядок величины (рис. 8). (На резкое увеличение энерговклада в SDBD при образовании филаментов указывали и измерения [48].)

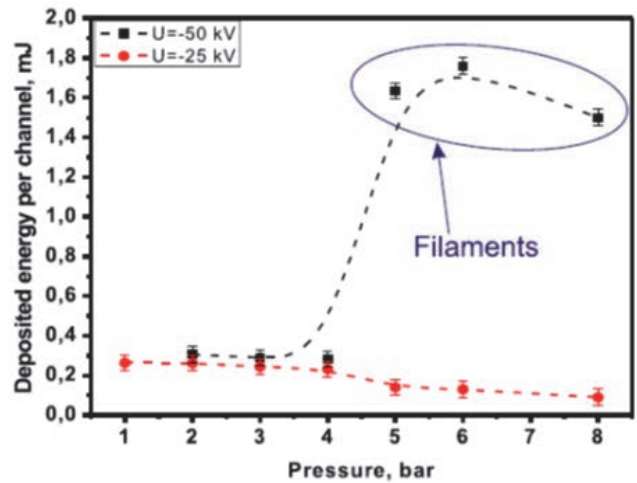

Рисунок 8. Энергия, вложенная в один плазменный канал SDBD (стример или филамента) в зависимости от давления воздуха [60]. Измерения для двух напряжений отрицательной полярности.

При этом удельный энерговклад на одну частицу аномально велик (~ 7 эВ), плотность электронов, оцененная по штарковскому уширению спектральных линий, и их температура достигают $5 \times 10^{18}$ см$^{-3}$ и ~3 эВ, соответственно. Характерное время распада плазмы составляет десятки наносекунд. В [60] предполагается, что плазма филаментов находится в локальном термодинамическом равновесии, поэтому ее относительно медленный распад связан с медленным процессом охлаждения филаментов. Главный вопрос связан с механизмом их образования. Согласно оценкам [60], большую роль в развитии филаментов играют ступенчатая ионизация и диссоциация молекул электронным ударом, а также нагрев газа при электрон-ионной рекомбинации. Сами филаменты зарождаются у поверхности высоковольтного электрода. Механизм зарождения в настоящее время не ясен. Например, при отрицательной полярности напряжения он может быть вызван автоэмиссией электронов с поверхности катода. Однако в настоящее время еще нет ясной физической картины явлений, приводящих к образованию филаментов в SDBD.

Моделирование развития всех структурных элементов SDBD представляет собой нестационарную трехмерную задачу, в которой все параметры разрядной плазмы резко меняются на малых временных и пространственных масштабах. В настоящее время количественно описать все эти явления на основе численных физических подходов не представляется возможным. Поэтому основной подход при моделировании развития SDBD основывается на двумерном (2D) приближении, в рамках которого учитывается неоднородность плазмы в направлении развития разряда и в направлении, перпендикулярном поверхности диэлектрика, а в третьем направлении плазма считается однородной [47, 64]. Тем самым рассматривается развитие плоского плазменного «листа» вдоль поверхности диэлектрика, а поперечной структурой разряда пренебрегается. Моделирование основывается на численном решении системы газодинамических уравнений для заряженных частиц (в рамках дрейфово-диффузионного приближения) и уравнении Пуассона для электрического потенциала, по аналогии с моделированием развития стримерного разряда в объеме [30]. При этом наряду с традиционными объемными процессами рождения и гибели заряженных частиц необходимо учесть процессы фотоэмиссии с поверхности, фотоионизации газа, приводящие к образованию затравочных электронов в нейтральном воздухе, а также эффекты нелокальности при описании процесса ионизации молекул электронным ударом. Остальные процессы, в том числе ионно-молекулярные процессы, описываются при атмосферном давлении в локальном приближении, когда все характеристики заряженных частиц определяются приведенным электрическим полем в данной точке и в данный момент времени. Однако при пониженных давлениях воздуха могут стать важными нелокальные эффекты и для этих процессов [61]. Для получения количественных характеристик, представляющих наибольший интерес в плазменной аэродинамике, необходимо самосогласованным образом моделировать развитие SDBD и производимых им газодинамических возмущений [62].

Из численного двумерного моделирования получается следующая качественная картина развития SDBD в воздухе [47, 64]. При положительной полярности напряжения на высоковольтном электроде разряд развивается в виде «плоского» стримера, у которого основная ионизация происходит во фронте. При отрицательной же полярности напряжения главная ионизация имеет место в катодном слое около края высоковольтного электрода, а сам разряд представляет собой особую форму нестационарного тлеющего разряда. При этом отрицательный разряд оказывается более однородным и диффузным по сравнению с положительным разрядом, у которого сильная продольная неоднородность плазмы присутствует на переднем фронте волны ионизации.

Двумерное моделирование SDBD позволяет получать разумное согласие экспериментальных данных с расчетом для интегральных разрядных характеристик: предельной длины развития разряда [62, 63], разрядного тока [62] и энерговклада [64]. Попытки сравнить расчет с экспериментом для более детальных характеристик разряда – эволюции во времени излучения из разрядной плазмы – оказываются менее успешными [63]. Это может указывать на несовершенство двумерных моделей SDBD, а также неучет в расчетах ряда важных эффектов, в частности – фотоэмиссии с поверхности диэлектрика и накопления поверхностного заряда от предыдущих разрядных импульсов.

В работах [33, 65] предпринимались попытки провести трехмерное численное моделирование развития SDBD. Однако эти попытки следует признать неудачными, поскольку при этом использовались слишком грубые расчетные сетки, что не позволило получить адекватные количественные характеристики разряда. Также на основе результатов двумерного моделирования SDBD была предложена приближенная аналитическая модель разряда [66], чтобы объяснить на качественном уровне процессы, происходящие в плазме. Эта модель позволила получить аналитические зависимости скорости распространения и длины зоны разряда от параметров диэлектрического слоя и приложенного напряжения в случае одиночного импульса.

# 3. Импульсные искровые разряды с быстрым нагревом газа

Для предотвращения перехода тлеющего и некоторых других типов газового разряда в дугу, обычно стараются ограничить разрядный ток с помощью включения в электрическую цепь высокоомных резисторов или диэлектрических слоев. Поддержание плазмы в воздухе атмосферного давления без образования дуги оказалось возможным также в высоковольтном наносекундном импульсно-периодическом разряде [67, 68]. Такой разряд развивается в сильном электрическом поле, создавая сильнонеравновесную плазму, а дуга не успевает образоваться из-за малой продолжительности импульса напряжения. При этом в каждом импульсе в сильном электрическом поле активно идет ионизация, и нарабатывается большое число метастабильных частиц, которые облегчают поддержание разряда. Кроме этого, что важно для управления потоками, происходит быстрый нагрев газа.

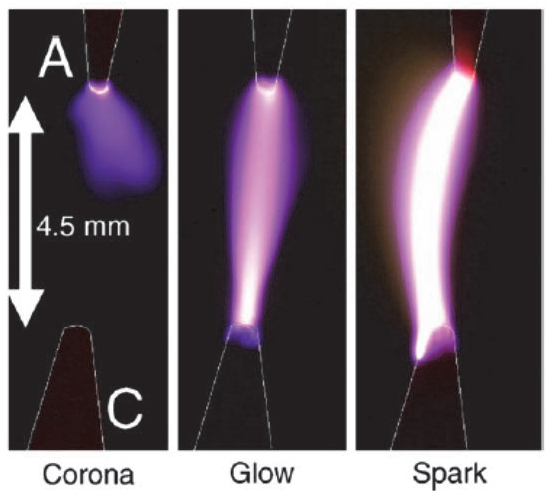

Рисунок 9. Фотографии наносекундного импульсно-периодического разряда (корона, тлеющий разряд, искра) в воздухе атмосферного давления при $T = 1000$ К и частоте $f = 10$ кГц [69]. Длина разрядного промежутка $d = 4.5$ мм. Анод и катод обозначены буквами A и C, соответственно. Напряжение на промежутке 5, 5.5 и 6 кВ (слева направо), соответственно.

В [67] экспериментально на примере разряда в промежутке длиной 1.5 мм было показано, что подача импульсов напряжения продолжительностью 10 нс, амплитудой $V = 5$ кВ и частотой следования импульсов 30 кГц позволяла зажечь импульсно-периодический разряд в воздухе при нормальных условиях. В первые несколько импульсов он по своим характеристикам был близок к импульсной стримерной короне, но потом устанавливался режим, в котором разрядный ток резко повышался до 100-150 А, энерговклад в одном импульсе возрастал до 2 мДж и температура газа до 3000 К. Оценки показали, что развитие разряда происходило при высоких значениях $E/n$, когда быстрый нагрев газа особенно эффективен. Такой режим развития разряда можно обозначить как наносекундная искра.

В воздухе, нагретом до 2000 К [68] и 1000 К [69], также наблюдался новый режим поддержания наносекундного импульсно-периодического разряда, который был назван тлеющим. На рис. 9 приведены фотографии разных форм наносекундного разряда – короны, тлеющего режима и искры – при наложении на промежуток длиной $d$ = 4.5 мм между точечными электродами импульсно-периодического напряжения $V$ с частотой $f$ = 10 кГц, длительностью 10 нс и временем роста и спада напряжения ~ 5 нс [69]. Разряд зажигался в воздухе атмосферного давления при температуре газа $T$ = 1000 К. Последовательность разных режимов при росте напряжения $V$ напоминает последовательность «корона – тлеющий разряд – дуга» в случае разряда постоянного тока [42]. При наложении наносекундного разряда с увеличением $V$ от 5 до 6 кВ происходил переход от короны к тлеющему режиму, который сменялся искрой. Во всех режимах наносекундного разряда излучение в плазме воздуха определялось второй положительной системой азота, что указывает на высокие значения приведенного электрического поля $E/n$ в разрядной плазме. В короне излучение присутствовало только у анода. При тлеющем режиме излучение становилось диффузным и занимало весь промежуток. Наконец, в искре интенсивность излучения резко росла, как и разрядный ток, который был на уровне 1 А для короны и тлеющего режима и возрастал до 20-40 А в случае искры. Энерговклад в импульсе также увеличивался: он составлял менее 1 мкДж в короне, 1-10 мкДж в тлеющем режиме и 200 мкДж – 1 мДж в искре [69]. Поскольку параметр $E/n$ был во всех режимах высоким (~ 150 - 600 Тд), то и нагрев становился заметным – он был согласно измерениям методами эмиссионной спектроскопии на уровне 200 К в первых двух режимах и достигал 2000-4000 К в искре. В пользу высоких полей здесь свидетельствовала наблюдаемая высокая интенсивность излучения второй положительной полосы $N_2$ и первой отрицательной полосы $N_2^+$. Механизмы быстрого нагрева в таких сильных полях соответствуют рассмотренным в разделе «Быстрый нагрев газа». При таких параметрах $E/n$ одним из главных каналов нагрева воздуха является выделение энергии при тушении триплетных состояний $N_2$(A, B, C) молекулами $O_2$.

Плотность электронов в плазме разряда была на уровне $10^{13}$ см$^{-3}$ в тлеющем режиме и достигала $10^{15}$ см$^{-3}$ в искре [70]. Именно поэтому искра характеризовалась намного более высокой интенсивностью излучения по сравнению с аналогичной величиной в тлеющем режиме разряда. Излучательный радиус плазменного канала в обоих случаях был ~ 1 мм. В случае искры плотность тока достигала 1 кА см$^{-2}$, что сравнимо с величинами для дугового разряда постоянного тока [42].

В тлеющем режиме сначала от анода развивался катодонаправленный стример. После замыкания промежутка он вызывал обратную волну потенциала, которая выравнивала электрическое поле в разрядном промежутке [69]. Напряжение выключалось еще до образования катодного слоя в нем, что принципиально отличает данный разряд от тлеющего разряда постоянного тока. Иная ситуация наблюдалась в наносекундной искре [70]. Здесь, как следует из анализа оптических наблюдений, разряд развивался однородно благодаря лавинной ионизации в объеме. Этот механизм пробоя принципиально отличается от развития обычной искры, инициируемой одиночным импульсом напряжения в воздухе атмосферного давления, где искровой канал обычно образуется после замыкания промежутка стримерным разрядом, развивающимся от одного из электродов [43]. Возможность однородного развития наносекундной искры при импульсно-периодическом напряжении связана с высокими ($\sim 10^{11}$ см$^{-3}$) концентрациями затравочных электронов, которые остаются от предыдущего импульса к началу следующего импульса. Оптические измерения показали, что нагрев на несколько тысяч градусов осуществляется в искровом режиме наносекундного импульсно-периодического разряда в воздухе за очень короткое время ~ 30 нс [70], что может быть важно для управления потоками с помощью плазмы наносекундного разряда.

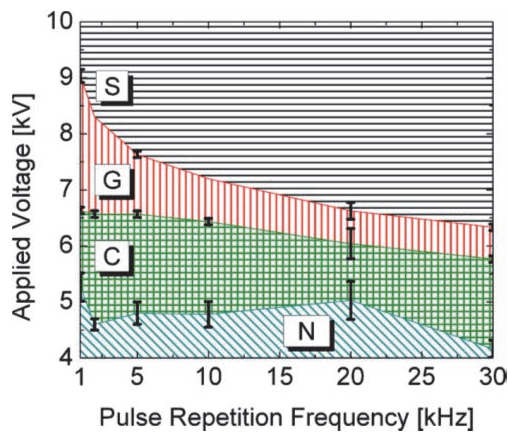

Рисунок 10. Области различных режимов наносекундного импульсно-периодического разряда в плоскости «напряжение-частота импульсов» для воздуха атмосферного давления при $T = 1000$ К и $d = 5$ мм [71]. S – искра, G – тлеющий разряд, C – корона, N – нет разряда.

В [71] подробно исследовались условия перехода между разными режимами наносекундного импульсно-периодического разряда в нагретом воздухе атмосферного давления. Экспериментально определенные области реализации разных режимов разряда в зависимости от амплитуды и частоты приложенного напряжения приведены на рис. 10 для $T = 1000$ К и межэлектродного промежутка длиной $d = 5$ мм. Уменьшение частоты следования импульсов от 30 до 1 кГц несколько расширяет и сдвигает к большим напряжениям области существования короны и тлеющего режима.

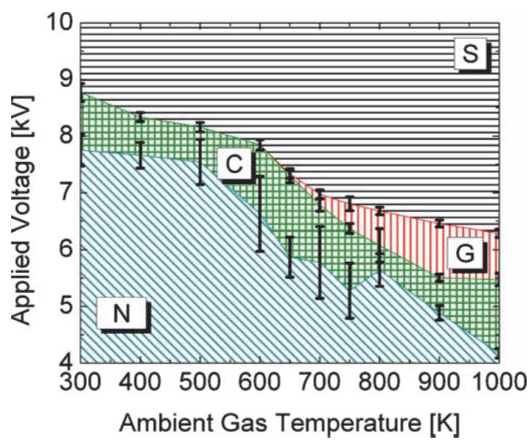

Рисунок 11. Области различных режимов наносекундного импульсно-периодического разряда в плоскости «напряжение-температура газа» для воздуха атмосферного давления при $f = 30$ кГц и $d = 5$ мм [71]. S – искра, G – тлеющий разряд, C – корона, N – нет разряда.

Кроме частоты и амплитуды импульсов напряжения в [71] также варьировались значения $T$ и $d$. На рис. 11 приведены области реализации разных режимов разряда в зависимости от напряжения и температуры газа для $f$ = 30 кГц и $d$ = 5 мм. При уменьшении $T$ увеличиваются напряжения, необходимые для зажигания короны из-за роста плотности газа при постоянном давлении. Важно то, что ниже 650 К в рассматриваемых условиях могут существовать только корона и искра, а тлеющий режим наблюдается только при $T$ > 650 К. Вариация $d$ при $T$ = 1000 К и $f$ = 30 кГц показала, что тлеющий режим может быть получен только в диапазоне 3 < $d$ < 9 мм. При других длинах промежутка наблюдались только корона и искра. Важный вопрос – можно ли получить тлеющий режим при комнатной температуре. Аналитические оценки [71] показали, что можно ожидать положительного ответа на этот вопрос при увеличении длительности высоковольтного импульса и/или при уменьшении радиуса кривизны на концах электродов.

Последующие экспериментальные исследования наносекундного импульсно периодического разряда в воздухе позволили определить характеристики создаваемой плазмы в диапазоне температур $T$ = 300 – 1000 К [72, 73], включая степень диссоциации $O_2$ (она достигала 50%), концентрации электронно-возбужденных молекул $N_2$ и динамику нагрева газа. Исследования проводились в режиме наносекундной искры. Из обработки данных по динамике нагрева и по энерговкладу в разряд удалось извлечь информацию по доле энергии, быстро передающейся в плазме воздуха в тепло, при различных значениях параметра $E/n$ (см. раздел «Быстрый нагрев газа»).

Впервые численное исследование наносекундного импульсно-периодического разряда в воздухе было выполнено в рамках 1D модели [74]. Расчеты были выполнены применительно к условиям эксперимента [67], когда в воздухе при нормальных условиях (1 атм, $T$ = 300 К) развивалась наносекундная искра. Полученные результаты качественно согласовались с экспериментом, позволив определить основные механизмы развития и поддержания разряда, а также механизмы нагрева газа. Было показано, что основную роль в разрядных процессах играл быстрый нагрев газа в сильном электрическом поле ($E/n$ = 150 – 300 Тд), связанный, прежде всего, с тушением электронно-возбужденных состояний $N_2$ на $O_2$. В результате из-за уменьшения $n$ между импульсами увеличивалось приведенное поле $E/n$ во время импульса напряжения, вызывая рост скорости ионизации, дальнейший нагрев газа и переход разряда в режим искры.

В [72] выполнено 2D параметрическое численное моделирование наносекундного импульсно-периодического разряда для условий, близких эксперименту [70]. При этом основное внимание было уделено тлеющему режиму и показано, что он реализуется, только если длительность импульса напряжения совсем немного превышает время замыкания промежутка стримерным разрядом. В противном случае в стадии после перекрытия промежутка происходят сильное энерговыделение и нагрев газа в канале, приводящие к переходу разряда в искру. Этого легко достичь в 5 мм промежутке при $T$ = 1000 К и затруднительно при комнатной температуре, что согласуется с наблюдениями [70]. Моделирование [73] последовательного развития нескольких импульсов разряда в тлеющем режиме для воздуха атмосферного давления при $T = 1000$ К показало его выход на «квазистационарный» уровень, соответствующий наблюдаемому в экспериментах с наносекундным импульсно-периодическим разрядом. При этом для развития последующих импульсов была важна концентрация заряженных частиц, оставшихся после распада плазмы к моменту начала следующего импульса. В нагретом газе она была больше из-за замедления распада плазмы с ростом T, что облегчало развитие разряда.

Еще одним типом импульсного искрового разряда является импульсная дуга, на основе которой был создан актуатор LAFPA (Localized Arc Filament Plasma Actuator). Этот тип разряда, в отличие от наносекундного импульсно-периодического разряда, с одной стороны – нашел широкое применение в плазменной аэродинамике [23-25], а с другой – почти не исследовался в отношении механизмов его развития. Впервые применять этот тип разряда для управления газодинамическими потоками было предложено в работе [75]. Разряд создавался в воздухе между двумя электродами, помещенными на расстояние в несколько миллиметров, при приложении импульсно-периодического напряжения частотой 10-300 кГц и длительностью импульсов от 1 мкс до 1 мс. Чтобы плазму не сносило высокоскоростным потоком, на поверхности тела, обтекаемого потоком, создавалось углубление, по которому и развивался разряд. Сначала при наложении напряжения происходил пробой промежутка, и образовывалась искра. После пробоя наступало резкое уменьшение напряжения на промежутке до нескольких сотен вольт, и это напряжение держалось вплоть до окончания импульса. При этом электрическое поле в канале было существенно меньше пробойного, а газ нагревался до температур 1000-2500 К. Все это указывает на то, что в основной стадии разряд поддерживался в форме импульсной дуги. Главным свойством разряда, которое оказалось наиболее ценным для управления газодинамическими потоками, был быстрый нагрев газа. Такой нагрев происходил именно во время пробоя разрядного промежутка на стадии наносекундной искры [23], когда электрическое поле характеризовалось высокими значениями $E/n$. При этом механизм быстрого нагрева определялся процессами, описанными в разделе «Быстрый нагрев газа» (прежде всего – выделением энергии при тушении возбужденных состояний $N_2(A, B, C)$ на молекулах $O_2$). Создаваемые при быстром локальном нагреве газа газодинамические возмущения и используются для воздействия на поток газа. Таким образом, хотя указанный искровой разряд внешне заметно отличается от описанного выше наносекундного импульсно-периодического разряда, но они близки в отношении процессов быстрого нагрева газа.

Еще одним типом искрового разряда, который используется для управления газодинамическими потоками благодаря быстрому локальному нагреву воздуха атмосферного давления, является филаментированный импульсный СВЧ разряд [17, 23]. Он имеет форму «клубка» плазменных филаментов, в которых температура газа поднимается до нескольких тысяч градусов на временах, лежащих в наносекундном диапазоне [76-78]. Быстрый нагрев газа в филаментах вызывает газодинамические возмущения типа слабых ударных волн, которые могут влиять на характеристики газового потока. Механизм быстрого нагрева газа в такой плазме не отличается от механизмов нагрева в описанных выше разрядах и разобранных в разделе «Быстрый нагрев газа». Отличие СВЧ разряда (как и оптического разряда) от других типов разряда, используемых для импульсного нагрева газа, состоит в том, что он позволяет производить нагрев дистанционно, в безэлектродном режиме, на значительном удалении от источника излучения электромагнитных волн.

Таким образом, с помощью описанных в данном разделе импульсных искровых разрядов в воздухе атмосферного давления можно создать в малых объемах плазму при высоких приведенных электрических полях $E/n$ и организовать быстрый нагрев газа в ней. В результате приводит генерация газодинамических возмущений, которые могут быть использованы для воздействия на основной поток.

## 4. Импульсные наносекундные оптические разряды

По сравнению с другими способами энерговклада в газ лазерный луч обладает рядом преимуществ. Это неинвазивный способ ввода энергии, позволяющий достигать высоких значений удельного энерговклада в нужном месте. При фокусировке лазерного луча происходит ионизация газа и образуется искра, что было продемонстрировано в широком диапазоне длин волн с использованием различных лазеров (газовые лазеры на $CO_2$ и эксимерных молекулах, твердотельные лазеры на рубине и неодиме) [79-81].

Образование лазерной искры, приводящее к поглощению энергии лазерного луча в газе, происходит при фокусировке луча в малом фокальном объеме. Сам процесс развивается через следующие последовательные стадии (рис. 12): появление первых электронов в результате многофотонной ионизации, лавинная ионизация газа в фокальной области, поглощение лазерной энергии газообразной плазмой, быстрое расширение плазмы, образование ударной волны и ее распространение в окружающий газ [79-81].

Имеется два механизма размножения электронов в газе под действием лазерного излучения. Первый механизм – многофотонная ионизация, в которой нейтральная частица одновременно поглощает достаточное для ионизации число фотонов m. Этот процесс записывается в виде

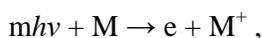
m$h\nu$ + M → e + M$^+$ ,

где М – нейтральная частица, $h\nu$ – квант света и m$h\nu$ – энергия, поглощаемая в этом процессе. При ионизации поглощаемая энергия должна превышать потенциал ионизации $I$: m$h\nu > I$. В рассматриваемом процессе плотность электронов растет со временем линейно. Ионизационный потенциал молекул в воздухе больше 12 эВ, а кванты видимого и ближнего ИК-диапазона обычно порядка или меньше 1 эВ. Поэтому ионизация молекул в лазерном луче должна быть многофотонной.

Второй механизм ионизации основан на поглощении лазерного излучения свободными электронами в процессах обратного тормозного рассеяния. Данные процессы являются обратными по отношению к тормозному испусканию квантов света при рассеянии высокоэнергичных электронов на нейтральных частицах. Если электроны приобретают энергию, превышающую потенциал ионизации нейтральных частиц, то столкновения между ними могут приводить к ударной ионизации

$$e + M \rightarrow 2e + M^+.$$

Этот процесс вызывает лавинную ионизацию, в которой плотность электронов увеличивается экспоненциально со временем, поскольку все новые электроны также нагреваются и начинают участвовать в ионизационных столкновениях.

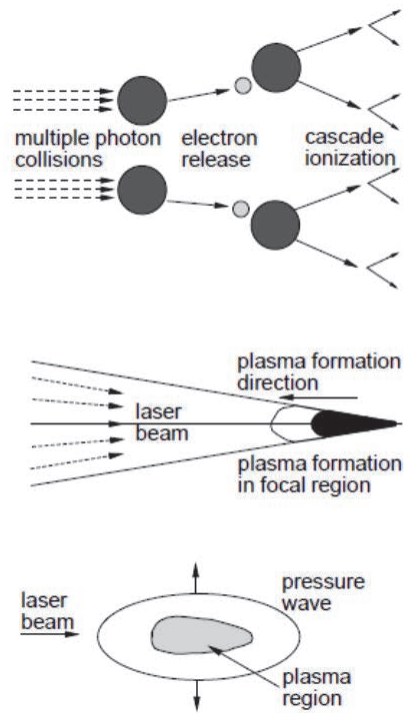

Рисунок 12. Процессы при оптическом пробое газа лазерным излучением [82].

При развитии лавинной ионизации образующаяся плазма начинает эффективно поглощать энергию лазерного луча. Сначала лавинная ионизация имеет место в малой фокальной области с максимальной интенсивностью излучения (рис. 12,б). Но поглощение лазерного излучения не ограничено фокальной областью, где идет первичная лавинная ионизация. Когда степень ионизации достигает большой величины в плазме, лавинная ионизация начинается и в областях с меньшей интенсивностью излучения, прилегающих к плазменной области. Новая область становится ионизованной и, переставая быть прозрачной для излучения, начинает поглощать его энергию. Таким образом, поглощающая область непрерывно смещается навстречу лазерному лучу, создавая волну нагрева газа. Тепловое излучение из сильно нагретой области поглощается холодным газом, который начинает в свою очередь поглощать свет [79]. Этот процесс называется «радиационным» механизмом распространения волны ионизации, который приводит к распределению лазерной энергии по большой области, что снижает максимальную температуру в газе. Кроме теплового излучения плазмы существуют и другие механизмы, приводящие к ионизации газа в слоях, примыкающих к уже образованной плазме. Это молекулярная теплопроводность и нагрев газа ударной волной. Указанные механизмы распространения плазмы могут реализоваться в различных комбинациях в зависимости от конкретных условий [79].

В финальной стадии поглощения энергии лазерный импульс заканчивается, а плазма распадается благодаря электрон-ионной рекомбинации. Фокальная область и прилегающие области оказываются существенно нагретыми. Здесь резко возрастает давление, что приводит к уменьшению плотности газа при распространении ударной волны и волн разрежения. В дальнейшем на месте, где происходил оптический пробой, может образоваться тороидальный вихрь, который создает поток газа в направлении распространения лазерного луча.

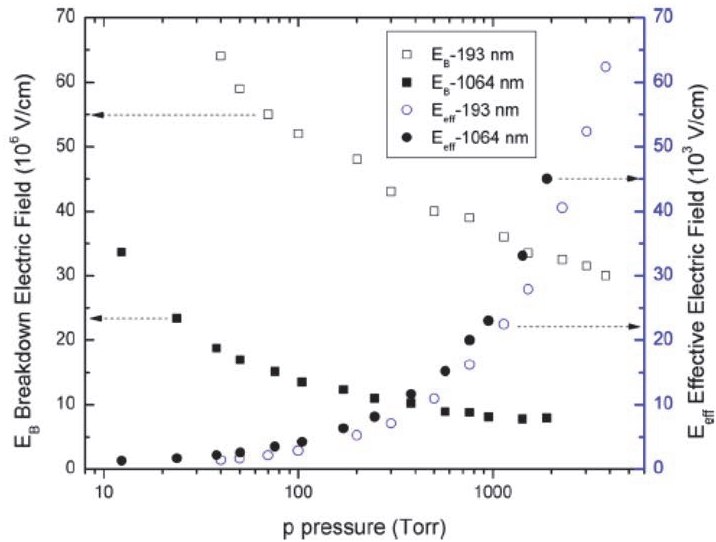

Рисунок 13. Зависимость порога пробоя ($E_B$) и эффективного электрического поля ($E_{eff}$) в воздухе от его давления для лазерного излучения с длинами волн 193 нм [83] и 1064 нм [84].

Порог пробоя газов лазерным излучением зависит от многих параметров [79-81]: характеристик газовой среды (состав газа, его давление, наличие примесей, в том числе – аэрозоля и твердых микрочастиц), характеристик излучения (длина волны, продолжительность импульса) и размеров фокальной области. В качестве примера на рис. 13 приведены результаты измерения порогового электрического поля $E_B$ для пробоя в воздухе в широком диапазоне давлений под действием излучения в УФ диапазоне (193 нм, ArF лазер, длительность импульса 20 нс) [83], и под действием излучения в ИК диапазоне (1064 нм, Nd:YAG лазер, длительность импульса 6 нс) [84]. На рисунке также приведено эффективное поле

$E_{eff} = E_B v_c/(v_c^2 + \omega^2)^{1/2}$,

которое характеризует эффективность нагрева электронов в переменном электрическом поле. Здесь $v_c$ – транспортная частота столкновений электронов с другими частицами (нейтральными частицами и ионами), а $\omega$ – частота электромагнитного поля. В рассматриваемых условиях $\omega \gg v_c$ и $E_{eff} \approx E_B v_c/\omega$. Как следует из рис. 13, порог пробоя растет с уменьшением давления и длины волны излучения. Высокие пороги пробоя при малых давлениях объясняются малым числом столкновений электронов с нейтральными частицами в этом случае [79]. Здесь для ионизации молекул электроны должны нагреться в лазерном поле до порога ионизации, а нагрев электронов происходит в столкновениях (обратное тормозное рассеяние).

Уменьшение длины волны на порядок величины приводит к увеличению порога пробоя примерно в 10 раз. В то же время поле $E_{eff}$ оказывается почти не зависящим от длины волны излучения. Отсюда следует, что рост порогового поля $E_B$ с уменьшением длины волны связан с менее эффективным нагревом электронов в лазерном поле на более высоких частотах. Зависимости порога пробоя воздуха от других характеристик газа и лазерного луча, их физическое объяснение и описание теоретических подходов для численного моделирования можно найти в [79-81].

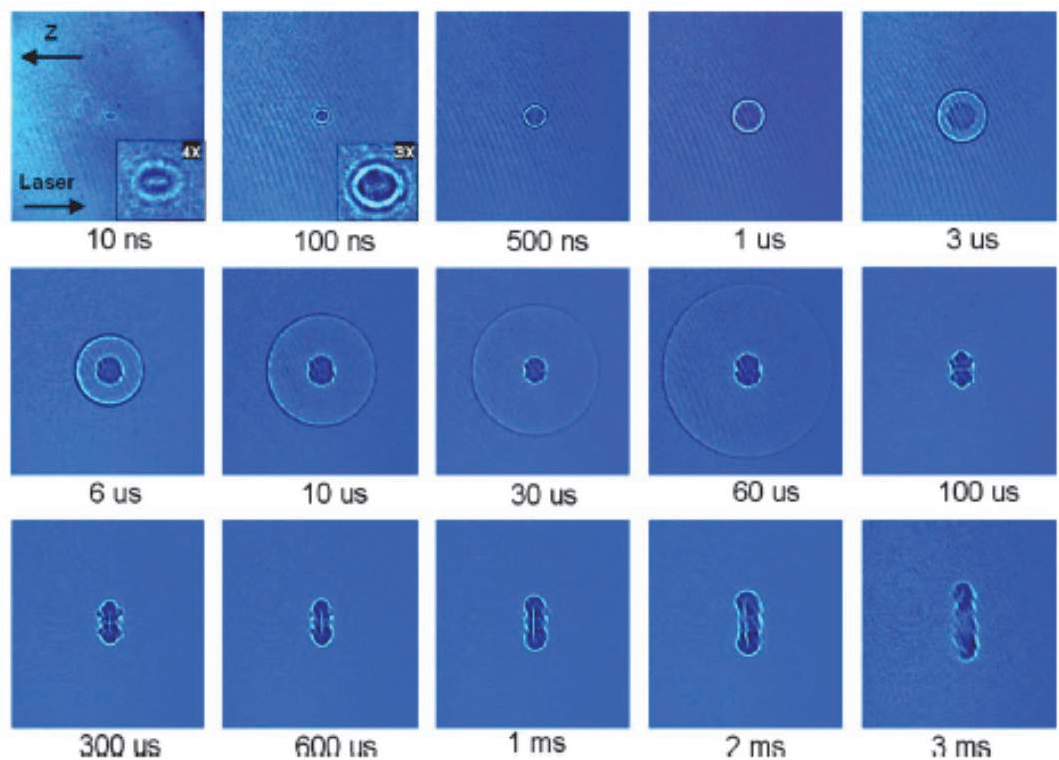

Рисунок 14. Динамика эволюции плазмы, создаваемой в воздухе лазерным лучом с энергией 135 мДж [83]. Размер каждого изображения 1.3×1.3 см$^2$.

На рис. 14 показаны фотографии, демонстрирующие развитие процессов, происходящих при лазерном пробое в воздухе атмосферного давления. Пробой осуществлялся под действием ArF лазера (193 нм) с энергией 135 мДж и длиной импульса 20 нс. Фотографии делались с помощью ICCD камеры (выдержка 10 нс) при подсветке объекта исследования непрерывным лазером. На рисунке направление распространения лазерного луча слева направо. Начало отсчета времени соответствовало началу лазерного импульса. Сразу после пробоя ($t < 25$ нс) наблюдалось интенсивное тормозное излучение из плазмы. На ранних временах ($t \leq 100$ нс) происходило расширение нагретой области из фокальной области, в результате чего плазма приобретала форму, слегка вытянутую вдоль распространения лазерного луча. На временах порядка 1 мкс ударная волна отделялась от нагретого плазменного ядра, и далее она уже расширялась отдельно. Распространение ударной волны наблюдалось на временах вплоть до 60 мкс, после чего эта волна слабела и выходила за пределы области наблюдения. Примерно в это же время происходила значительная деформация излучающей плазменной области. Отсюда можно заключить, что ударная волна играла важную роль в устойчивости расширяющегося нагретого плазменного ядра. Когда волна удалялась от плазменного ядра достаточно далеко, более холодный воздух проникал в нагретое ядро в направлении распространения лазерного луча. Дальнейшее расширение плазменной области приводило на временах порядка 2 мс к формированию тороидальной структуры вихревого типа. Такая эволюция плазменной области и создаваемой ею ударной волны при развитии в потоке газа может оказывать значительное влияние на его характеристики.

Образование лазерной искры в газе и вызываемые ею газодинамические явления представляют собой сложную комбинацию различных физических процессов, приводящих к генерации на наносекундных временах полностью или частично ионизованной равновесной плазмы с температурой в диапазоне $10^4 – 10^5$ К [79-81]. Их теоретическое описание обычно ограничивается рассмотрением какой-либо одной-двух стадий формирования лазерной искры, в то время как остальные стадии либо не рассматриваются, либо учитываются в сильно упрощенном виде. Например, моделирование газодинамических эффектов, вызываемых энерговкладом от лазерной искры в потоке воздуха, часто моделируется в предположении о сверхбыстром распределенном энерговкладе в заданном объеме газа [17].

Энергию лазерного излучения можно вводить в газ не только в одном импульсе, но и более сложным образом. В [85, 86] была предложена новая концепция ввода лазерной энергии при комбинации двух лазерных импульсов, одного – в УФ диапазоне (266 нм), а второго – в ближнем ИК (1064 нм). Это позволило избежать оптического пробоя, управляя одновременно и энерговкладом, и нагревом газа. Здесь первичные электроны создавались за счет многофотонной ионизации в поле УФ импульса. После этого с заданной (~10 нс) задержкой прикладывался ИК импульс, в котором электроны нагревались так, чтобы обеспечить лавинную ионизацию газа. Плазма, создаваемая двойным лазерным импульсом, оказывается неравновесной и требует для своего создания намного меньше энергии, чем в случае с генерацией равновесной плазмы одним лазерным импульсом. Кроме того, использование двух лазерных импульсов позволяет эффективно управлять плотностью плазмы и нагревом газа. Например, при обычном лазерном пробое воздуха под действием одного лазерного импульса образуется полностью ионизованная плазма с температурой, превышающей 10000 К [87, 88]. Использование же двух импульсов в УФ и ИК диапазонах позволяет создавать плазму с разной ($10^{-4} – 10^{-2}$) степенью ионизации и в широком (400 – 10000 К) диапазоне температур [88]. Возможность конроля за вкладываемой энергией может оказаться особенно ценной в задачах управления потоками газа. Кроме того, этот подход оказался востребованным и для стимулированного плазмой воспламенения и горения [86, 89].

В связи с прогрессом в развитии фемтосекундных лазеров [90] в последнее время также появились предложения воздействовать на сверхзвуковой поток с помощью тонких плазменных филаментов, образуемых при пробое воздуха под действием тераваттного лазерного излучения фемтосекундной длительности [91]. При этом создается сильнонеравновесная плазма высокой (~$10^{17}$ см$^{-3}$) концентрации, при распаде которой в наносекундном диапазоне происходит импульсный нагрев газа [92].

## 5. Распад неравновесной плазмы воздуха

При высоких приведенных электрических полях $E/n$ значительная часть энергии, вкладываемой в разряд, тратится на ионизацию газа. Этот канал передачи энергии от электронов становится основным при $E/n$ ~ 1000 Тд и выше. Во время распада плазмы в послесвечении наносекундного разряда происходит рекомбинация заряженных частиц. В результате выделяется значительная энергия, часть которой идет в тепло. Чтобы количественно описать этот процесс, приводящий к быстрому нагреву газа, необходимо иметь представление о столкновительных процессах, благодаря которым и происходит распад плазмы воздуха [93, 94].

В наносекундном разряде в воздухе во время высоковольтного импульса при ионизации молекул электронным ударом образуются только первичные ионы. В случае сухого воздуха это ионы $N_2^+$ и $O_2^+$, а во влажном воздухе к ним добавляются ионы $H_2O^+$. Первичные ионы вступают в ионно-молекулярные процессы, что приводит к образованию новых ионов. Это происходит во время распада плазмы, а если давление газа достаточно велико, то может происходить и во время наносекундного разряда. В стадии разряда все заряженные частицы нагреваются в сильном электрическом поле. Средняя энергия электронов значительно больше средней энергии ионов. Например, в плазме воздуха при $E/n$ = 200 Тд средняя энергия электронов $\varepsilon_e$ ~ 6 эВ [42], а средняя энергия ионов $\varepsilon_i$ ~ 1 эВ [61]. При этом энергетические распределения и электронов, и ионов в слабоионизованной плазме являются неравновесными.

При выключении электрического поля, когда происходит распад плазмы, заряженные частицы также начинают остывать, и их средняя энергия стремится к средней энергии нейтральных частиц, а энергетические распределения стремятся к равновесному максвелловскому распределению. Ионы в воздухе охлаждаются намного быстрее электронов, поскольку для ионов, в отличие от электронов, каждое упругое столкновение с нейтральной частицей приводит к сильному уменьшению кинетической энергии. При атмосферном давлении характерное время релаксации средней энергии ионов в воздухе составляет доли наносекунды [61]. Время релаксации средней энергии электронов значительно больше. Важно, что на охлаждение электронов влияют и процессы электрон-ионной рекомбинации из-за так называемого эффекта рекомбинационного нагрева [95]. Поскольку процессы релаксации энергии электронов и их плотности оказываются связанными друг с другом, то в общем случае их моделирование надо производить самосогласованным образом. Процессы электрон-ионной рекомбинации являются непороговыми, и коэффициенты рекомбинации обычно являются степенными функциями от эффективной температуры электронов $T_e = 2/3\ \varepsilon_e$. Поэтому при теоретическом изучении распада плазмы обычно достаточно следить за релаксацией $T_e$, и нет необходимости численно моделировать релаксацию энергетического распределения электронов.

В типичных для плазменной аэродинамики условиях распад сильнонеравновесной слабоионизованной плазмы воздуха происходит при давлениях порядка атмосферного от начальных концентраций электронов в диапазоне $10^{14}$-$10^{15}$ см$^{-3}$. Экспериментально исследовать распад плазмы в таких условиях сложно, поскольку разрядная плазма, как правило, является сильно неоднородной, а наиболее интересными является наносекундный и микросекундный временные диапазоны. Здесь оказываются малопригодными стандартные подходы измерения плотности плазмы типа зондового или СВЧ. В качестве возможного подхода здесь можно рассматривать экспериментальное исследование распада плазмы воздуха после высоковольтного разряда при пониженных давлениях (1 – 10 Торр), где разряд развивается однородно. Разработка кинетических схем для процессов, определяющих распад плазмы в этих условиях, и их верификация на экспериментальных данных позволяет с помощью этих схем проводить численное моделирование свойств распадающейся плазмы в практически важных условиях при повышенных давлениях воздуха. Важно, чтобы удельный энерговклад в высоковольтный разряд был достаточно мал. При этом генерация заряженных частиц не сопровождается наработкой большого количества возбужденных частиц, способных сильно повлиять на кинетику распада плазмы. Выполнение этого условия сильно упрощает теоретический анализ экспериментальных данных. Именно такие условия реализовывались, например, в экспериментах [96, 97], где с помощью СВЧ интерферометрии изучалась динамика изменения плотности электронов при распаде плазмы в воздухе (1 – 10 Торр) после высоковольтного наносекундного разряда с удельным энерговкладом 0.002 – 0.02 эВ/мол. Численное моделирование [96, 97] показало, что распад плазмы в этих условиях можно описать на основе достаточно простой кинетической схемы, в которой основным каналом гибели заряженных частиц является электрон-ионная рекомбинация с положительными ионами, в том числе с кластерными ионами, образующимися в послесвечении разряда.

В литературе имеются подробные кинетические схемы, позволяющие описать гибель заряженных частиц и ионно-молекулярные процессы при распаде плазмы в сухом и влажном воздухе [98-104]. Современные данные по коэффициентам диссоциативной электрон-ионной рекомбинации, которая обычно является основным каналом гибели заряженных частиц в рассматриваемых условиях, приведены в [105].

Анализ имеющихся данных по константам скорости элементарных процессов указывает на следующие особенности, которые должны быть учтены при моделировании распада плазмы высоковольтного наносекундного разряда в воздухе атмосферного давления.

1. Поскольку коэффициенты электрон-ионной рекомбинации зависят от эффективной электронной температуры, то надо самосогласованным образом моделировать гибель заряженных частиц и релаксацию средней энергии электронов после прекращения разряда при их столкновениях с молекулами.

2. Коэффициент электрон-ионной рекомбинации для кластерных ионов на порядок величины больше, чем для простых молекулярных ионов. Поэтому надо учесть образование кластерных ионов во время распада плазмы и рекомбинацию электронов с этими ионами.

3. Образование кластерных ионов – тройной процесс, скорость которого пропорциональна $N^2$ в пределе малых давлений газа и пропорциональна $n$ в пределе больших давлений. При атмосферном давлении для образования кластерных азотных и кислородных ионов реализуется режим малых давлений, а для образования гидратированных ионов $H_3O^+(H_2O)_k$ при этом имеет место режим высоких давлений [102, 104].

4. В случае гидратированных ионов $H_3O^+(H_2O)_k$ (доминирующие положительные ионы при распаде плазмы во влажном воздухе) имеется два типа экспериментов по измерению коэффициентов рекомбинации, которые дают сильно (до порядка величины) различающиеся результаты. Это прямые измерения коэффициентов рекомбинации по скорости распада плазмы с гидратированными ионами при не слишком малых (> 1 Торр) давлениях [106-109] и эксперименты при аномально малых ($10^{-11}$ Торр) давлениях в ионных накопительных кольцах [110-112], где измеряются сечения рекомбинации, а коэффициенты рекомбинации получаются при интегрировании сечений по энергетическому распределению электронов. Нет сомнений в надежности измерений обоих типов. В настоящее время есть единственное объяснение этого различия – предполагается, что диссоциативная рекомбинация электронов с гидратированными ионами идет по разным механизмам при очень малых и повышенных давлениях [113]. Отсюда следует практический вывод – при моделировании распада плазмы во влажном воздухе атмосферного давления необходимо использовать коэффициенты рекомбинации для гидратированных ионов, полученные и верифицированные в [106-109].

5. При достаточно высокой (> $10^{12}$ см$^{-3}$) плотности электронов и не слишком больших значениях их температуры $T_e$ становится важным процесс тройной электрон-ионной рекомбинации с третьим телом – электроном. Этот процесс достаточно хорошо изучен для атомарных ионов [114, 115] и практически не исследован для молекулярных ионов. В то же время высказывалось предположение о том, что скорость тройной рекомбинации для молекулярных ионов (в частности, для ионов $O_2^+$ [96, 97, 116]) может быть значительно больше, чем для атомарных ионов, и что в этих случаях имеются разные зависимости коэффициентов рекомбинации от $T_e$.

Для иллюстрации особенностей распада плазмы высоковольтного разряда на рис. 15 и 16 приведены результаты численного нульмерного моделирования эволюции эффективной температуры электронов и плотностей заряженных частиц в послесвечении разряда в сухом и влажном воздухе атмосферного давления. Кинетическая схема столкновительных процессов, описывающих распад плазмы в этом случае, взята из [96, 97, 109]. При этом были учтены описанные выше особенности в столкновительных процессах. Как и в [96, 97, 109], считалось, что плазма является однородной, и что энерговклад в разряд настолько мал, чтобы можно было пренебречь влиянием возбужденных частиц на кинетику распада плазмы. Полагалось, что электрическое поле в конце разряда исчезает на временах, гораздо меньших времени релаксации $T_e$ и плотности плазмы. Расчеты были выполнены при начальной плотности электронов $n_e(0) = 10^{15}$ см$^{-3}$ (типичные значения для высоковольтных наносекундных разрядов типа SDBD) и $T = 300$ К.

Релаксация $T_e$ при распаде плазмы моделировалась на основе численного решения уравнения [95-97]

$$\frac{dT_e}{dt} = -\nu_\varepsilon(T_e - T) - \frac{2}{3}T_e^2\left(\frac{dk_3}{dT_e}n_e n_i + \frac{dk_2}{dT_e}n_i\right) - \frac{2}{3}I k_3 n_e n_i \qquad (1)$$

совместно с уравнениями баланса для электронов и ионов в нульмерном приближении. Здесь $\nu_\varepsilon$ - частота релаксации энергии электронов в столкновениях с молекулами, $n_i$ – плотность ионов, $k_3$ – коэффициент тройной электрон-ионной рекомбинации, $k_2$ – коэффициент парной диссоциативной рекомбинации электронов с ионами, $I$ – энергия, приобретаемая свободными электронами при тройной электрон-ионной рекомбинации. Уравнение (1) записано для случая одного сорта положительных ионов. Если их много, то по соответствующим членам в правой части (1) производится суммирование. Значение $I$ полагалось равным 0.136 эВ [117]. Использованные в расчете частоты $\nu_\varepsilon$ для $O_2$, $N_2$ и $H_2O$ взяты из [109]. Предполагалось, что из-за малого энерговклада в разряд колебательное возбуждение молекул мало. Если это нет так, то в уравнении (1) необходимо учитывать передачу энергии между электронами и колебательными степенями свободы молекул, что может привести к дополнительному нагреву электронов в послесвечении разряда [118, 119].

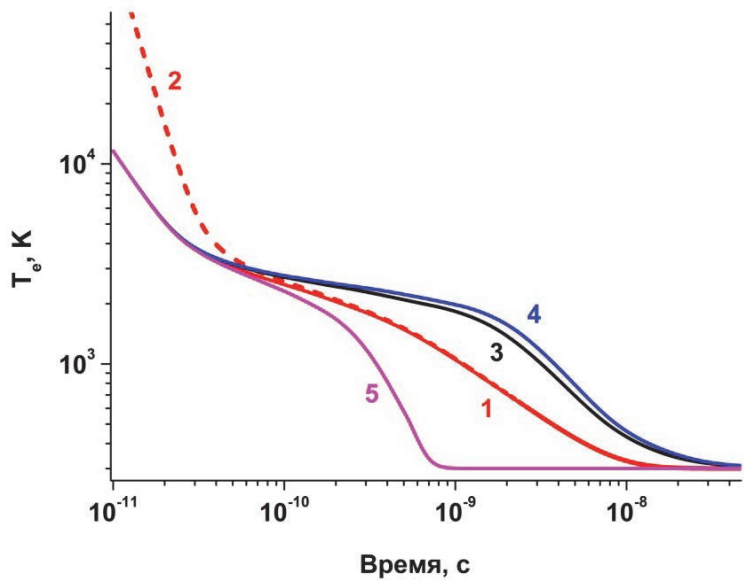

Рисунок 15. Динамика релаксации эффективной температуры электронов в воздухе атмосферного давления. 1 – сухой воздух, $T_e(0) = 1$ эВ, без рекомбинационного нагрева; 2 – сухой воздух, $T_e(0) = 10$ эВ, без рекомбинационного нагрева; 3 – сухой воздух, $T_e(0) = 1$ эВ, с учетом рекомбинационного нагрева при $I = 0$; 4 – сухой воздух, $T_e(0) = 1$ эВ, с учетом рекомбинационного нагрева при $I = 0.136$ эВ; 5 – воздух с 2% $H_2O$, $T_e(0) = 1$ эВ, без рекомбинационного нагрева. Расчеты для $n_e(0) = 10^{15}$ см$^{-3}$.

Второй и третий члены в правой части (1) описывают рекомбинационный нагрев, который вызван следующими причинами. Во-первых, при диссоциативной и тройной электрон-ионной рекомбинации более эффективно вступают в реакцию электроны с меньшей энергией. В результате из-за убыли «холодных» электронов средняя энергия электронов повышается, что и характеризует второй член в правой части (1). Во-вторых, в процессе тройной электрон-ионной рекомбинации часть выделяемой при этом процессе энергии передается свободным электронам, нагревая их (третий член в правой части (1)).

а)

б)

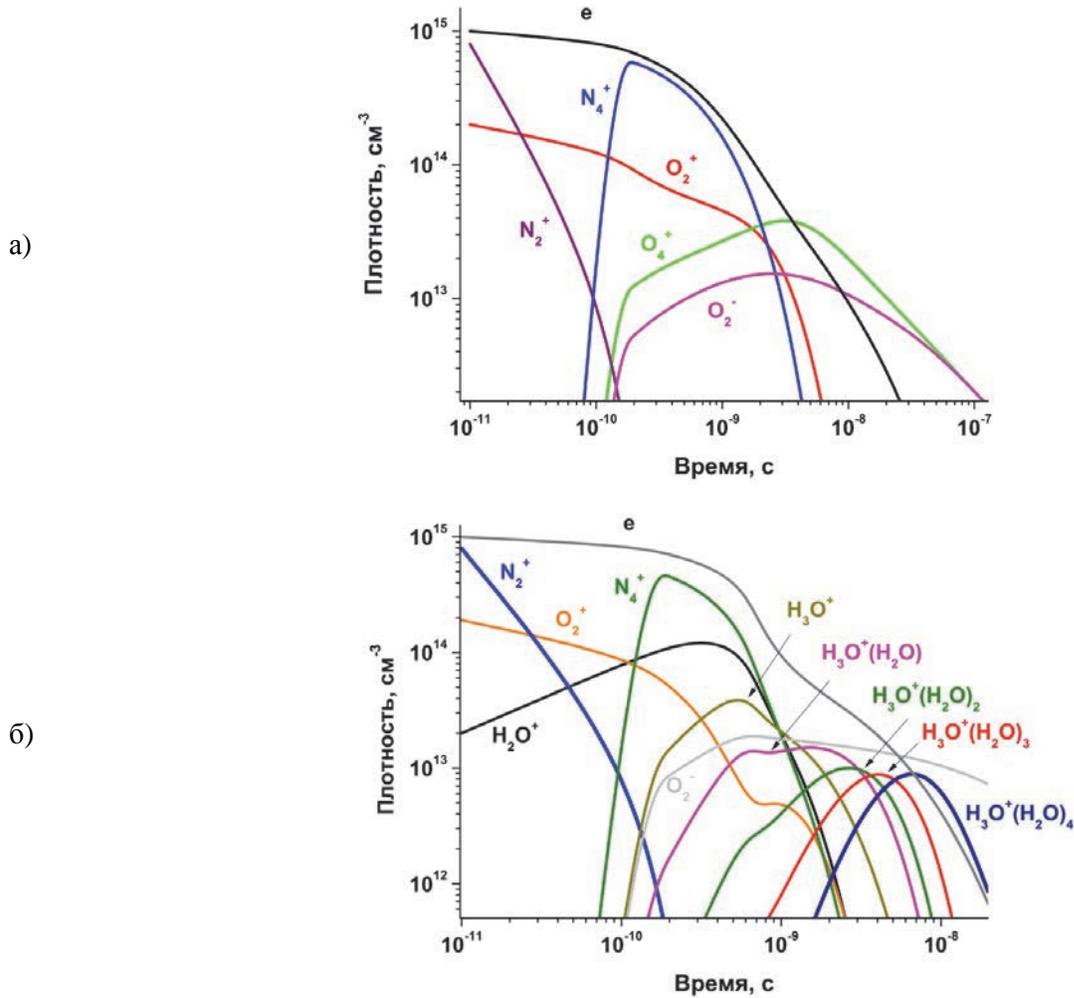

Рисунок 16. Эволюция во времени плотности электронов и ионов при распаде плазмы в сухом (а) и влажном (2% $H_2O$) воздухе (б) при атмосферном давлении. Расчеты для $n_e(0) = 10^{15}$ см$^{-3}$.

На рис. 15 приведены результаты расчета эволюции во времени эффективной температуры электронов $T_e$ в воздухе атмосферного давления. Поскольку начальные значения $T_e$ в высоковольтном разряде могут меняться в широком диапазоне, то расчеты были проведены для $T_e(0) = 1$ и 10 эВ. Как следует из рисунка, значение $T_e(0)$ важно только в начальный промежуток времени ($t < 0.05$ нс), а дальнейшее охлаждение электронов не зависит от того, насколько они нагреты вначале. Это связано с тем, что при высоких энергиях релаксация электронов определяется их неупругими столкновениями с возбуждением электронных состояний молекул, для которых частоты $\nu_\varepsilon$ велики. Эти процессы эффективны в отношении охлаждения электронов. При меньших же энергиях становятся основными процессы колебательного и вращательного возбуждения молекул электронным ударом, в которых охлаждение электронов при столкновении с $N_2$ и $O_2$ не так эффективно. Поэтому основное время охлаждения электронов до комнатной температуры определяется именно релаксацией электронов с малой энергией.

Как следует из рис. 15, в сухом воздухе время релаксации $T_e$ составляет 10 нс, если рекомбинационный нагрев не учитывать. При учете этого эффекта релаксация $T_e$ существенно замедляется при $T_e < 3000$ К, когда из-за падения частоты $v_\varepsilon$ первый член в правой части уравнения (1) резко уменьшается. В результате время полной термализации электронов, когда $T_e = T$, увеличивается и достигает 30 нс, что в несколько раз больше времени термализации при отсутствии рекомбинационного нагрева. Во влажном воздухе существенно увеличивается скорость релаксации энергии электронов именно при $T_e < 3000$ К из-за эффективного колебательного и вращательного возбуждения электронным ударом молекул $H_2O$ с большим постоянным дипольным моментом. Это приводит к уменьшению времени термализации более чем на порядок величины – до 0.7 нс.

Данные по эволюции плотности заряженных частиц при распаде плазмы в сухом воздухе (рис. 16 а) свидетельствуют о том, что хотя в начале этой стадии присутствуют только простые ионы $N_2^+$ и $O_2^+$, но распад плазмы происходит уже тогда, когда доминирующими являются кластерные ионы $N_4^+$ и $O_4^+$. Именно с ними происходит рекомбинация электронов в послесвечении высоковольтного разряда в сухом воздухе атмосферного давления. При этом первичные молекулярные ионы превращаются в кластерные ионы в процессах

$$N_2^+ + N_2 + M \rightarrow N_4^+ + M, \qquad (2)$$

$$N_2^+ + O_2 \rightarrow N_2 + O_2^+, \qquad (3)$$

$$N_4^+ + O_2 \rightarrow 2N_2 + O_2^+, \qquad (4)$$

$$O_2^+ + O_2 + M \rightarrow O_4^+ + M, \qquad (5)$$

а гибель электронов при уменьшении их плотности на два порядка величины осуществляется при диссоциативной электрон-ионной рекомбинации

$$e + N_4^+ \rightarrow N_2 + N_2, \qquad (6)$$

$$e + O_4^+ \rightarrow O_2 + O_2, \qquad (7)$$

$$e + O_2^+ \rightarrow O + O. \qquad (8)$$

Заметный вклад в гибель заряженных частиц дают и процессы тройной электрон-ионной рекомбинации типа

$$e + A^+ + e \rightarrow A + e. \qquad (9)$$

Когда плотность электронов падает до $10^{13}$ см$^{-3}$, то становится важным тройное прилипание электронов к молекулам $O_2$

$$e + O_2 + M \rightarrow O_2^- + M, \qquad (10)$$

где M = O$_2$. На поздних стадиях распада включается кинетика отрицательных ионов, которые гибнут в процессах ион-ионной рекомбинации. Если во время разрядной стадии нарабатывается большое количество атомов и возбужденных частиц, то при столкновении с ними становятся возможными процессы отлипания электронов от отрицательных ионов, что существенно замедляет распад плазмы [32].

Присутствие молекул H$_2$O во влажном воздухе приводит к изменению состава положительных ионов в послесвечении разряда – здесь начинают господствовать кластерные (гидратированные) ионы H$_3$O$^+$(H$_2$O)$_k$ (рис. 16,б), скорость диссоциативной рекомбинации для которых быстро растет с $k$ и становится заметно большей, чем у кластерных ионов N$_4^+$ и O$_4^+$. Поэтому увеличение влажности воздуха приводит к ускорению распада плазмы. Этому способствует и эффективное прилипание электронов в процессе (10) для M = H$_2$O.

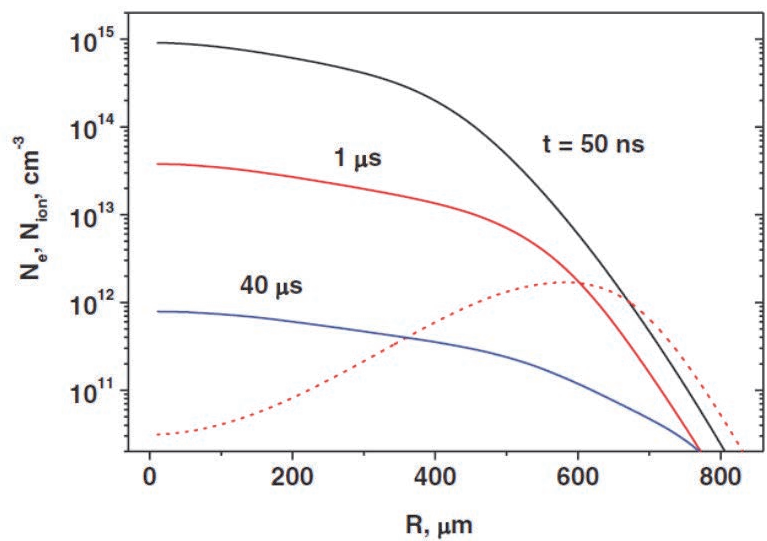

Рисунок 17. Радиальные профили электронов (сплошные линии) и отрицательных ионов (штриховая линия) при распаде плазмы наносекундного искрового разряда в воздухе атмосферного давления, нагретого до 1500 К. профиль ионов соответствует 1 мкс. Расчеты [127] для условий эксперимента [128].

Энергия диссоциации кластерных ионов и энергия связи электронов в отрицательных ионах (энергия сродства нейтральных частиц к электрону) малы (~ 1 эВ и меньше) по сравнению с потенциалами ионизации нейтральных частиц. Поэтому при развитии разряда в нагретом воздухе или при существенном нагреве последнего во время разрядной стадии резко замедляется образование кластерных ионов и ускоряется их диссоциация и отлипание электронов от отрицательных ионов. Наиболее хорошо этот эффект для импульсных наносекундных разрядов в воздухе изучен применительно к стримерному разряду в длинных промежутках (см. [120-123] и ссылки там). Перечисленные эффекты в нагретом воздухе для стримерной плазмы и ее распада получили как экспериментальное, так и численное подтверждение на основе имеющихся кинетических моделей для положительных и отрицательных моделей.

В наносекундных разрядах с высоким (~ 1 эВ на молекулу) удельным энерговкладом степень диссоциации молекул $O_2$ в воздухе оказывается больше 10% и может достигать 80%. Это происходит в наносекундном импульсном [124, 125] и импульсном периодическом (частота повторения 10-30 кГц) искровом разрядах [72, 70] в коротких (1-7 мм) разрядных промежутках, а также в капиллярном разрядах [126]. Присутствие большого количества атомов O в газе может существенно замедлять распад плазмы в послесвечении таких разрядов из-за разрушения положительных кластерных ионов и отлипания электронов от отрицательных ионов при столкновении с атомами. В качестве примера на рис. 17 приведены результаты численного 1D моделирования [127] распада плазмы в условиях эксперимента [128], где изучался распад плазмы наносекундной искры. Здесь вблизи оси разряда гибель заряженных частиц определялась электрон-ионной рекомбинацией, в то время как гибель электронов за счет их прилипания к молекулам $O_2$ была компенсирована быстрым отлипанием на атомах O. И только вдали от оси, где атомов O мало, концентрация отрицательных ионов заметно возрастала. В результате радиальный профиль для этих ионов оказывался немонотонным.

## 6. Быстрый нагрев воздуха в сильном электрическом поле

В слабоионизованной газоразрядной плазме энергия от электрического поля передается основным носителям электрического тока – электронам. Далее эта энергия в столкновениях с нейтральными частицами, в основном неупругих, перераспределяется по разным степеням свободы этих частиц. Большая часть энергии в конце концов переходит в тепло. Но скорость нагрева газа сильно зависит от того, через какие степени свободы происходит этот переход. Например, релаксация энергии, запасенной в колебаниях молекул, осуществляется в воздухе атмосферного давления на временах миллисекундного диапазона и более. В задачах управления потоками с помощью плазмы наибольший интерес представляет гораздо более быстрый нагрев газа – на наносекундном масштабе времен. Такой тип нагрева, называемый по устоявшейся в литературе терминологии быстрым нагревом [129-134], применительно к плазме воздуха и будет рассмотрен ниже.

Распределение энергии по разным степеням свободы молекул и механизмы быстрого нагрева в разрядной плазме воздуха зависят от приведенного электрического поля *E/n*. При *E/n* < 20 Тд быстрый нагрев определяется упругими столкновениями электронов с молекулами и возбуждением их вращательных состояний с последующей быстрой (за несколько столкновений) *RT* релаксацией. При более высоких значениях *E/n* этот механизм дает вклад в быстрый нагрев не более 3% от общего энерговклада. Еще первые эксперименты показали, что доля энергии, быстро переходящая в тепло в плазме воздуха, растет с *E/n* и достигает *10-15%* при *E/n* порядка 100 Тд (см. ссылки в [135, 136]).

Для задач плазменной аэродинамики наибольший интерес представляет быстрый нагрев в более высоких полях, вплоть до 1000 Тд. Количественное исследование быстрого нагрева в наносекундных высоковольтных разрядах в воздухе высокого (~ 1 атм) давления представляет собой сложную задачу, поскольку плазма в этом случае обычно сильно неоднородна.

В экспериментах [137-139] исследовался быстрый (на временах менее 1 мкс) нагрев воздуха при инициировании наносекундного скользящего поверхностного разряда. На основе обработки результатов измерений был сделан вывод о том, что доля разрядной энергии, быстро превращающейся в тепло, растет с увеличением давления от 25 до 230 Торр примерно от 15 до 60%. При этом электрическое поле в разряде не измерялось, что не позволяет определить каналы, по которым происходил быстрый нагрев в этих экспериментах. Оценочные значения приведенного электрического поля лежали в диапазоне 100-1000 Тд [139].

Более последовательные подходы одновременного измерения доли энергии, быстро передаваемой в тепло, и электрического поля при высоких значениях $E/n$ были использованы в [140, 141]. В этих работах, во-первых, исследовался быстрый нагрев воздуха по измерению скорости ударной волны при ее распространении в плазме высоковольтного импульсного наносекундного разряда при давлении $p$ = 20 Торр, когда разряд был однородным. Из обработки результатов был сделан вывод о том, что доля энергии, быстро передаваемая в тепло, составляет 36-40% на временах порядка $\tau$ = 50 мкс, что соответствует произведению $p\tau$ ~ 1 атм×мкс. Энерговклад в разряд, согласно оценкам, осуществлялся при $E/n$ ~ 600 Тд. Во-вторых, с помощью методов эмиссионной спектроскопии изучался быстрый нагрев в импульсном наносекундном SDBD при $p$ = 1 атм, который в этом случае создавал неоднородную плазму. Измерения делались как во время наносекундного разряда, так и через 1 мкс после его окончания, для чего на электроды подавался дополнительный слабый диагностический импульс напряжения, который инициировал свечение из плазмы, но почти не приводил к дополнительному нагреву газа. Из обработки полученных результатов было получено, что во время разряда доля энергии, быстро передаваемая в тепло, составляла 30-40%, а через 1 мкс после разряда она достигала 55-65%. При этом приведенное электрическое поле, оцененное из измерений методом эмиссионной спектроскопии, составляло 800-900 Тд. Результаты измерений [141] в SDBD для воздуха атмосферного давления были обобщены в [142] на диапазон давлений 300-750 Торр и на другие смеси $N_2:O_2$. Доля энергии, быстро передаваемой в тепло, оставалась постоянной в пределах ошибки измерений при уменьшении давления воздуха до 300 Торр, и заметно снижалась при уменьшении содержания кислорода в смесях $N_2:O_2$.

Высоковольтный импульсный наносекундный разряд, развивающийся в виде быстрой ионизационной волны в воздухе пониженного (2-7 Торр) давления однородным образом, также использовался в [143] для изучения эффективности быстрого нагрева в его послесвечении. При этом температура газа измерялась по излучению второй положительной системы азота. Из обработки результатов был сделан вывод о том, что на временах 50-100 мкс ($p\tau$ = 0.15 – 0.9 атм×мкс) выделяется в тепло примерно 24% от вложенной в разряд энергии. Основной энерговклад в разряд здесь осуществлялся при $E/n$ = 200 – 400 Тд.

Быстрый нагрев воздуха атмосферного давления при температуре газа $T$ = 1000-1500 К экспериментально исследовался в повторяющемся наносекундном импульсном разряде в коротком (4 мм) разрядном промежутке между двумя острыми электродами [72]. Электрическое поле в промежутке во время основного энерговклада менялось в диапазоне 200-300 Тд. Динамика изменения газовой температуры контролировалась методами эмиссионной спектроскопии по второй и первой положительным системам азота. Анализ полученных данных показал, что доля вложенной в разряд энергии, превращающаяся в тепло за 20 нс, составляла 21%.

Похожая экспериментальная установка была использована в [73] для изучения быстрого нагрева в воздухе с помощью шлирен-съемки, показывающей развития радиальных неоднородностей в разрядном промежутке при наносекундном повторяющемся разряде и создаваемой им расходящейся цилиндрической ударной волне. Из сравнения результатов одномерного моделирования газодинамических эффектов и шлирен-фотографий была определена доля энергии, быстро передаваемая в тепло. Через 50 нс после разряда она составила 25% при $E/n$ = 164 Тд и возросла до 75% при 270 Тд. При этом начальное давление было атмосферным, а начальная газовая температура лежала в диапазоне 300-1000 К.

Таким образом, многочисленные эксперименты свидетельствуют о том, что эффективность быстрого (на временах менее 1 мкс при 1 атм) нагрева воздуха в разрядной плазме увеличивается с ростом приведенного электрического поля $E/n$ и давления газа. При высоких (> 400 Тд) электрических полях энергия, быстро передаваемая в тепло, становится в плазме воздуха того же порядка, что и энергия, вложенная в разряд.

Кинетические модели, описывающие быстрый нагрев в воздухе при высоких значениях $E/n$, типичных для SDBD, рассмотрены в [141, 144]. Обе эти модели являются экстраполяцией модели [136], предложенной для умеренных приведенных электрических полей, на область высоких $E/n$. В [136] (см. также [145]) быстрый нагрев газа для $E/n$ в диапазоне 80-200 Тд был объяснен в основном выделением избыточной энергии при диссоциации $O_2$ электронным ударом

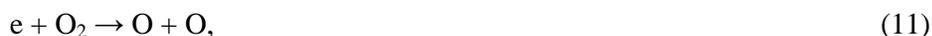
$e + O_2 \rightarrow O + O,$ \hfill (11)

при тушении электронно-возбужденных молекул $N_2$ в столкновениях с $O_2$

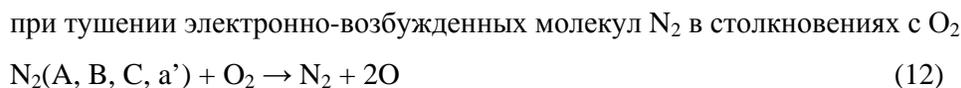
$N_2(A, B, C, a') + O_2 \rightarrow N_2 + 2O$ \hfill (12)

и при тушении состояния $O(^1D)$

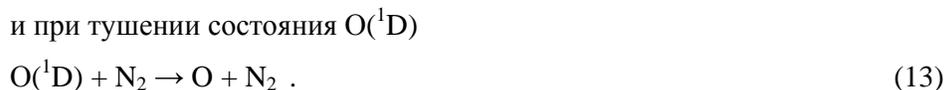
$O(^1D) + N_2 \rightarrow O + N_2$ . \hfill (13)

Для задач плазменной аэродинамики наибольший интерес представляет быстрый нагрев в полях, типичных для SDBD, где они достигают 1000 Тд. При таких высоких полях становится важным возбуждение более высоких электронных состояний молекул и их ионизация электронным ударом [136, 141, 144]. Здесь энергия быстро переходит в тепло при тушении этих состояний

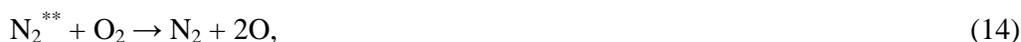
$N_2^{**} + O_2 \rightarrow N_2 + 2O,$ \hfill (14)

при диссоциации молекул азота через эти состояния

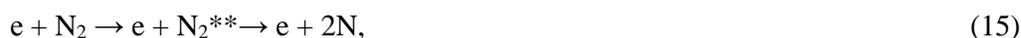
$e + N_2 \rightarrow e + N_2^{**} \rightarrow e + 2N,$ \hfill (15)

и в процессах рекомбинации заряженных частиц между собой. Спектр положительных ионов в сухом воздухе достаточно сложный, и он меняется в процессе распада плазмы наносекундного разряда (см. раздел о распаде плазмы). Преобладающими здесь являются ионы $O_2^+$, $N_2^+$, $O_4^+$ и $N_4^+$, а основным процессом рекомбинации электронов с ионами является диссоциативная рекомбинация:

$$e + O_2^+ \rightarrow O + O, \qquad (16)$$

$$e + N_2^+ \rightarrow N + N, \qquad (17)$$

$$e + O_4^+ \rightarrow O_2 + O_2, \qquad (18)$$

$$e + N_4^+ \rightarrow N_2 + N_2. \qquad (19)$$

Во влажном воздухе ионный состав усложняется из-за их гидратации. На поздних стадиях распада плазмы в воздухе атмосферного давления электроны гибнут в результате прилипания электронов к молекулам $O_2$. Обычно это тройное прилипание с образованием ионов $O_2^-$:

$$e + 2O_2 \rightarrow O_2^- + O_2. \qquad (20)$$

В дальнейшем отрицательные ионы рекомбинируют с положительными в процессах типа

$$O_2^- + AB^+ (+M) \rightarrow O_2 + AB (+M) \qquad (21)$$

с большим выделением энергии, которая может переходить в тепло, а частично – и на внутренние степени свободы молекул или тратиться на их диссоциацию. Выделяемая при рекомбинации заряженных частиц энергия в значительной степени зависит от продуктов этого процесса.

Модели [141, 144] и моделирование на их основе быстрого нагрева в воздухе при высоких *E/n* отличаются друг от друга главным образом тем, какие берутся конечные продукты в реакциях (18), (19) и (21), а также состоянием продуктов для некоторых процессов тушения электронно-возбужденных молекул. В [141] был использован подход, в котором при отсутствии надежной информации по продуктам реакции предполагалось, что вся выделяемая при этом энергия переходит в тепло, а возбуждением и диссоциацией продуктов пренебрегалось. При этом модель [141] позволяла получить оценку сверху для эффективности быстрого нагрева воздуха, которая подтверждается существующими экспериментальными данными по эффективности быстрого нагрева при высоких приведенных полях и давлениях [141, 142, 225]. В [144] использовался подход, где при отсутствии надежных данных по продуктам реакций степень возбуждения и диссоциации этих продуктов определялась из дополнительных соображений, иногда не очень обоснованных. Это уменьшило эффективность быстрого нагрева воздуха за счет образования возбужденных частиц и диссоциации молекул в продуктах рекомбинации заряженных частиц и некоторых процессов тушения электронно-возбужденных молекул. Однако ряд используемых при этом допущений требует дополнительной проверки.

На рис. 18 приведена вычисленная в [141] доля вложенной в разряд энергии, быстро передаваемой в тепло в плазме воздуха, в зависимости от приведенного электрического поля. Там же для сравнения приведены имеющиеся экспериментальные данные, а также результаты расчета этой величины из [145], и по простой аппроксимационной формуле, предложенной в [144]. Согласно этой формуле доля энергии, быстро переходящей тепло, равна в воздухе 30% от доли электронной энергии, затрачиваемой на возбуждение электронных состояний, диссоциацию и ионизацию молекул. Между всеми расчетами наблюдается хорошее согласие при не слишком больших (< 400 Тд) $E/n$. При более высоких полях результаты расчета [141] при 1 атм оказываются заметно выше других показанных на рис. 18 расчетов. Модель [144] многократно использовалась для расчета быстрого нагрева в наносекундных разрядах в воздухе при различных экспериментальных условиях [143, 144, 127]. При этом по динамике нагрева воздуха получалось хорошее согласие расчетов с экспериментом. Но в этих условиях всегда основной энерговклад происходил при $E/n$ < 400 Тд, когда, согласно рис. 19, разные модели дают близкие результаты.

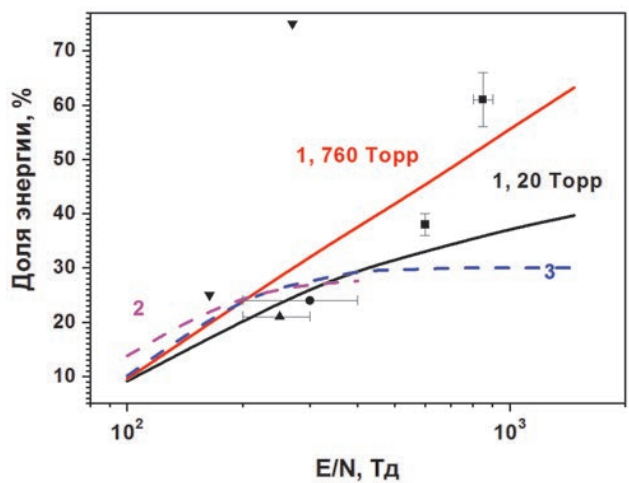

Рисунок 18. Результаты расчета доли энергии, быстро передаваемой в тепло, в воздухе в зависимости от приведенного электрического поля [141] (кривые 1). Расчеты выполнены при разных давлениях для плотности электронов $n_{ef} = 10^{14}$ см$^{-3}$. Для сравнения даны результаты расчета [145] (кривая 2) и по аппроксимационной формуле, предложенной в [144] (кривая 3). Также символами приведены экспериментальные данные: квадраты [141], круги [143], треугольники [72] и перевернутые треугольники [73].

Экспериментальные данные, относящиеся к быстрому нагреву воздуха при более высоких $E/n$, получены только в работах [141, 142, 225]. Сравнение результатов этого эксперимента и расчета по предложенной в этой работе модели для эволюции во времени доли энергии разряда, быстро переходящей в тепло, показывает разумное согласие при разных давлениях с учетом того, что плазма при 1 атм является сильнонеоднородной, а расчетная модель является нульмерной и дает лишь оценку сверху, поскольку в ней для ряда процессов не учитываются возможные возбуждение и диссоциация продуктов. Отметим, что модель быстрого нагрева, предложенная в [144], не позволяет описать единственные имеющиеся на данный момент в диапазоне высоких приведенных полей и давлений экспериментальные данные [141, 142, 225], что, по-видимому, свидетельствует о низкой надежности модели [144] в этих условиях.

Выше рассматривался разряд в воздухе, состав которого моделировался смесью $N_2:O_2$ в случае сухого воздуха и смесью $N_2:O_2:H_2O$ в случае влажного. Ситуация сильно меняется в разрядах с высоким удельным энерговкладом, когда степень диссоциации молекул $O_2$ оказывается больше 10%. Такие условия реализуются в наносекундном импульсном [124, 125] и импульсном периодическом (частота повторения 10-30 кГц) искровом разрядах [72, 70] в коротких (1-7 мм) разрядных промежутках, а также в капиллярном разрядах [126], где исследовался, в том числе и быстрый нагрев холодного или предварительно нагретого воздуха. При этом типичные приведенные электрические поля лежали в диапазоне 150-400 Тд, удельный энерговклад составлял ~ 1 эВ на молекулу, а степень диссоциации $O_2$ лежала в диапазоне 10-80%. В этом случае атомы O становились одним из основных компонентов газовой смеси ($N_2:O_2:O$ для сухого воздуха). Моделирование [127] для этих условий быстрого нагрева воздуха при различных (300-1500 К) начальных газовых температурах показало, что существенный вклад в быстрый нагрев в рассматриваемых условиях дают процессы тушения возбужденных молекул $N_2$ на атомах O

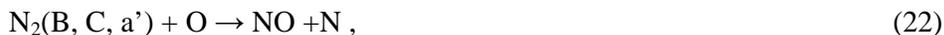

$$N_2(B, C, a') + O \rightarrow NO + N,  \qquad (22)$$

а также

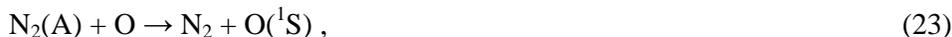

$$N_2(A) + O \rightarrow N_2 + O(^1S),  \qquad (23)$$

и

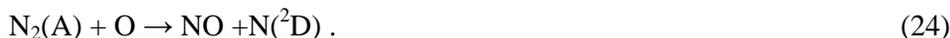

$$N_2(A) + O \rightarrow NO + N(^2D).  \qquad (24)$$

Важно, что здесь и доля молекул NO также оказывалась высокой, что сказывалось на быстром нагреве, наработке других окислов азота и колебательном возбуждении молекул.

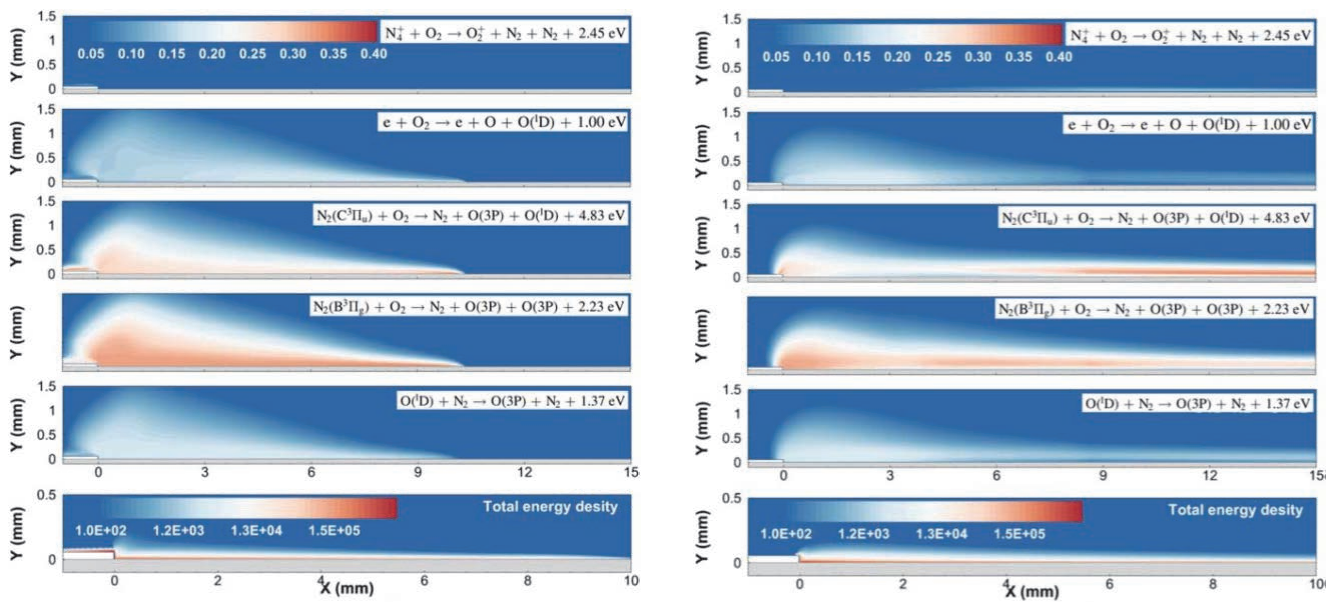

Рисунок 19. Пространственное распределение долей быстрого нагрева по основным пяти каналам (по отношению к быстрому нагреву по всем каналам) для отрицательного (слева, $V = -24$ кВ) и положительного (справа, $V = +24$ кВ) разрядов [146]. На последнем рисунке в каждой группе приведено распределение удельного энерговклада (Дж/м$^3$).

Поскольку в реальных разрядах приведенное электрическое поле *E/n* сильно неоднородно и нестационарно, то представляет особый интерес провести численное моделирование кинетики нагрева газа с учетом изменения этого параметра, когда одновременно меняются и электрическое поле, и плотность газа (за счет газодинамических эффектов). Такие расчеты были выполнены в [146] для SDBD в воздухе атмосферного давления, где для этого использовалась 2D модель, ранее развитая в [62], а быстрый нагрев газа рассматривался в рамках модели [144]. На рис. 19 приведены полученные в [146] пространственные распределения вкладов в быстрый нагрев в плазме SDBD для пяти основных каналов. Расчеты сделаны для напряжения 24 кВ как положительной, так и отрицательной полярности. Там же приведены пространственные распределения удельного энерговклада в разряд. Для обеих полярностей основными каналами быстрого нагрева являлось тушение триплетных состояний $N_2(C, B)$ на молекулах $O_2$ (процесс (12)), а удельный энерговклад был наибольшим около высоковольтного электрода и уменьшался по мере удаления от него. Плазма положительного разряда распространялась дальше, и в ней осуществлялся быстрый нагрев газа на больших расстояниях от высоковольтного электрода. Такое распределение энерговклада качественно согласуется с экспериментальными данными [53], где была исследована динамика развития наносекундного SDBD в диапазоне 12-760 Торр при напряжении на высоковольтном электроде ±10 и ±20 кВ.

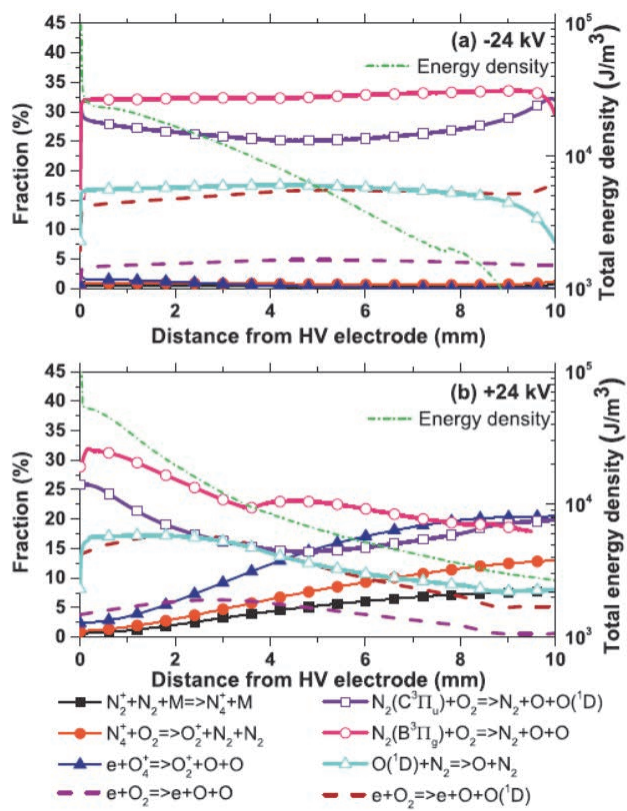

Рисунок 20. Доли быстрого нагрева по основным пяти каналам (по отношению к быстрому нагреву по всем каналам) и удельный энерговклад для отрицательного (*V* = - 24 кВ) и положительного (*V* = +24 кВ) разрядов в зависимости от расстояния от высоковольтного электрода [146]. Данные соответствуют высоте 25 мкм над поверхностью диэлектрика.

Таким образом, имеющиеся экспериментальные исследования свидетельствуют о большой эффективности быстрого нагрева в разрядной плазме воздуха при высоких (вплоть до ~ 1000 Тд) электрических полях и давлениях. Основные механизмы, обеспечивающие быстрый нагрев воздуха в рассматриваемых условиях, в основном поняты, что позволяет осуществлять детальное моделирование процессов, контролирующих управление газовыми потоками с помощью плазмы.

# 7. Управление конфигурацией ударных волн и распределением нагрузок

## 7.1. Основные механизмы управления обтеканием сверхзвукового объекта

Концепции методов снижения аэродинамического сопротивления и теплообмена с помощью энерговыделения перед головной ударной волной на осесимметричных затупленных телах, движущихся в атмосфере со сверхзвуковой скоростью, обсуждались и изучались в течение долгого времени [147, 148].

Имеются два подхода для объяснения природы взаимодействия сверхзвуковых потоков с плазмой. Первый подход описывает процесс на основе теплового механизма и связан с выделением энергии в разряде. Во втором механизме предполагается, что основной причиной воздействия плазмы является передача импульса от заряженных частиц, ускоренных электрическим полем, нейтральной компоненте газовой среды. При этом важную роль играет нескомпенсированный объемный заряд, который и приводит к появлению объемной силы.

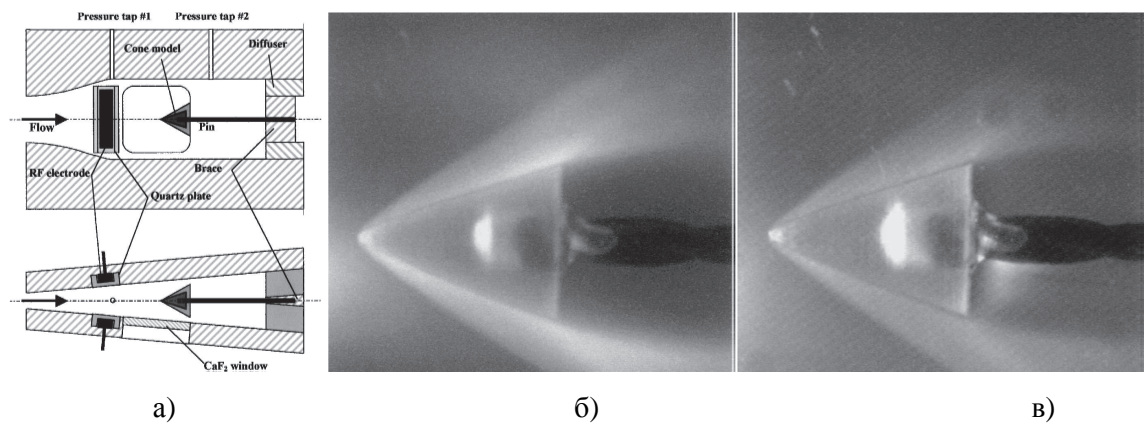

а) б) в)

Рисунок 21. а) Схема рабочей секции ВЧ разряда в сверхзвуковом потоке: вид сбоку и вид сверху; б) – в) фотографии сверхзвукового обтекания конуса при визуализации с помощью плазмы, созданной ВЧ разрядом и разрядом постоянного тока в потоке смеси 30% $N_2$ -70% He при $P_0$ = 250 Тор: б) включен только разряд постоянного тока мощностью 175 Вт; в) добавлен ВЧ разряд мощностью 250 Вт [149].

В вышеупомянутой работе [7] показано, что при распространении ударной волны в разрядной трубке скорость волны увеличивается, а амплитуда падает. Авторы сравнивали полученное увеличение скорости с расчетными значениями, соответствующими тепловыделению разряда. Полученное различие (1200-1300 м/с вместо 900 м/с) было объяснено дополнительным нагревом за счет релаксации колебательной энергии и тушения электронно-возбужденных состояний нейтральных частиц, а также образованием двойного слоя перед ударной волной. Однако в более поздних работах такая интерпретация была поставлена под сомнение [149, 150]. В качестве основной причины воздействия плазмы на газовый поток рассматривался его нагрев в разряде.

В работе [149] обсуждалось воздействие неравновесной плазмы на ударную волну, образующуюся перед конусом в сверхзвуковом потоке при числе Маха $M = 2.5$. Эксперименты были выполнены в аэродинамической трубе с использованием сверхзвукового потока плазмы вокруг квазидвумерного клина (рисунок 21). Из наблюдений следовало, что косая ударная волна может быть существенно ослаблена при воздействии плазмы поперечного ВЧ-разряда. Полученное ослабление ударной волны оказалось соответствующим повышению температуры в пограничных слоях из-за нагрева в разряде.

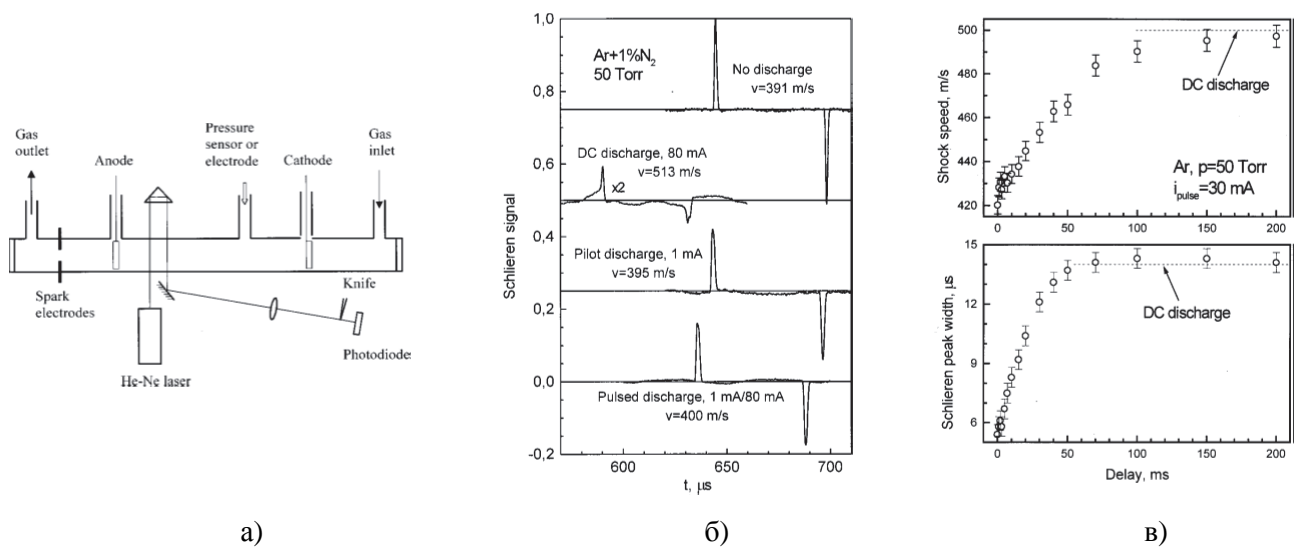

а) б) в)

Рисунок 22. а) Схема экспериментальной установки для исследования распространения ударной волны в тлеющих разрядах; б) шлирен сигналы и скорость ударной волны в тлеющем разряде в Ar + 11% $N_2$ при 50 Торр: без разряда, разряд постоянного тока при 80 мА, слабый (~ 1 мА) инициирующий разряд и импульсный разряд; в) эволюция скорости ударной волны и ширина шлирен сигнала в фиксированной точке внутри разряда с временной задержкой между началом разряда и запуском ударной волны [151].

В работе [150] изучалось распространение ударных волн от искры в плазме тлеющего разряда в аргоне и аргон-азотных смесях. Было получено прямое доказательство теплового механизма взаимодействия ударной волны с плазмой для импульсных разрядов. При субмиллисекундной задержке между началом разряда и запуском ударной волны параметры плазмы достигали стационарных значений, но повышение температуры на протяжении нескольких миллисекунд оставалось малым из-за низкой мощности разряда. Было показано, что шлирен сигналы оказались практически идентичными тем, что наблюдались без разряда, значительно отличаясь от сигналов в разрядах с полностью установленными профилями температуры (рисунок 22).

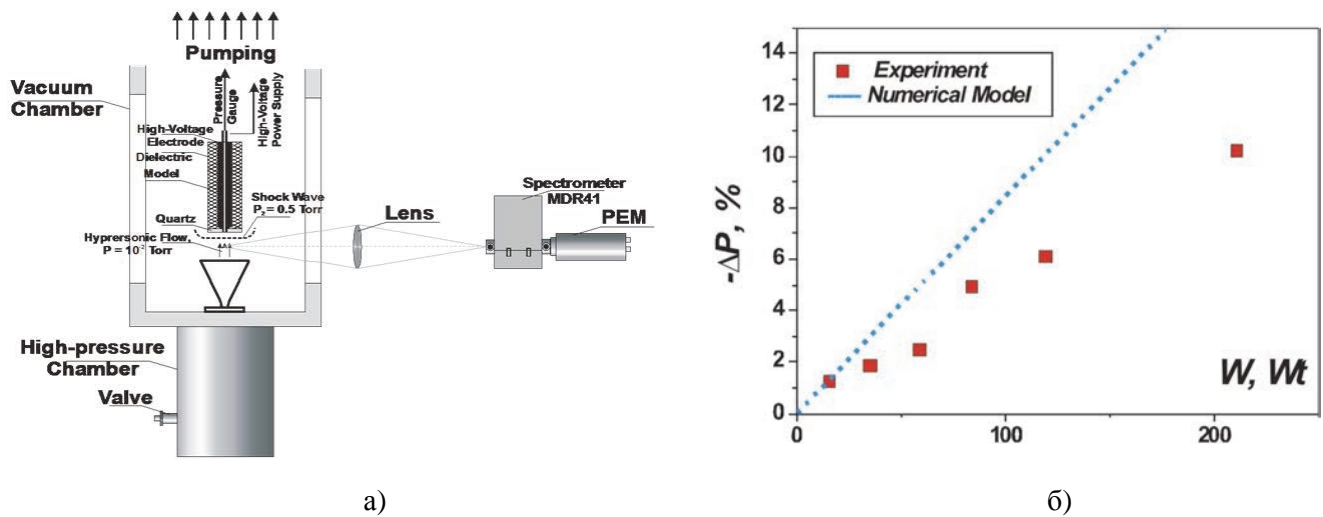

Рисунок 23. а) Организация гиперзвукового потока плазмы; б) Сравнение результатов эксперимента и численного моделирования для изменения давления торможения в зависимости от мощности разряда в потоке воздуха при $M = 8.2$ [154].

Тем не менее, остается важный вопрос о возможных особенностях разряда и свойств плазмы в молекулярных газах. Значительное возбуждение внутренних степеней свободы молекулярных газов, прилипание электронов к молекулам и более низкие температуры электронов (сильное влияние двойного слоя при малых ионных числах Маха) могут привести к значительному влиянию неравновесного возбуждения на взаимодействие разряда с потоком газа. Эти вопросы обсуждались в работах [152-154, 140].

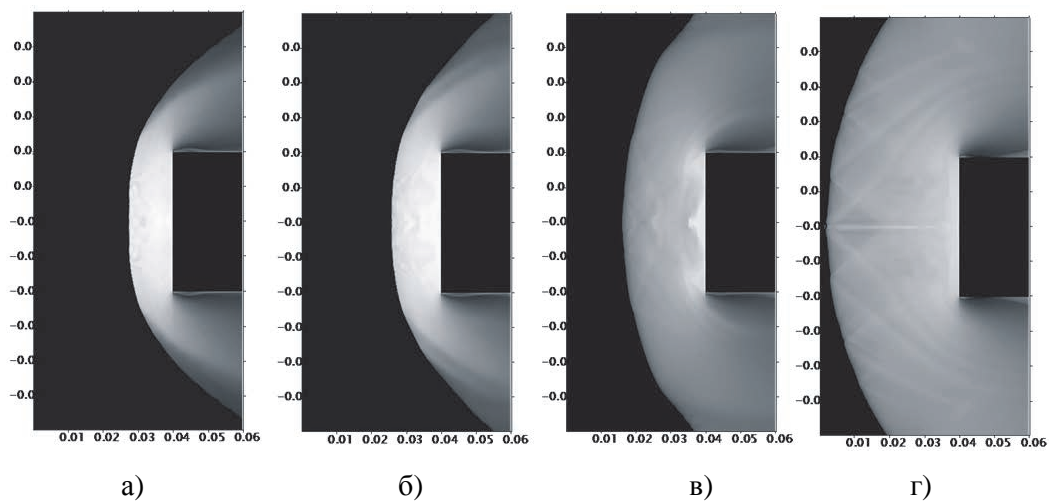

Рисунок 24. Результаты расчетов поля потока для различных значений энерговклада в разряд. а) без разряда; б) $E = 0.044$ эВ/молек; в) $E = 0.049$ эВ/молек; г) $E = 0.054$ эВ/молек. Воздух, $M = 8.2$, Температура торможения $T_0 = 300$ К, статическая температура потока $T_1 = 21$ К, статическое давление $P_1 = 5\times10^{-3}$ Торр [154].

В работах [152-154] описаны результаты исследования структуры сильных ударных волн в плазме с сильными приведенными электрическими полями ($E/n$ ~10 кТд). Авторы получили профили вращательной температуры газа, распределение плотности электронов и электрическое поле в потоке воздуха при $M = 8.2$. Сравнение результатов эксперимента и численного моделирования показало, что изменение сопротивления может быть полностью объяснено нагревом газа при термализации в газоразрядной плазме (рисунок 23).

На рисунке 24 показаны результаты вычисления изменения поля течения для различных значений энерговклада в разряд. Основное энерговыделение происходит за фронтом ударной волны из-за более высокой плотности газа и, как следствие, более высокой плотности энерговклада и нагрева в процессах электрон-ионной рекомбинации.

Аналогичная конфигурация разряда, но при низких $E/n$, была исследована в работе [155] в потоке при $M = 3$. Было обнаружено, что разряд, развивающийся при относительно низких величинах приведенного поля, вкладывает энергию в основном в колебательные степени свободы газа, а не в тепло. На основе численного моделирования была показана критическая роль быстрой термализации энергии для эффективного управления конфигурацией ударной волны и сопротивлением. Таким образом, эти режимы принципиально отличаются от режимов, исследованных в [154], где из-за высоких значений $E/n$ (~10 кТд) основным направлением энерговклада была ионизация газа с быстрой рекомбинацией и последующим нагревом.

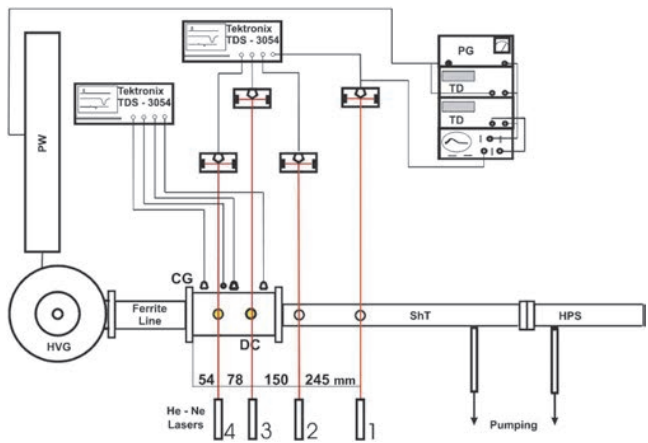 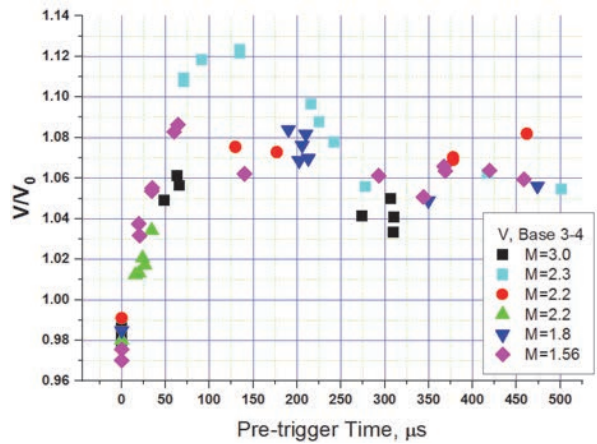

а) б)

Рисунок 25. а) Экспериментальная схема для наблюдения взаимодействия ударной волны с плазмой высоковольтного наносекундного разряда; б) относительное изменение скорости ударной волны в области плазмы в зависимости от времени релаксации плазмы [140].

В работе [140] представлены исследования распространения ударной волны через распадающуюся плазму воздуха после импульсного наносекундного разряда ($E/n$ ~ 600-800 Тд). Было обнаружено, что скорость распространения ударной волны увеличивается при увеличении времени задержки между выключением разряда и приходом ударной волны в область плазмы. Это наблюдение позволило подтвердить термический характер взаимодействия ударной волны с плазмой и оценить время термализации энергии в плазме (рисунок 25).

На рисунке 26 показан результат взаимодействия наносекундного SDBD разряда и отошедшей ударной волны перед цилиндром в потоке при $M = 5$ [156, 157]. Температура газа в плазменном слое была всего на несколько десятков градусов выше температуры торможения потока ($T = 340\pm30$ К). Взаимодействие волны сжатия, созданной расширяющимся газом из области разряда, и головной отошедшей ударной волны вызывает ее смещение в направлении вверх по потоку, тем самым увеличивая расстояние отхода волны от поверхности тела до 25%.

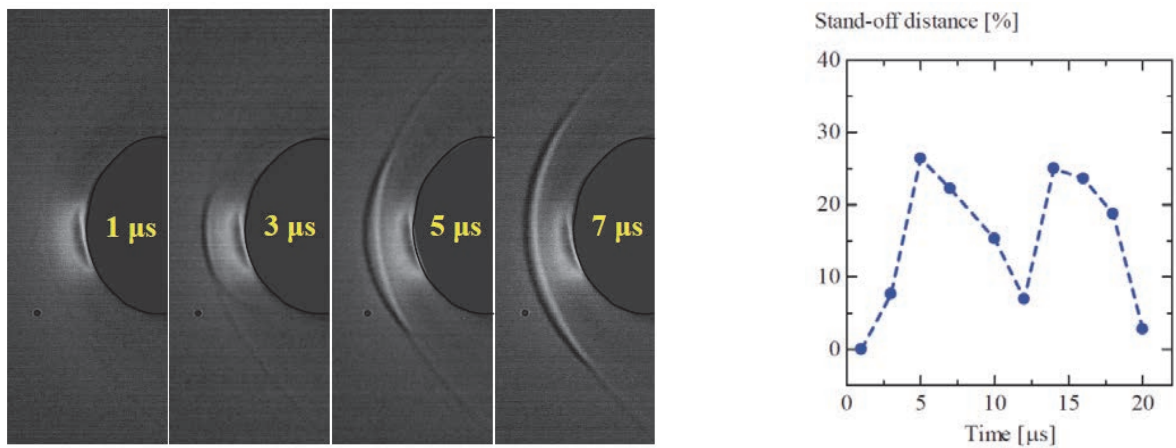

Рисунок 26. Взаимодействие отошедшей ударной волны с возмущением, созданным поверхностным разрядом. Слева показано отличие шлирен-сигнала от исходного. Сухой воздух, $P_0 = 370$ торр. Справа показан процент увеличения расстояния отхода ударной волны после первого и второго импульсов разряда [156].

Таким образом, результаты работ [140, 149, 150, 152-154, 156] ясно демонстрируют тепловой характер взаимодействия ударной волны с неравновесной плазмой. Теория взаимодействия ударных волн с горячими слоями (по сути, с энтропийными слоями любой природы) была разработана много лет назад. Установлено, что распространение возмущения в горячем слое происходит с большей скоростью, чем в основном течении, что приводит к формированию перед фронтом волны так называемого «предвестника».

В работах группы Немчинова (например, [158]) показано, что длина такого предвестника при достаточном перегреве слоя линейно зависит от длины пробега ударной волны вдоль горячего слоя: $l \approx U_s t$, где $t$ – время взаимодействия волны с нагретым слоем, $U_s$ – скорость волны. Данное соотношение было проверено до значений $U_s t = (80\text{-}90)\,h$ ($h$ – толщина слоя) для плоских нагретых слоев (двумерные течения) и до $U_s t = (300\text{-}400)\,d$ ($d$ – диаметр слоя) для нагретых каналов (аксиально симметричный случай). Это означает, что должен существовать механизм, который обеспечивает накопление газа в зоне предвестника как из горячего слоя, так и из невозмущенной области потока. Одним из возможных вариантов такой ситуации является отрыв горячего слоя от оси или плоскости симметрии и формирование вихревой структуры.

В работе [158] было установлено, что вплоть до отношения температуры газа в нагретом слое к температуре основного потока $T'/T \cong 2.5$ отношение длины предвестника к длине пробега ударной волны в горячем слое находится в диапазоне $\xi = 0.02\text{-}0.05$ ($M = 3$, $\gamma = 1.4$). Только начиная с перегрева $T'/T > 2.5$ наблюдалась зависимость коэффициента $\xi$ от величины перегрева $T'/T$. При этом значение $\xi = 0.29$ достигалось при $T'/T > 5.5$.

Критерий перестройки потока был предложен в [159] и [160] сразу после первых испытаний сверхмощных бомб, в которых излучение из эпицентра взрыва приводило к образованию горячих слоев над земной поверхностью и интенсивному взаимодействию взрывной волны с этими слоями. Критерий Марка-Гриффитса был основан на сравнении давления торможения газа из энтропийного слоя и статического давления газа в основном потоке. Предполагалось, что скорости ударной волны в энтропийном слое и вне его совпадают, по крайней мере, до момента, когда критические условия начинают выполняться. Можно показать, что условие неизменности скорости фронта ударной волны во всех областях де-факто запрещает перестройку потока даже тогда, когда критерий Марка-Гриффитса выполняется. Это указывает на противоречивость модели. В самом деле, оценка на основе критерия Марка-Гриффитса при $M = 3$ и $\gamma = 1.4$ указывает на начало перестройки потока при $T'/T > 1.18$, что значительно отличается от наблюдаемых значений: $T'/T(crit) \cong 2.5$.

В работах [161, 162] предложена модификация критерия Марка-Гриффитса. В качестве условий перестройки течения от первоначального стабильного распространения ударной волны к автомодельному росту предвестника были предложены следующие требования:

а) статическое давление в энтропийном слое за фронтом ударной волны должно быть равным статическому давлению за ударной волной в основном потоке;

б) скорость газа в энтропийном слое за фронтом ударной волны должна быть равна скорости основной ударной волны.

Условия а) и б) означают полное торможение газа из энтропийного слоя по отношению к фронту главной ударной волны. Математические детали можно найти в [161, 162]. Ниже приведем окончательное выражение для критического значения перегрева продольного теплового слоя, при котором инициируется автомодельный рост предвестника перед ударной волной:

$$\left(\frac{T'}{T}\right)_{crit} = \left(\frac{M^2}{M^2-1}\frac{\gamma+1}{2}\right)^2 \qquad (25)$$

Этот критерий количественно воспроизводит существующие экспериментальные результаты по взаимодействию ударной волны с горячим слоем и позволяет определять энергию, необходимую для управления конфигурацией ударных волн в сверхзвуковых потоках с помощью нагретых слоев (рисунок 27).

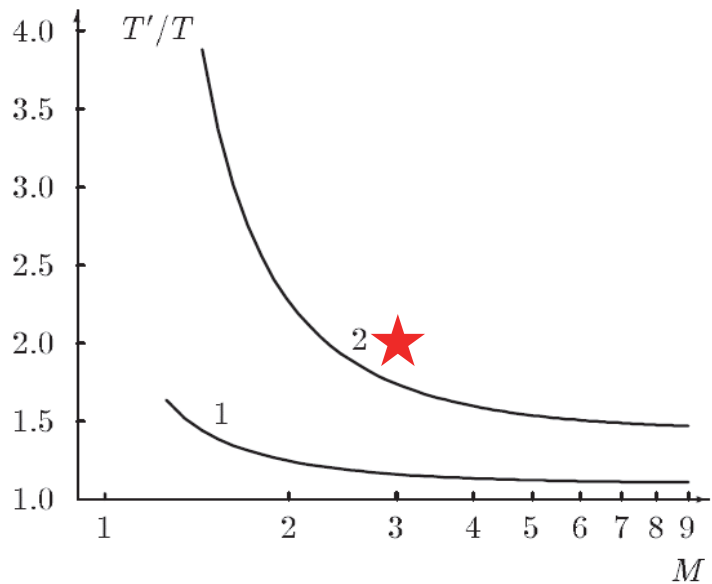

Рисунок 27. Зависимость критического значения перегрева слоя *(T'/T)* $_{crit}$ от числа Маха в ударной волне при *γ* = 1.4. Кривая 1 – критерий Марка-Гриффитса; кривая 2 – уравнение (25); точка – эксперимент [163].

Другим важным вопросом является минимально возможная толщина энтропийного слоя, при которой достигается стабильное взаимодействие ударной волны со слоем. Минимизация толщины слоя означает минимизацию потребления энергии и чрезвычайно важно с точки зрения практического управления потоком. Согласно результатам [158] автомодельное решение существует, по крайней мере, до длин пробега волны вдоль слоя $dx$ = (300-400)$d$ с увеличением длины предвестника по линейному закону, близкому к $l \sim dx$.

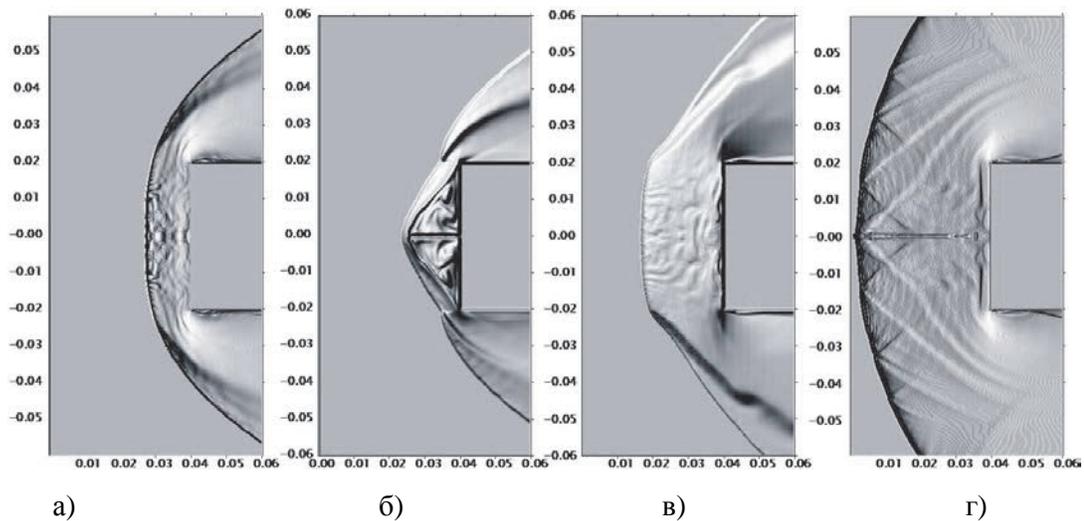

а) б) в) г)

Рисунок 28. Численное моделирование взаимодействия ударной волны с нагретым слоем различной толщины для воздуха при $M$ = 8.2, $T_0$ = 300 K и $P_0$ = 35 Тор. Энергия возбуждения $E$ = 0.054 эВ/мол, толщина слоя: а) 0%; б) 10%; в) 100%; г) весь поток [164].

Эти соотношения показывают, что можно ожидать изменения формы ударной волны даже при использовании энтропийного слоя с поперечным размером, составляющим около 1% от размера объекта. На рисунке 28 приведен пример взаимодействия энтропийного слоя различных размеров с головной ударной волной. При относительной толщине слоя в 10 и 100% и при возбуждении всего потока уменьшение давления торможения относительно невозбужденного потока составляло 21%, 34% и 29%, соответственно. Ясно, что уменьшение размера энтропийного слоя позволяет значительно увеличить эффективность управления потоком с помощью такого слоя.

Примерно равное влияние тонкого и толстого энтропийных слоев на конфигурацию ударных волн позволяет использовать сфокусированное относительно слабое возбуждение газа перед фронтом ударной волны. Именно поэтому многие эксперименты по плазменному контролю над уменьшением сопротивления проводились при генерации тонких плазменных филамент [165]. В качестве примера можно привести выделение энергии импульсного лазера [166], генерацию филамент с помощью СВЧ излучения [167] и инжекцию встречных струй [168].

В работе [168] была показана возможность управления структурой ударной волны и снижения сопротивления при нагреве газа посредством введения горячей струи плазмы в основной поток. Измерения лобового сопротивления были выполнены при инжекции как газовой, так и плазменной струи. Эксперименты проводились при числе Маха 6. С помощью плазменного факела в основной поток был введен встречный поток при числе Маха 3.28. Инжекция плазменной струи от сферически затупленного цилиндра позволила значительно уменьшить лобовое сопротивление. Шлирен фотография и расчетная картина в конфигурации со встречным потоком показаны на рисунке 29.

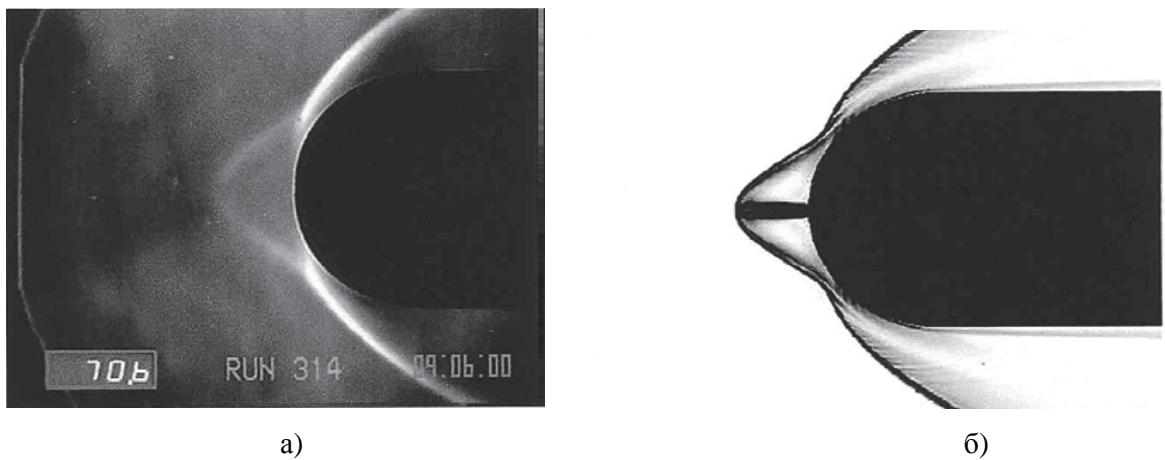

а) б)

Рисунок 29. а) Шлирен фотография в эксперименте со встречным потоком; б) расчетная картина потока в случае со встречной струей [168].

В качестве вывода по этой части можно сделать заключение, что основным механизмом воздействия плазмы на сверхзвуковые потоки является нагрев. Роль ионного ветра в таких условиях, как правило, пренебрежимо мала и может быть исключена из анализа.

Эксперименты демонстрируют высокую эффективность при создании небольших областей нагретого газа и управлении конфигурацией ударных волн и потока. Плазменные эффекты, такие как пространственные заряды и образование двойных слоев, не должны существенно влиять на распространение сильных ударных волн. Основная проблема управления сверхзвуковыми и гиперзвуковыми потоками с помощью энтропийных слоев сводится к технологии создания горячей филаменты перед головной ударной волной или передаче воздействия от разряда за фронтом ударной волны вверх по потоку. Для создания возмущения перед волной необходимо создавать плазменную филаменту с заданными свойствами и расположением в разреженной области перед фронтом волны.

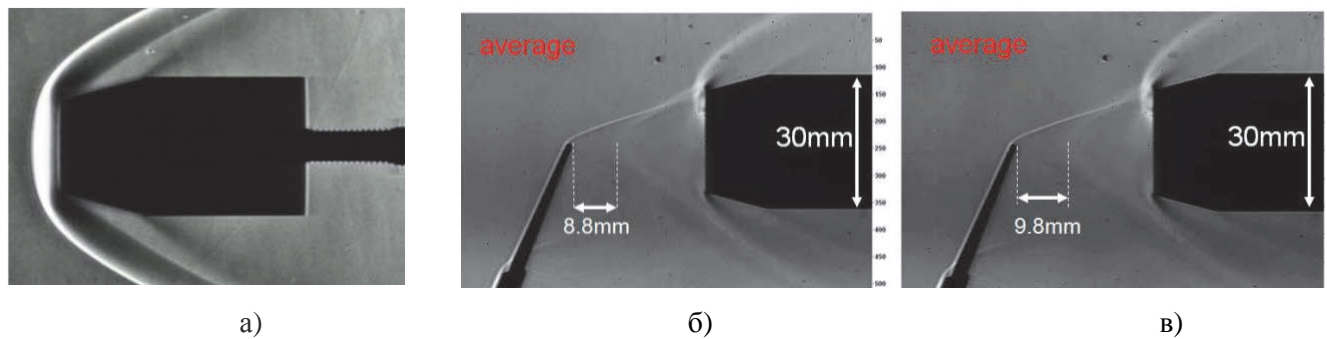

Рисунок 30. Шлирен фотографии обтекания усеченного конуса-цилиндра а) без электрода; б) без разряда и в) с разрядом [169].

Эта проблема явно продемонстрирована в работе [169], где изучено влияние энерговклада посредством дугового разряда на обтекание потоком с $M = 5$ моделей, представляющих собой усеченный конус-цилиндр с полууглом раствора $15^0$. Для инициирования дугового разряда авторы должны были поместить высоковольтный вольфрамовый электрод впереди модели (рисунок 30). Влияние дугового разряда оказалось значительно слабее, чем влияние самого электрода перед моделью. Кроме того, система таких электродов будет снесена потоком во всех практически интересных условиях.

## 7.2. Создание возмущений перед сверхзвуковым объектом

В качестве методов создания плазмы перед телом в потоке в настоящее время рассматриваются СВЧ излучение, лазеры, электронные пучки [150, 151]. При этом может быть перспективной комбинация нескольких методов генерации плазмы, например – использование струй или лазерных и электронных пучков для локализации энерговклада СВЧ излучения или коротких высоковольтных импульсов [170]. Ниже мы рассмотрим несколько наиболее важных работ с использованием таких методов.

## 7.3. СВЧ-разряд

В работе [167] были представлены результаты управления структурой головной ударной волны при формировании тонкого плазменного слоя с помощью СВЧ разряда. Плазменная филамента образовывалась за 1-2 нс и имела поперечный размер не более $3\times10^{-3}$ см при удельном энерговкладе около 7 эВ на частицу с пиковой плотностью электронов $n_e \cong 5\times10^{16}$ см$^{-3}$.

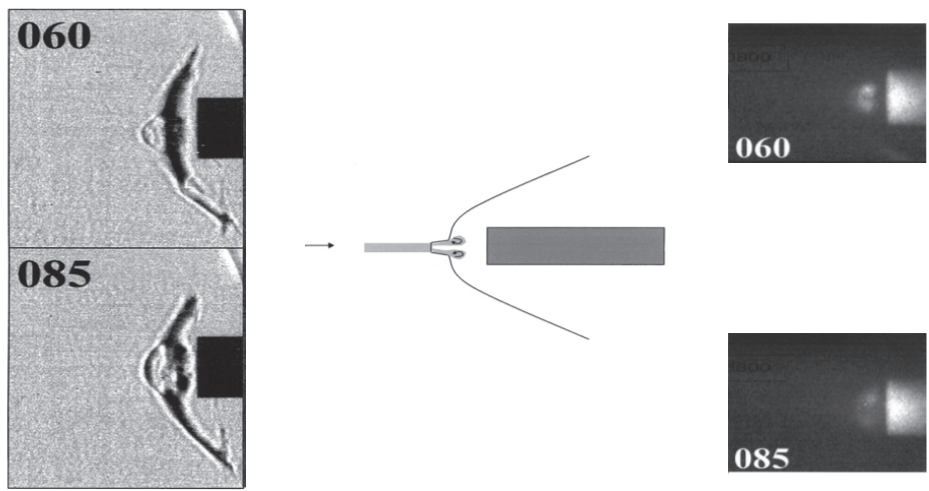

Рисунок 31. Взаимодействие СВЧ филаменты с потоком около тупого тела. Слева – шлирен изображения, справа – поля распределения светимости через 60 и 85 мкс после СВЧ разряда [167].

Нагрев газа в филаменте составлял примерно 2800 К при скорости нагрева ~ 2000-3000 К/мкс. Измерения с временным разрешением показали, что разряд существенно изменяет конфигурацию головной ударной волны благодаря формированию горячей филаменты. На рисунке 31 показано распространение головной ударной волны сквозь нагретую зону с образованием тех же структур, что наблюдались при взаимодействии ударной волны с типичными нагретыми слоями (смотри рисунок 28). Взаимодействие филамент с головной ударной волной через 60 и 85 мкс показано на шлирен фотографиях (рисунок 31, слева). При этом можно видеть образование вихрей, проявляющихся на рисунке в виде светлых областей на полях распределения светимости (рисунок 31, справа).

В работе [167] было показано, что энергоэффективность снижения давления торможения прямо пропорциональна отношению $D/d$, где $D$ и $d$ – диаметры модели и плазменной филаменты, соответственно. Справедливость этого соотношения была доказана для диаметров модели от 8 до 30 мм. Эти результаты подтверждают вывод работ [158, 161, 162] об увеличении эффективности взаимодействия и снижения аэродинамического сопротивления при уменьшении толщины нагретого слоя.

Развитием этого направления являются СВЧ-разряды, инициируемые лазерной плазмой [171]. Относительно слабая ионизация лазерным излучением может инициировать выделение микроволновой энергии и вызвать быстрый нагрев газа. В этом варианте возможна точная локализация энерговыделения СВЧ-разряда, и даже одновременное создание нескольких локализованных областей плазмы с использованием одного и того же микроволнового импульса. Взаимодействие между плазмой, созданной импульсом затравочного лазера и импульсом микроволнового излучения показано на рис. 32. Увеличение энергии фемтосекундного импульса затравочного лазера приводило к значительному изменению характера процесса распространения разряда. При увеличении энергии импульса в 3 раза зажигание разряда наблюдалось вдоль всей длины перетяжки вблизи точки фокусировки луча затравочного лазера. В этом случае значительно увеличивалась область сильной связи лазерной плазмы с энергией СВЧ-импульса (рис. 32, [172]).

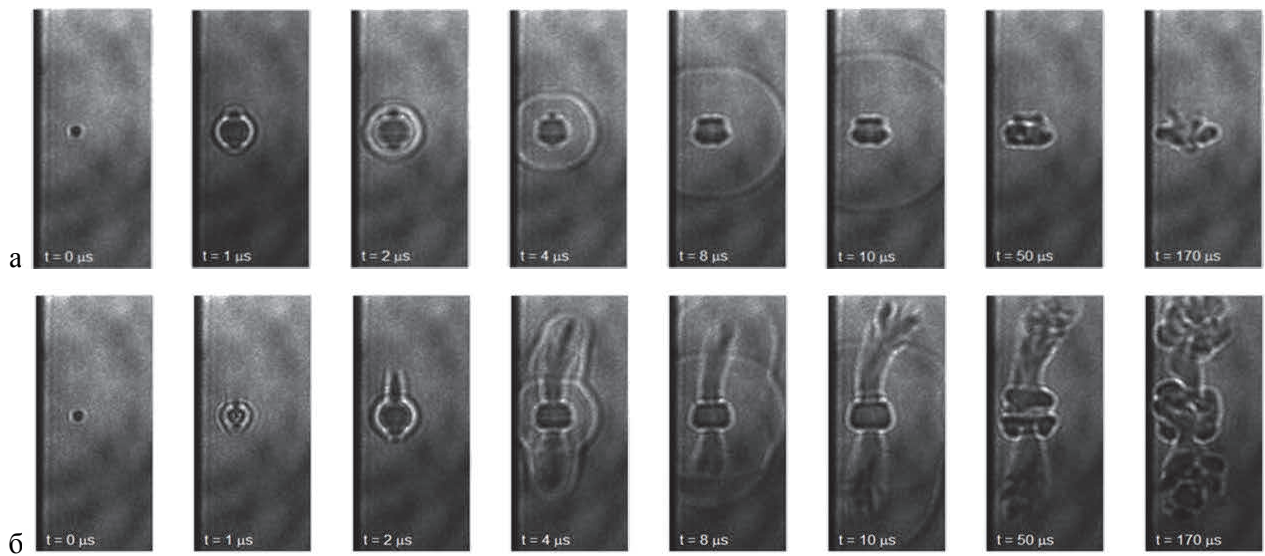

Рисунок 32. Эволюция лазерного импульса и СВЧ-разряда в воздухе. (а) Эволюция оптического разряда без СВЧ-поля; (б) Эволюция оптического разряда в СВЧ-поле [172].

### 7.4. Лазерная искра

Первые упоминания о возможности управления потоком с помощью импульсного энерговыделения от оптического разряда, инициируемого лазерным излучением, появились еще в 80-х годах [173].

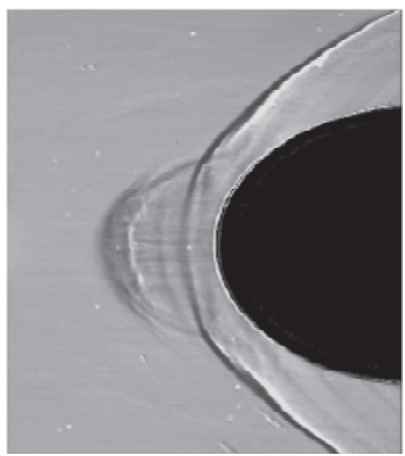
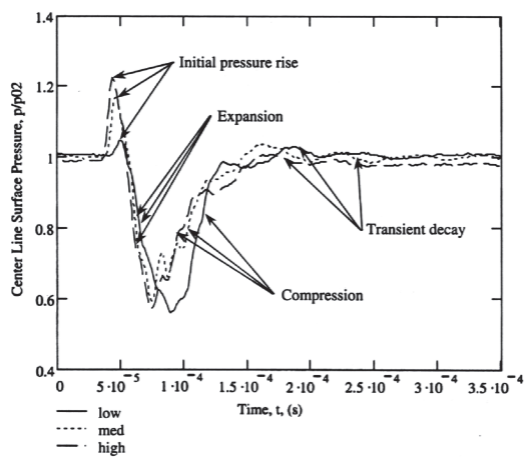

а)     б)

Рисунок 33. а) Взаимодействие головной ударной волны с нагретым лазером пятном; б) изменение во времени давления торможения на центральной линии тела при энергии лазерного импульса 13, 127 и 258 мДж [166].

В работе [166] исследовалось влияние импульсного энерговклада на обтекание сферы при $M$ = 3-4.5. Во всех экспериментах энерговклад обеспечивался импульсным Nd:YAG лазером (532 нм, длительность импульса 10 нс, частота повторения 10 Гц) с уровнями энергии от 12 до 300 мДж. Одиночный лазерный импульс создавал нагретое пятно перед головной ударной волной от одиночной сферы. На рисунке 33 показано взаимодействие этого пятна с ударной волной. Результаты (рисунок 33,б) подтверждают вывод работ [158, 161, 162] о слабой зависимости динамики взаимодействия от энергии, если интенсивность возмущения находится выше порога сильного взаимодействия.

В работе [91] представлены результаты исследования взаимодействия ударной волны $M$ = 3 с плазмой, сформированной излучением фемтосекундного лазера. Фемтосекундный импульс отличается от наносекундного принципиально другим механизмом развития плазмы – в этом случае большая часть электрон-ионных пар образуется не в результате лавинной ионизации электронным ударом, а за счет многофотонной ионизации молекул. При этом нагрев газа за время лазерного импульса пренебрежимо мал, и основной рост температуры тяжелой компоненты происходит в процессе рекомбинации образовавшейся плазмы. Было обнаружено, что выделение энергии при таком распаде лазерной плазмы перед телом в сверхзвуковом потоке вызывает значительное, хотя и кратковременное, снижение сопротивления. Рисунок 34 показывает шлирен-картинки развития течения при возбуждении потока перед фронтом головной ударной волны с помощью тераваттного импульса лазера, который может излучать лазерные импульсы длительностью ~50 фс. Средняя длина волны импульс лазера была 800 нм, а максимальная энергия импульса, измеренная на границе тестовой модели, составляла 150 мДж/импульс.

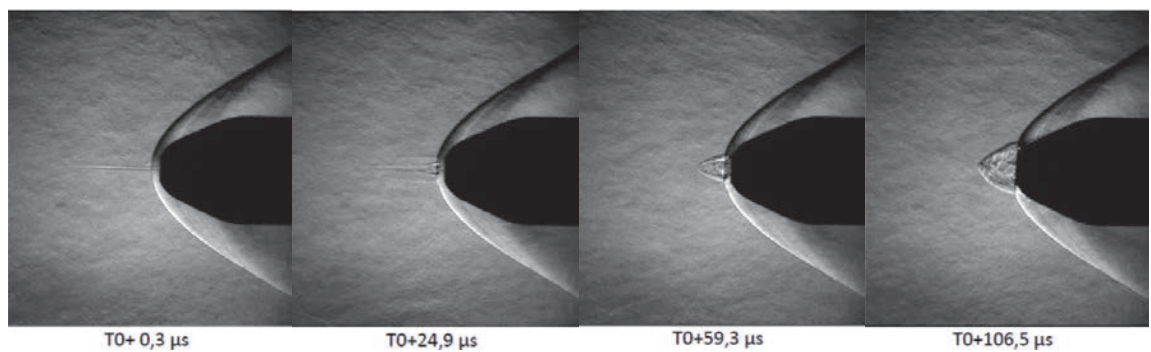

Рисунок 34. Шлирен-картинки потока при взаимодействии с распадающейся лазерной плазмой фемтосекундного лазера [91].

В работе [174] представлена попытка описания взаимодействия лазерной плазмы с ударной волной из первых принципов – начиная с формирования лазерной искры до момента развития газодинамического возмущения. Газодинамические эффекты были описаны в рамках уравнений Навье-Стокса с учетом неравновесных химических реакций в двухтемпературном приближении (для тяжелых частиц и свободных электронов). Расчеты были выполнены в сверхзвуковом и гиперзвуковом потоках воздуха для сферического затупленного тела и тела в форме двойного конуса (рис. 35). Моделирование развития оптического разряда осуществлялось с помощью кинетического подхода для фотонов с учетом процесса обратного тормозного рассеяния, приводящего к нагреву электронов, однако без учета рассеяния излучения лазера формирующейся плазменной областью, обладающей отрицательным показателем преломления при $\omega < \omega_{pe}$, что приводило к значительному усилению эффекта. Результаты моделирования качественно подтверждают, что точечное энерговыделение может эффективно использоваться для управления ударными волнами.

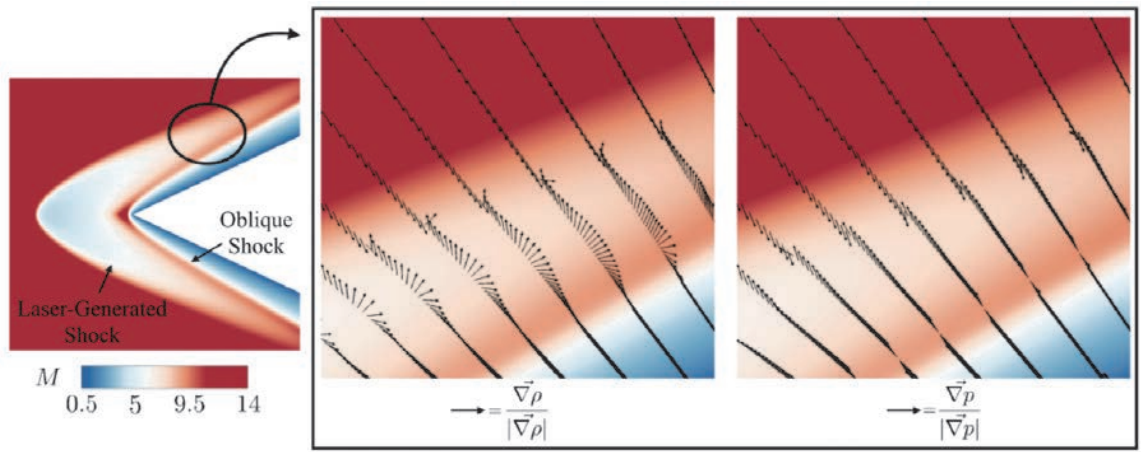

Рисунок 35. Контуры изолиний по числу Маха через 2 мкс после 600 мДж-импульса лазера, сфокусированного на 5 мм выше по потоку. Справа – увеличенная картина взаимодействия лазерной ударной волны и основной конической ударной волны с градиентом плотности и градиентом давления [174].

Таким образом, за последнее время убедительно показано, что основным механизмом взаимодействия лазерной плазмы с потоком – независимо от длительности лазерного импульса – является нагрев газа в результате распада плазмы. С другой стороны, снижение сопротивления, полученное в таких экспериментах и расчетах, как правило, становится существенным только для затупленных тел. Если объект имеет высокое аэродинамическое качество, получить существенное снижение коэффициента сопротивления, как правило, не удается [175]. Однако нецентральное воздействие на быстролетящий объект может привести к нарушению симметрии распределения сил и существенно поменять его траекторию.

## 8. Управление траекторией движения сверхзвукового объекта

В работе [176] был проведен экспериментальный и теоретический анализ управления траекторией быстровращающихся сверхзвуковых объектов с помощью лазерной искры. Для проведения таких экспериментов был сконструирован специальный пневмо-подвес, обеспечивавший закрутку модели в сверхзвуковом потоке и возможность изменения оси вращения модели без трения. Лазерная искра генерировалась Nd-YAG лазером (длина волны излучения 1.06 мкм, энергия в импульсе 300 мДж, длительность импульса 5 нс). Было показано, что искра генерирует сильную ударную волну, пятно горячего газа и медленную воздушную струю на поздней стадии распада возмущения, направленную к источнику излучения. Нагрев газа рассматривался как основной механизм перераспределения давления вдоль поверхности объекта и изменения его положения в пространстве. Были проведены три серии экспериментов.

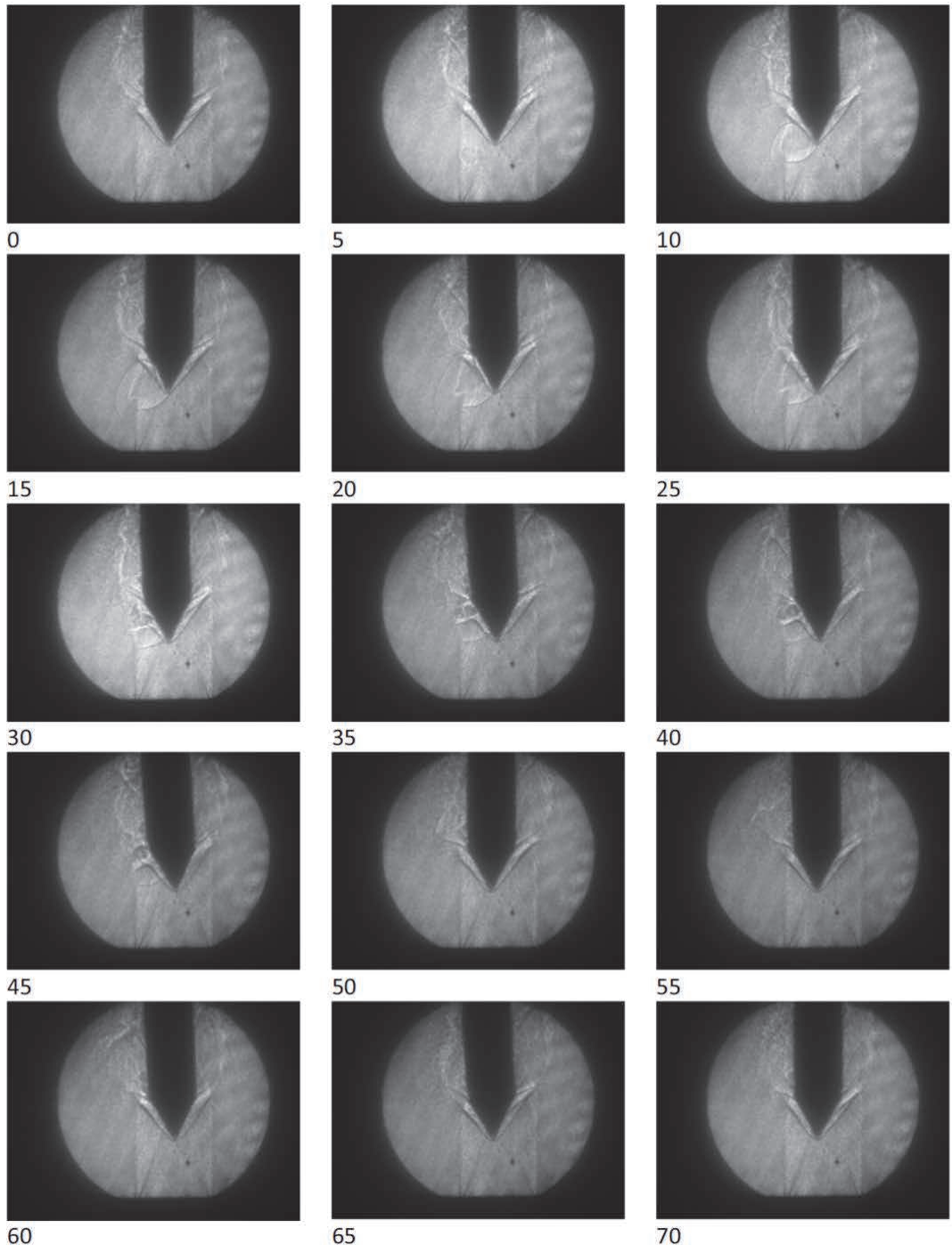

Рисунок 36. Взаимодействие лазерной искры с ударной волной перед коническим вращающимся объектом. Частота вращения $f$ = 400 Гц. Воздух, $M$ = 3. Статическое давление 1 атм. Числа – время после лазерной искры в мкс. Размер изображения 46×34 мм.

Первая серия демонстрировала дестабилизацию вращающегося объекта единичной лазерной искрой. Во второй и третьей сериях проанализирована динамика изменения угла атаки объекта с помощью 3-х последовательных лазерных искр, следующих друг за другом с интервалом от 50 до 100 мкс. Во всех трех случаях была продемонстрирована возможность дестабилизации траектории вращающегося объекта (рис.36, одиночная искра). Продолжительность возмущения положения оси объекта соответствует времени распространения горячей точки вдоль его поверхности (~50 микросекунд). Газовая область низкой плотности вызывает перераспределение давления и поддерживает развитие возмущения траектории (рис. 36). После того, как горячая точка, сформированная лазерной искрой, покидает область взаимодействия, возмущение оси вращения объекта уменьшается. Увеличение количества последовательных импульсов увеличивает продолжительность неустойчивого движения объекта. Таким образом, возможность контроля траектории вращающегося объекта сильно зависит от продолжительности взаимодействия, которая ограничена длиной искры лазера в направлении траектории движения объекта.

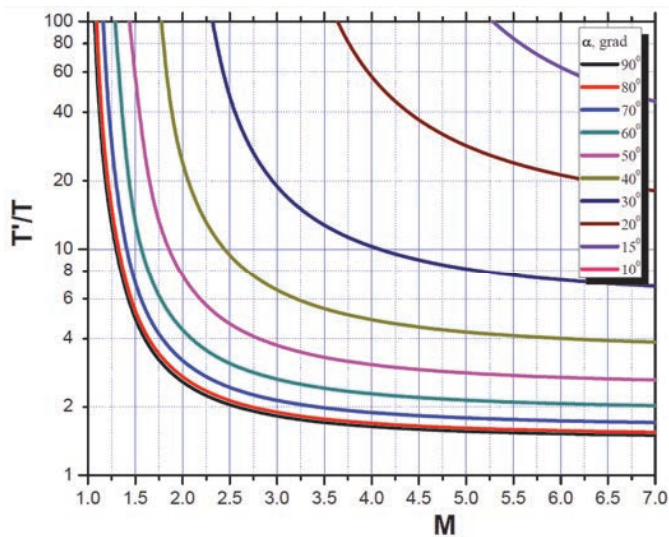

Рисунок 37. Критическая температура в горячем слое в зависимости от числа Маха и угла наклона ударной волны относительно направления потока [177].

Проведенный в [177] теоретический анализ позволил расширить аналитический критерий (1) сильного взаимодействия между ударной волной и горячим слоем на случай произвольного угла наклона ударной волны α:

$$\left(\frac{T'}{T}\right)_{crit} = \left(\left[\frac{M^2 \sin(\alpha)}{M^2 \sin^2(\alpha) - 1}\right] \times \left[\frac{\gamma + 1}{2}\right]\right)^2 \qquad (26)$$

Было показано, что критическая температура горячего слоя увеличивается с уменьшением угла наклона волны относительно направления потока и уменьшается с увеличением числа Маха ударной волны (рис.37).

Таким образом, для плоской ударной волны ($\alpha = 90^0$) при $M = 3$ безразмерная критическая температура горячего слоя равна $T'/T = 1.8$. Для наклонной ударной волны ($\alpha = 40^0$) режим сильного взаимодействия требует $T'/T = 6.5$. Численное моделирование демонстрирует хорошее согласие с теоретическим прогнозом.

На основании полученных экспериментальных и теоретических данных был сделан вывод о высокой эффективности управления траекторией вращающихся объектов лазерными разрядами. Энергия, требующаяся для отклонения траектории объекта на $\Delta\alpha = 1^0$, определяется эффективностью нагрева слоя газа перед объектом, длиной взаимодействия и критическим диаметром возникающей горячей точки. Оценки показывают, что минимальная энергия лазерной искры, необходимая для изменения направления траектории стандартной пули 7.62-мм на $\Delta\alpha = 1^0$ для длины взаимодействия ~ 3.5 см составляет около 2 мДж.

В работе [176] данное исследование было расширено на объекты калибром 30 мм в потоке при $M = 4$. Длинная лазерная искра перед объектом создавалась с помощью Nd-YAG лазера с энергией импульса $E = 2.84$ Дж на длине волны 1064 нм. Было показано, что создаваемое боковое усиле достаточно для отклонения траектории движения такого тяжелого объекта более чем на 10 метров на дистанции 1 км.

Позднее аналогичные эксперименты были проведены в работе [178]. Эксперименты проводились в сверхзвуковой аэродинамической трубе университета Ратгерс для количественной оценки влияния выделения энергии лазерной искры на поле обтекания заостренного цилиндра при числе Маха в свободном потоке 3.4.

В работе [179] численно исследован эффект внеосевого оптического разряда на динамику полета тела при М = 3.4. Получены данные о снижение сопротивления, боковом усилии и моментах, рассмотрена структура и динамика потока. Показано, что чрезмерное увеличение радиального удаления разряда от оси объекта уменьшает эффект воздействия. Высказано предположение о наличии оптимального удаления оптического разряда от оси объекта, которое дает максимальную боковую силу для заданной удельной поглощенной энергии.

# 9. Управление квазистационарными отрывными течениями и слоями

## 9.1. Управление слоями смешения и шумом скоростной струи

Управление высокоскоростными слоями смешения с помощью импульсного энерговыделения в локализованном наносекундном искровом разряде (LAFPA) было предложено в 2009-м году в работе [180]. Система из восьми синхронизированных искровых актуаторов, расположенных по периметру круглой затопленной струи, создавала локальные возмущения в слое смешения. При этом различные комбинации сдвига фаз зажигания отдельных разрядников позволяли возбуждать в слое смешения разные моды неустойчивости, управляя скоростью смешения и шумом затопленной струи. В работе [181] было показано, что струя реагирует на управление с помощью такого импульсного энерговыделения в большом диапазоне чисел Струхаля и азимутальных мод (рис. 38) при работе в широком диапазоне чисел Маха (от 0.9 до 1.65) и чисел Рейнольдса (от $0.2\times10^6$ до $1.65\times10^6$).

Воздействие таких синхронизированных искровых разрядов на смешение струи численно исследовано в работе [182]. Было показано, что несимметричное воздействие актуаторов разрушает симметрию вихрей в слое, значительно усиливая смешение.

Влияние плазменных ns-SDBD актуаторов на турбулентные сдвиговые слои (как в случае обычного слоя смешения, так и течения около стенки с обратной ступенькой) было экспериментально исследовано в цикле работ [183-187]. Цель этих работ – дать представление о фундаментальных процессах при управлении слоем смешения затопленной струи с помощью тепловых возмущений. Из полученных результатов следует, что тепловые возмущения вызывают локальные возмущения потока с большой амплитудой, воздействие которых на поток сильно зависит от соотношения между пространственным масштабом тепловых возмущений и толщиной слоя смешения. Этот результат полностью подтверждает выводы работ по управлению отрывом пограничного слоя с помощью импульсного энерговыделения в ns-SDBD [33], где была обнаружена сильная зависимость эффективности актуаторов от числа Рейнольдса потока.

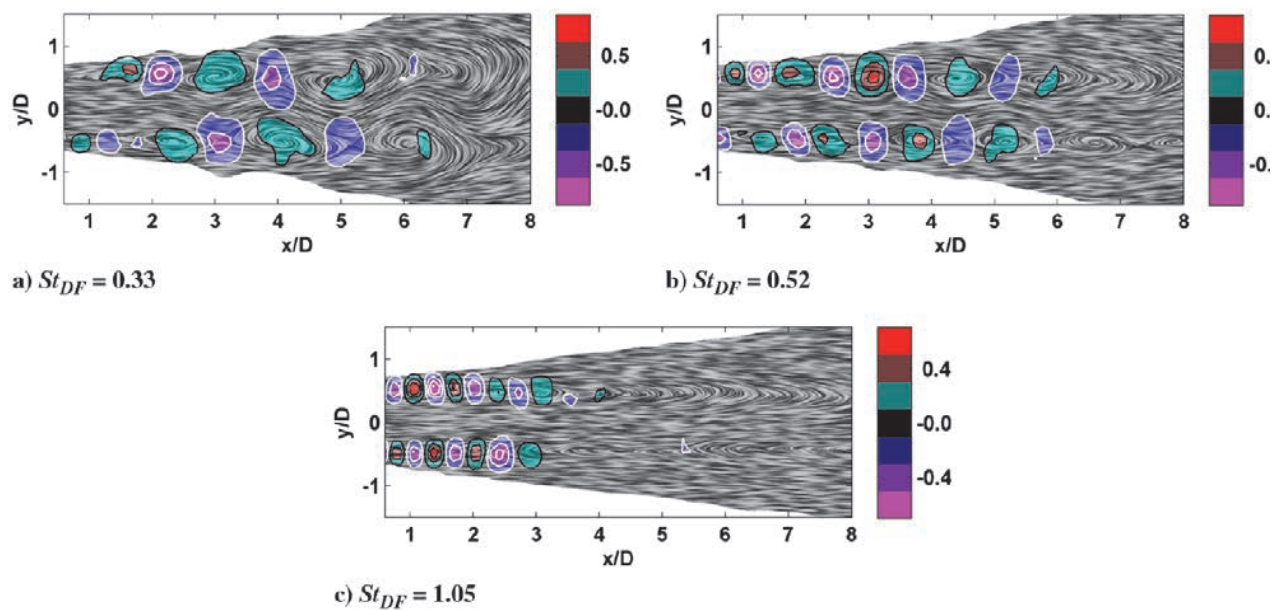

Рис.38. Влияние изменения числа Струхаля по частоте срабатывания наносекундных искровых разрядников на структуру потока: $M = 1.3$ [180].

Этот результат полностью подтверждает выводы работ по управлению отрывом пограничного слоя с помощью импульсного энерговыделения в ns-SDBD [33], где была обнаружена сильная зависимость эффективности актуаторов от числа Рейнольдса потока.

## 9.2. Управление взаимодействием ударной волны с пограничным слоем

Управление индуцированным ударной волной отрывом пограничного слоя (Shock Wave – Boundary Layer Interaction, SWBLI) с помощью плазменных актуаторов является важной проблемой сверхзвуковой внутренней аэродинамики и исследуется достаточно давно [188, 189]. В этих работах разряд постоянного тока в магнитном поле 1-5 Тл вызывал как нагрев газа в пограничном слое, так и ускорение или замедление потока в зависимости от направления силы Ампера. Было показано, что разряд может инициировать отрыв погранслоя, усилить или ослабить его. Увеличение температуры газа в результате энерговыделения в разряде усиливает отрыв пограничного слоя, в то время как разгон газа магнитогазодинамическими силами уменьшает размер отрывного «пузыря» и смещает его вниз по потоку.

Управление взаимодействием ударной волны с пограничным слоем с помощью только импульсного нагрева приводит, таким образом, к противоречию. Повышение температуры газа в пограничном слое легко приводит к усилению такого взаимодействия и увеличению размеров отрывной зоны. Как правило, этот эффект нежелателен, поскольку может привести к запиранию потока в сверхзвуковом воздухозаборнике. Чтобы разрешить это противоречие, в работе [190] было предложено использовать локализованные импульсные искровые разряды для создания локальных возмущений в пограничном слое, приводящих к образованию крупных вихрей и внесению дополнительного импульса из основного потока в пограничный слой. Однако в последующих работах (смотри, например, [191]), было обнаружено, что такие локализованные искровые разряды смещают отраженную ударную волну вверх по течению примерно на толщину пограничного слоя. На основе результатов измерений методом PIV (Particle Image Velocimetry) с фазовой синхронизацией, а также параметрических исследований, был сделан вывод о том, что воздействие искровых разрядов на поток сводилось к нагреву пограничного слоя выше точки взаимодействия с ударной волной, что, как известно, является нежелательным эффектом.

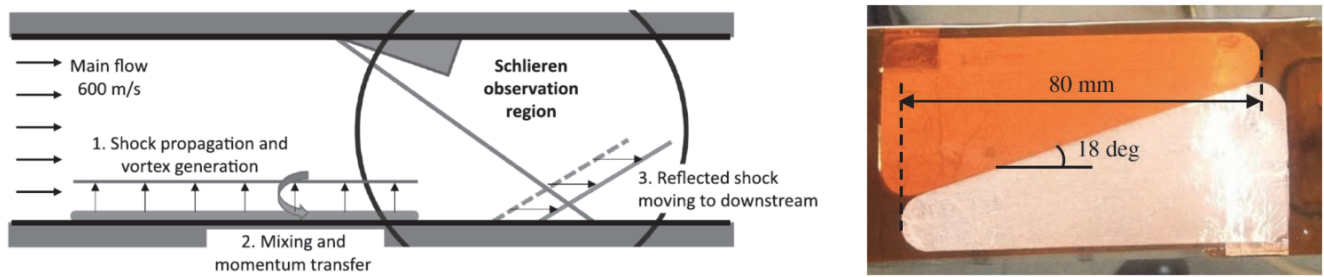

Рисунок 39. Сверхзвуковая аэродинамическая труба для исследования SWBLI и схема расположения актуатора и генератора ударной волны. (Рисунок слева). Скошенный под углом $18^0$ относительно потока несимметричный актуатор. Сверху – закрытый электрод, снизу – открытый. Направление потока слева направо. (Рисунок справа). [193]

Решение было предложено в работе [29], где для активации пограничного слоя были использованы не высокотемпературные искровые разряды, а поверхностные диэлектрические разряды с ограниченным энерговкладом. При этом нагрев газа в пограничном слое уменьшился в десятки раз, а газодинамические возмущения, вносимые разрядом в поток, остались достаточно интенсивными.

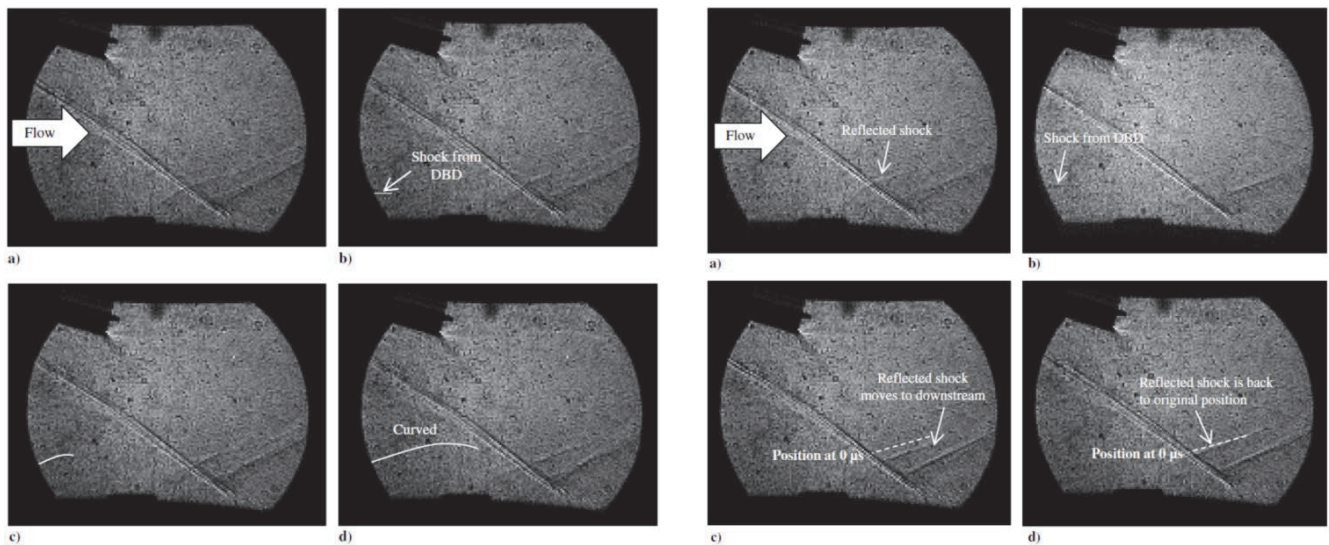

Рисунок 40. 4 кадра слева: распространение ударной волны, созданной импульсным разрядом: a) 8 мкс, b) 16 мкс, c) 24 мкс и d) 32 мкс после импульса. 4 кадра справа: движение отраженной ударной волны показывает уменьшение размера отрывной зоны: a) 0 мкс, b) 16 мкс, c) 224 мкс и d) 520 мкс после импульса. [193]

В работах [192, 193] было проведено детальное исследование механизма управления взаимодействием ударной волны с пограничным слоем с помощью поверхностного наносекундного барьерного разряда (рис. 39).

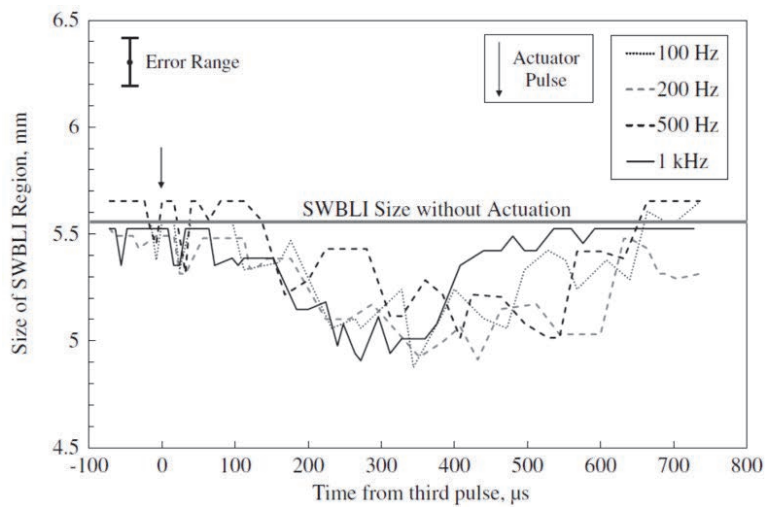

Рисунок 41. Вертикальный размер отрывной зоны при SWBLI в зависимости от частоты следования высоковольтных импульсов [193].

Рисунок 40 показывает динамику распространения ударной волны, возникшей из-за импульсного энерговыделения в плазме. За 32 микросекунды ударная волна переходит в звуковое возмущение, и возмущение от нее становится слабым. Однако, возмущение течения в пограничном слое, вызванное появлением косой ступеньки газа низкой плотности, провоцирует перемешивание пограничного слоя с основным потоком, разгоняет его, и резко уменьшает высоту отрывной зоны в области взаимодействия (рис. 40). Такое уменьшение зоны взаимодействия свидетельствует о том, что перенос импульса в пограничный слой в результате образования вихря на косой зоне энерговыделения становится более существенным, чем влияние нагрева газа в пограничном слое, которое ведет к усилению взаимодействия.

Рисунок 41 показывает, что при увеличении частоты следования высоковольтных импульсов влияние нагрева становится сильнее. При частоте следования 5 кГц влияние перемешивания и нагрева практически компенсировали друг друга, а на больших частотах преобладало влияние нагрева, что приводило к увеличению размера отрывной зоны [193].

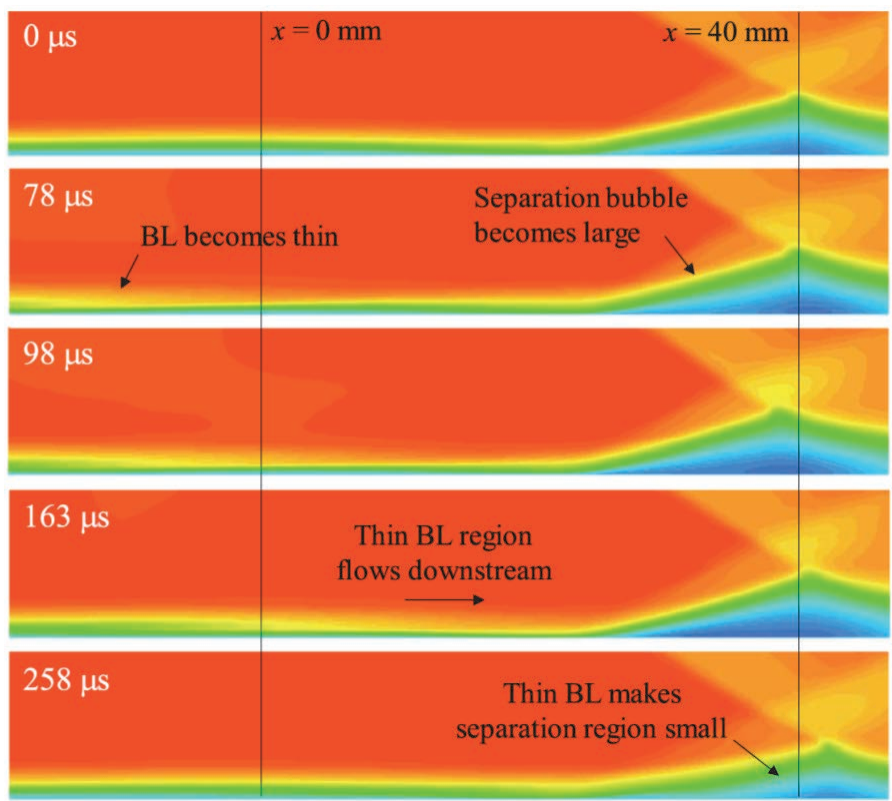

Рисунок 42. Эволюция скорости в направлении по оси X в случае наклонного электрода. Геометрия соответствует эксперименту [193], схема которого приведена на рис.39. [194].

В работе [194] было проведено численное моделирование развития возмущения в пограничном слое и взаимодействия такого модифицированного слоя с ударной волной. При этом были использованы данные о распределении энергии в разряде, полученные в работе [53]. Рисунок 42 показывает результат такого численного анализа. Видно, что пограничный слой становится значительно тоньше в результате воздействия возмущения (78-я мкс), затем область тонкого погранслоя сносится потоком в область взаимодействия (98 и 163 мкс), что приводит к значительному уменьшению размера отрывной области (258-я мкс). Таким образом, созданная тепловая косая ступенька в пограничном слое приводит к существенной перестройке потока и области взаимодействия с ударной волной.

## 9.3. Управление ламинарно-турбулентным переходом и турбулентными погранслоями

В работе [195] наблюдалось подавление волн Толмина-Шлихтинга под действием плазменных актуаторов, работающих в импульсном режиме. Эти волны, являющиеся предвестником турбулизации потока, искусственно возбуждались в эксперименте специальной колеблющейся пластиной, установленной заподлицо с поверхностью канала, и усиливались специально созданным градиентом давления в газе. Плазменный актуатор располагался вниз по потоку от области возбуждения волн. Действие актуатора приводило к появлению в пограничном слое нестационарной силы, которая вызывала затухание волн. В результате амплитуда колебаний скорости на частоте возбуждения значительно уменьшалась. Было изучено влияние различных параметров актуатора на затухание волн. Исследования были проведены в открытой аэродинамической трубе с рабочей секцией сечением $0.45 \times 0.45$ м$^2$ и длиной 2 м. Вставки на крышке рабочей секции создавали обратный градиент давления 25 Па/м. Толщина пограничного слоя была равной 5 мм в точке $x = 590$ мм, давая в результате число Рейнольдса $Re = 1100$ [195]. На рисунке 43,а показана рабочая секция, а на рисунке 43,б – особенности расположения актуатора.

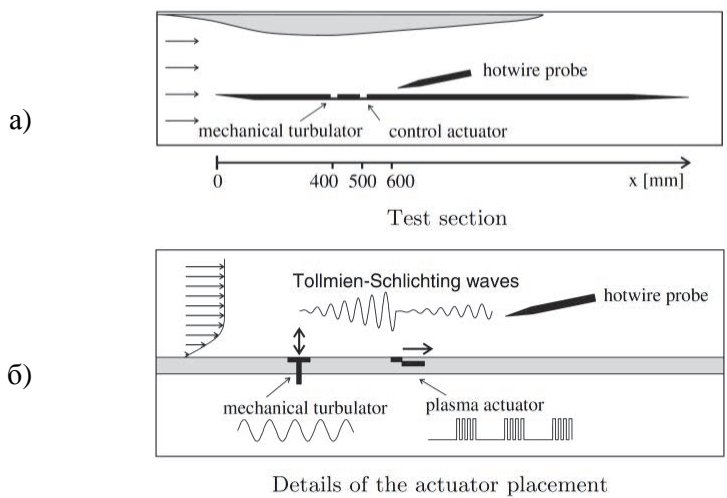

Рисунок 43. (а) рабочая секция, (б) особенности расположения актуатора [195].

На рисунке 44 приведена более подробная информация о возбуждаемых частотах и форме колебаний при наличии и в отсутствии плазменного актуатора. Рисунки 44,а показывают спектры мощности для колебаний скорости, и рисунок 44,б дает временные маркеры этих колебаний. Под действием актуатора амплитуда колебаний на основной частоте заметно уменьшается, а моды $f_2$ и $f_3$ остаются неизменными. Мода же $f_4$ пропадает на фоне шума.

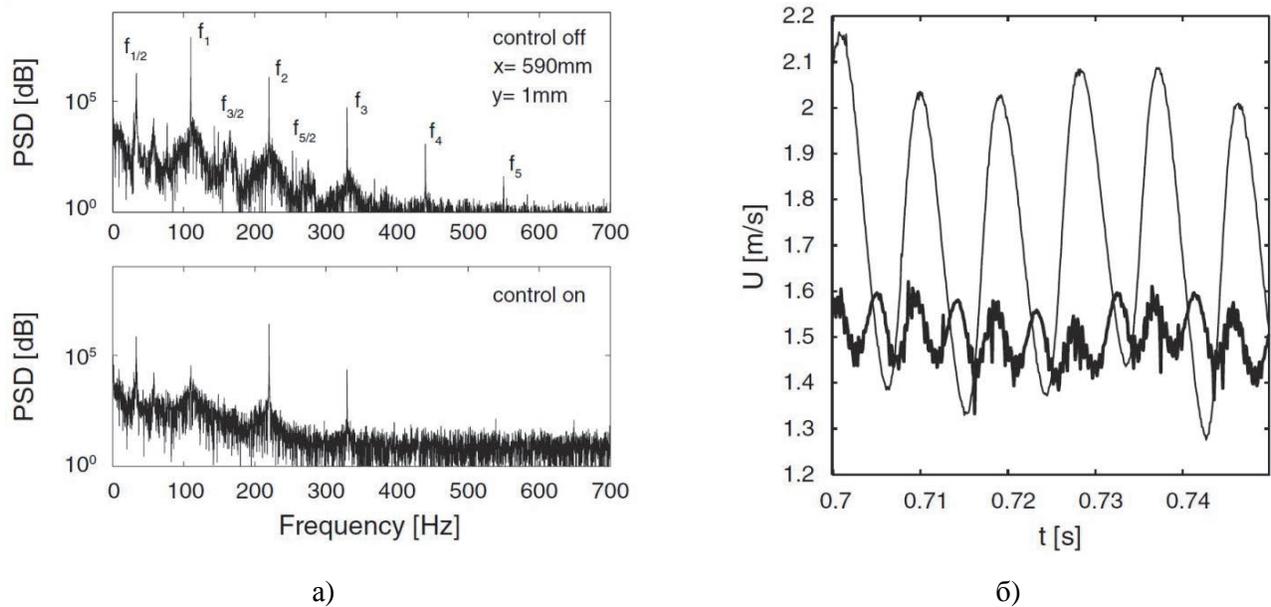

Рисунок 44. Спектры мощности пульсаций (а) и осциллограммы (б) с включением (толстые линии) и без включения (тонкие линии) актуатора. Точка $x = 590$ мм, $y = 1$ мм [195].

Недавно интересный результат был опубликован в работе [196]. Установленные на поверхности плоской стенки сборки плазменных актуаторов с импульсным питанием были использованы для снижения сопротивления трения на поверхности в турбулентных пограничных слоях. Сборки были предназначены для создания пристеночного потока, индуцированного плазмой, в направлении, перпендикулярном основному потоку, с целью предотвращения подъема низкоскоростных пристеночных вихрей в основной поток [196]. Использовались две различные конструкции сборок (рис. 45).

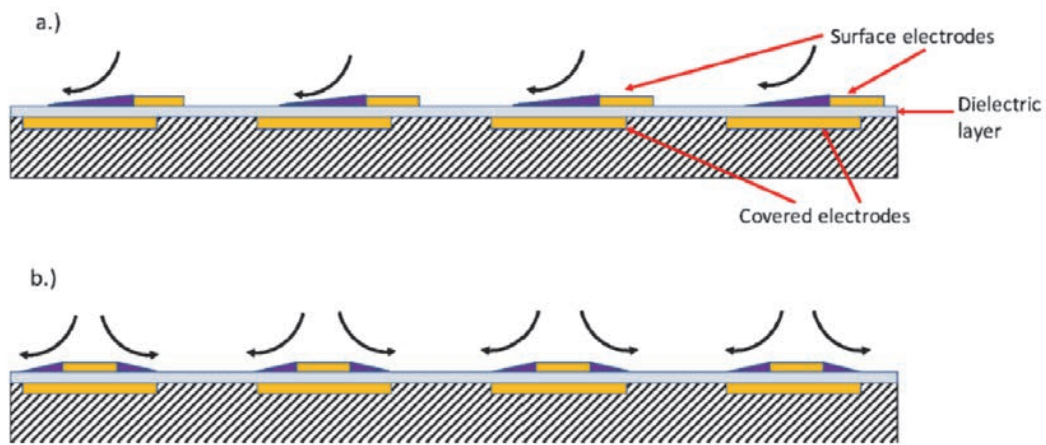

Рисунок 45. Схема двух конфигураций сборок импульсных плазменных актуаторов, использованных в экспериментах по снижению сопротивления трения [196]. Электроды ориентированы по направлению основного потока (направление вектора скорости сквозь плоскость рисунка), при этом индуцированный поток направлен поперек основного. (а) однонаправленный поток, (б) противоположно-направленные струи.

Первая из них (а) создает однонаправленный продольный поток, а вторая (б) – серию противоположно-направленных струй. Авторы утверждают, что обе конфигурации показывают беспрецедентные уровни снижения сопротивления, превышающие 70%. Такое утверждение заслуживает очень детального рассмотрения. Приведем данные работы [196] по снижению сопротивления для обоих типов использованных актуаторов (рис. 46).

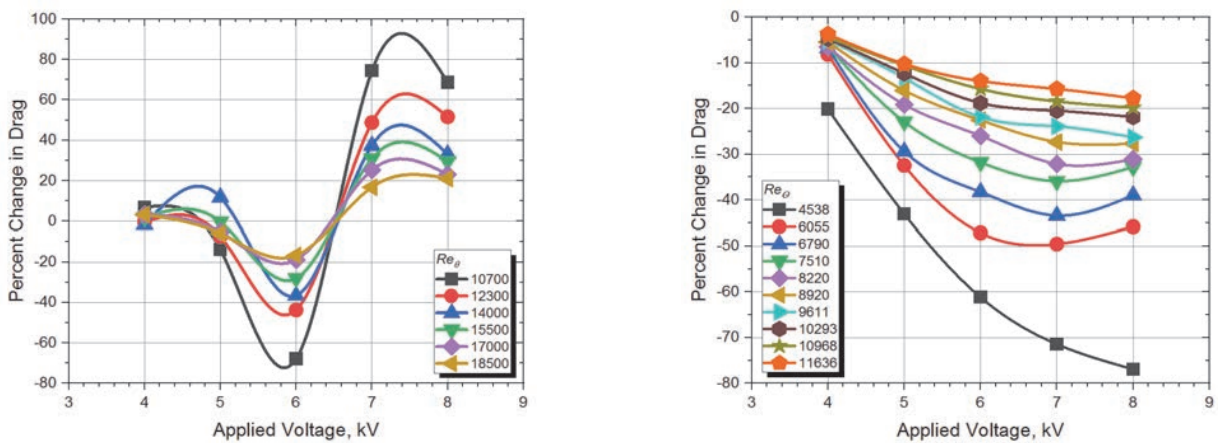

Рисунок 46. Процентное изменение сопротивления относительно базовой линии в зависимости от приложенного напряжения и числа Рейнольдса. Слева: данные для симметричной сборки актуаторов с противоположно-направленными потоками; справа: данные для асимметричной сборки актуаторов с однонаправленным потоком. По данным [196].

Авторы [196, 197] считают, что величина достигнутого снижения сопротивления зависит от количества элементарных вихрей между соседними поверхностными электродами. Это снижение сопротивления экспоненциально увеличивается при уменьшении $Re_\Theta$ ($Re_\Theta$ - число Рейнольдса, рассчитанное по толщине вытеснения импульса в пограничном слое) как следствие уменьшения количества элементарных вихрей под одновременным контролем. Данное утверждение позволяет провести анализ полученных в [196, 197] зависимостей в системе координат «снижение сопротивления – $Re_\Theta$» (рис. 47).

Рисунок 47, аналоги которого приведены в работе [196] в несколько усеченном варианте, показывает проблему данных работ. С одной стороны, данные действительно показывают падающую экспоненциальную зависимость между снижением сопротивления и числом Рейнольдса потока. С другой стороны, небольшая экстраполяция зависимостей, приведенных в [196, 197], предсказывает наличие области отрицательных абсолютных сопротивлений при низких числах Рейнольдса.

Таким образом, данные [196, 197] содержат в себе некоторое противоречие. Авторы настоящего обзора не берутся однозначно определить природу этого противоречия и могут только предположить, что вносимые актуатором возмущения в пристеночной области приводили в условиях [196] к глобальному отрыву потока. Актуатор и измерительная пластина попадали в область обратных пристеночных течений, которые и вызывали изменение нагрузки на весы. Необходимо отметить, что окончательного решения данной проблемы пока нет, и, по-видимому, необходимы дальнейшие исследования, чтобы подтвердить или опровергнуть возможность столь значительного снижения турбулентного сопротивления с помощью импульсных приповерхностных разрядов. В частности, ответом на вопрос о природе явления могли бы стать синхронные измерения силы сопротивления не только непосредственно на пластине с актуаторами, но и на окружающих ее участках стенки.

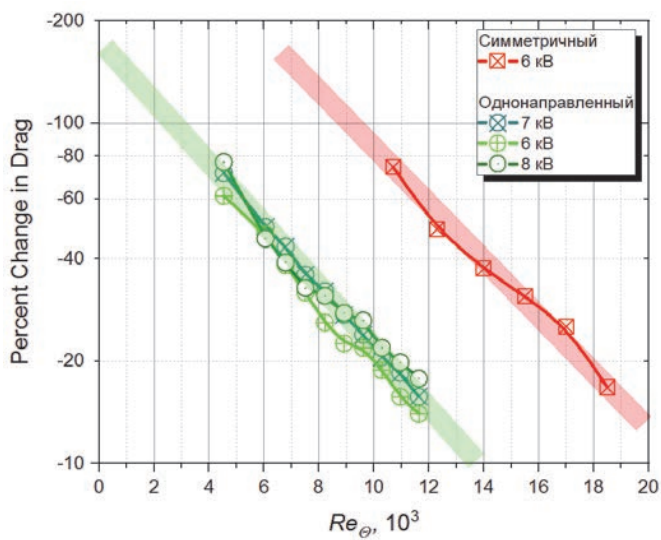

Рисунок 47. Процентное изменение сопротивления относительно базовой точки в зависимости от числа Рейнольдса потока. Точки – данные [196] для симметричного и однонаправленного актуаторов при различных напряжениях на электродах.

# 10. Управление отрывом пограничного слоя на больших углах атаки

## 10.1. Управление отрывом потока с помощью ионного ветра

В отличие от случая с сильными ударными волнами, в большинстве работ, посвященных управлению медленными дозвуковыми течениями, указывается на роль плазменных эффектов (ионного ветра, в частности) в ускорении потока в пограничном слое. Такое ускорение приводит к генерации пристеночных струй газа высокой скорости, что подавляет отрыв пограничного слоя и может быть использовано для управления положением ламинарно-турбулентного перехода [2].

Плазменный слой вблизи поверхности может быть создан различными способами. Например, в работах [198, 199] и более поздних публикациях [200, 201] использовались разряды постоянного тока с электродами, расположенными выше или на поверхности профиля. При этом создаваемый в разряде ионный ветер может ускорить поток в пограничном слое до 3-5 м/с [174].

В [8-12] предложен другой подход для создания плазменного слоя вблизи поверхности с целью управления потоком. Этот подход основан на инициировании поверхностного DBD разряда при наложении переменного синусоидального напряжения. Разряд развивался по поверхности в виде тонких стримеров над нижним закрытым электродом [203]. Этот тип плазменного актуатора и его модификации интенсивно исследуются в настоящее время [2]. В работе [204] продемонстрировано создание несимметричным актуатором силы на уровне 0.2 мН/Вт. Практически тот же результат (0.3 мН/Вт) был получен в [205]. Скорость потока под действием таких актуаторов может достигать 5 м/с [10]. В [202] получены значения индуцированной скорости до 8 м/с. Такое ускорение потока обеспечивает эффективный контроль профиля скорости в пограничном слое, а также его отрыва при скоростях потока в несколько десятков метров в секунду.

Представляется, что физические ограничения на предельную скорость ионного ветра в разряде [206] из-за невозможности повышения постоянного электрического поля выше порога пробоя газа и сильных потерь на трение в пристеночном слое не позволяют значительно продвинуться в улучшении эффективности актуаторов, использующих этот механизм управления потоком. В то же время, в дозвуковой аэродинамике очень важен диапазон скоростей от 100 м/с (скорость взлета и посадки самолетов) до 250 м/с (крейсерская скорость). Таким образом, продвижение в область более высоких скоростей является важной и актуальной задачей.

### 10.2. Управление отрывом потока с помощью импульсного энерговыделения

В работе [207] было предложено использовать в плазменном актуаторе импульсный наносекундный разряд. Значения $E/n$ в разрядах такого типа могут превышать порог пробоя в несколько раз. Высокие значения приведенного электрического поля представляются очевидным преимуществом этого разряда. К достоинствам предложенной идеи также относятся сравнительно низкое энергопотребление в разряде и возможность инициирования таких разрядов в широком диапазоне давлений, скоростей потока и состава газов, в том числе – с повышенной влажностью. Первые эксперименты [207] показали, что с помощью наносекундного импульсного разряда можно надежно управлять отрывом пограничного слоя при скоростях до 75 м/с и погонной потребляемой мощности менее 1 Вт/см.

Далее, воздействие импульсного скользящего разряда на отрыв потока было экспериментально исследовано в [208]. Здесь высокая эффективность импульсного разряда была продемонстрирована вплоть до скоростей 110 м/с. При этом был сделан вывод о том, что основным механизмом воздействия плазмы является возмущение, вносимое в пограничный слой, а не ускорение газа. Было показано, что изменение частоты импульсов в актуаторе позволяет оптимизировать его воздействие на силу сопротивления, подъемную силу и присоединение потока. Оптимальная частота оказалось равной $f_{opt} \sim U_\infty /L$, где $U_\infty$ – скорость основного потока и $L$ – типичное расстояние вдоль поверхности до зоны отрыва. Позднее этот результат был подтвержден в эксперименте [209] для чисел Рейнольдса (по хорде) до $10^6$ и максимальной скорости свободного потока 60 м/с.

Масштабирование воздействия наносекундного импульсного плазменного актуатора было исследовано в [210, 211]. Эксперименты по управления отрывом потока выполнялись на прямоугольном (размеры 0.5×1 m$^2$) крыле при использовании DBD разряда в дозвуковом потоке с числами Рейнольдса $Re = (0.35-0.875)\times10^6$ по хорде. Измерения давления на поверхности и визуализация потока показали, что с помощью плазменных актуаторов можно существенно уменьшить или полностью устранить отрыв потока от крыла, что приводит к восстановлению отрицательного пика давления около передней кромки на верхней поверхности профиля (рис. 48).

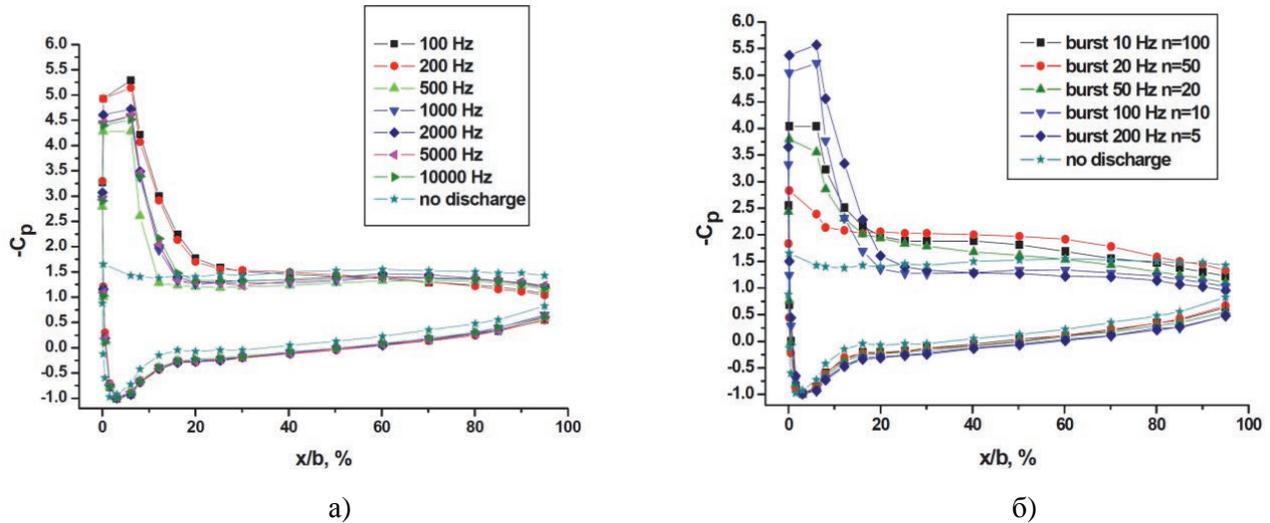

Рисунок 48. Распределение давления по поверхности профиля [210]. Верхняя группа кривых – данные для верхней поверхности, нижняя – для нижней. α = 19$^0$; $U_\infty$ = 19 м/с, $V$ = 24 кВ; $Re = 0.8\times10^6$. а) – периодический режим, б) – режим пачек импульсов.

Данные были получены в широком диапазоне углов атаки, скоростей потока, частот плазменного возбуждения и подводимой мощности. В работе также обсуждалась возможность использования разных импульсов напряжения, в том числе – микросекундные и наносекундные импульсы. Как и в [208], было показано, что эффективность актуатора сильно зависит от частоты разряда (рисунок 49).

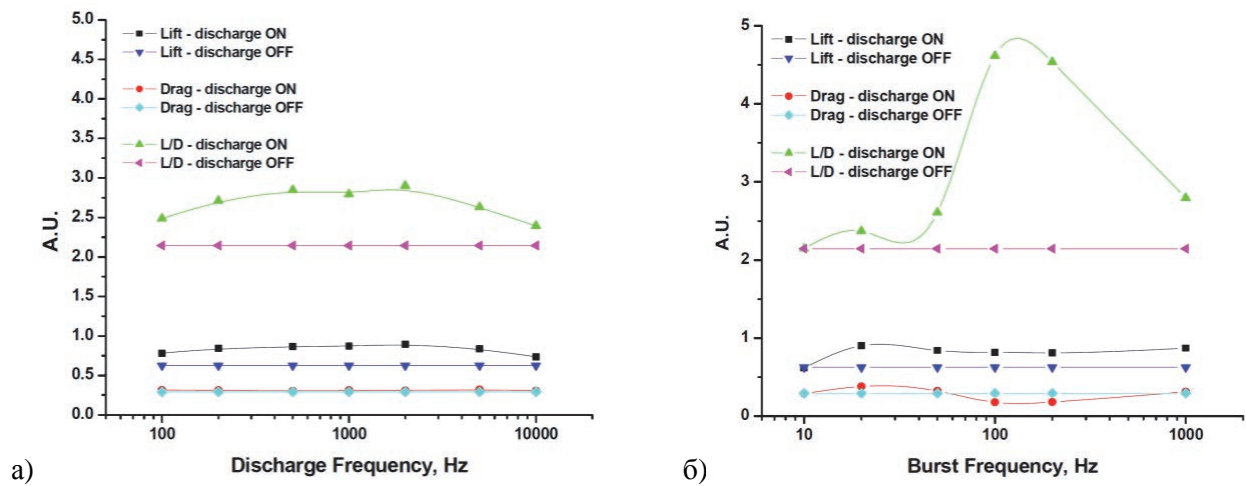

Рисунок 49. Подъемная сила, сила сопротивления и их отношение в зависимости от частоты действия актуатора (α = 22$^0$, $U_\infty$ = 17.4 м/с). а) – периодический режим, $P$ = 2.5-250 Вт при $f = 10^2$-$10^4$ Гц, соответственно; б) – режим пачек импульсов при $P = 25$ Вт [210].

В [212] экспериментально изучался отрыв потока на прямоугольном крыле при использовании наносекундного SDBD разряда в случае трансзвукового течения при числах Рейнольдса (по хорде) в диапазоне (0.5-2)×10$^6$. Эксперимент выполнялся в аэродинамической трубе на трансзвуковой скорости. Система была модернизирована для экспериментов в импульсном режиме. Сопло с рабочей камерой работало при числах Маха от $M = 0.6$ до $M = 0.9$. Фото модели и схема измерения давления приведены на рисунке 50. Было исследовано влияние разряда на картину течения вблизи поверхности. Частота разряда в экспериментах была $f = 5$ кГц. Использовались высоковольтные импульсы амплитудой 25 кВ и длительностью 12 нс. Энергия в одном импульсе составила 10 мДж. Угол атаки менялся от 0 до 30$^0$.

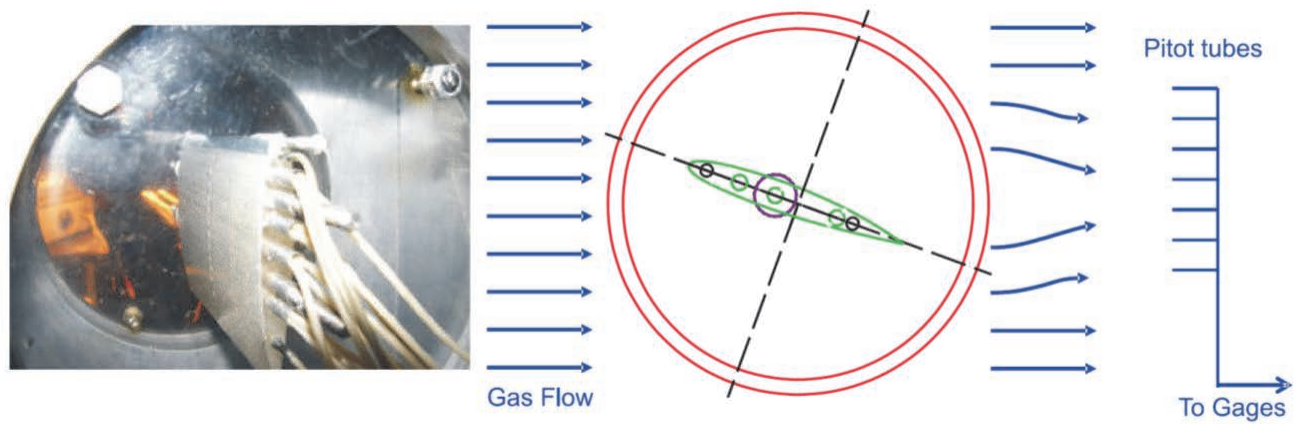

Рисунок 50. Фотография модели и схема измерения давления в следе. Хорда модели 15 см, размах по передней кромке 30 см. Распределение давления измерялось в следе модели и на ее поверхности [212].

При малых углах атаки наблюдался режим течения без отрыва, но образовывалась сверхзвуковая зона и формировалась висячая ударная волна. Эти режимы четко фиксировались по скачку давления в середине поверхности профиля. Такой скачок связан с положением висячего скачка уплотнения (рисунок 51). Влияние разряда при малых углах атаки в диапазоне α = 0-15$^0$ было незначительным. При более высоких значениях угла атаки наблюдался отрыв потока, и характер распределения давления менялся (рисунок 51). В этом случае включение разряда возвращало поток в режим безотрывного обтекания с образованием висячего скачка уплотнения над верхней поверхностью и соответствующим ему скачкообразным изменением давления. Плазменный актуатор также позволил удалить высокочастотные пульсации давления в следе модели. Представленные на рисунке 51,б данные демонстрируют снижение уровня пульсаций при числе Маха M = 0.7. Датчик №1 регистрировал давление на верхней поверхности модели и служил индикатором изменения угла атаки. Датчики №2-4 были расположены в следе модели и регистрировали исчезновение пульсаций давления при включении разряда. Этот эффект наблюдался при больших углах атаки (начиная с α = 24$^0$) для чисел Маха $M =$ 0.65-0.75. Среднее значение давления вблизи поверхности модели при этом существенно не менялось, в то время как высокочастотные пульсации амплитуды резко снижались.

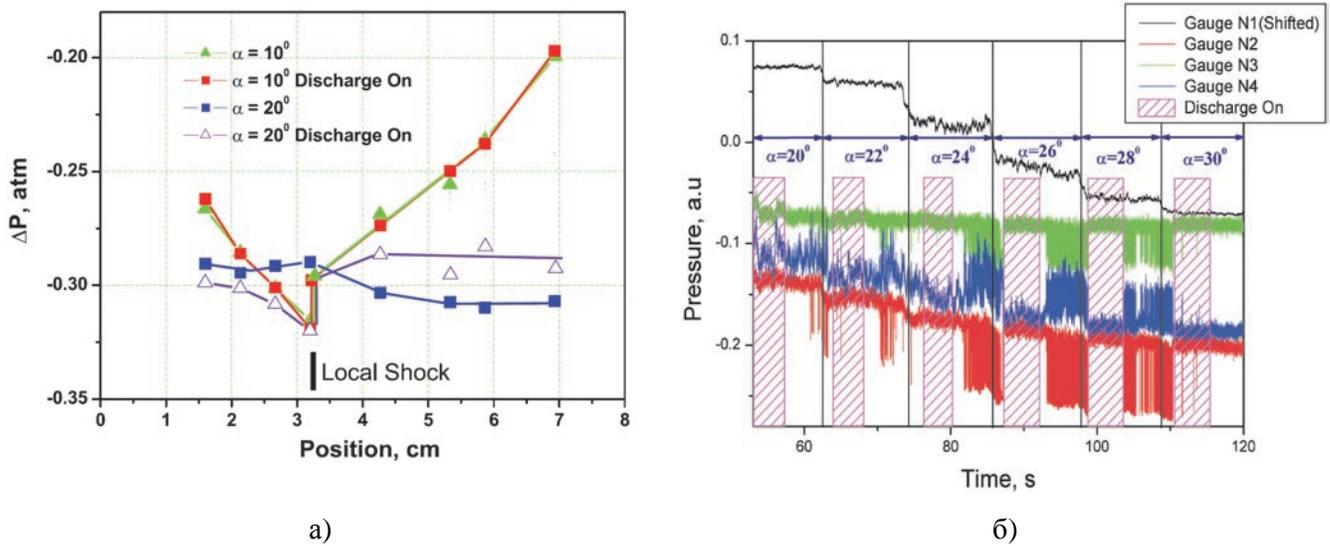

а) б)

Рисунок 51. (а) Распределение давления на поверхности модели с включенным разрядом и без разряда. Число Маха $M = 0.74$; (б) Снижение уровня шума в следе модели. Число Маха $M = 0.7$. Статическое давление $P = 1$ атм [212].

Таким образом, анализ управления отрывом потока с помощью SDBD актуатора для профиля C-141 при трансзвуковых скоростях ($M = 0.65 – 0.75$) привел к выводу об эффективном подавлении отрыва потока при углах атаки от $18^0$ до $30^0$. Разряд уменьшал как отрыв потока, так и высокочастотные пульсации давления в следе. Эти эксперименты продемонстрировали возможность управления отрывом трансзвукового потока с помощью низкоэнергичных импульсных наносекундных разрядов. Важно, что наносекундный импульсный разряд продемонстрировал чрезвычайно высокую эффективность работы в аэродинамических плазменных актуаторах для широкого диапазона скоростей ($M = 0.03 – 0.75$) и чисел Рейнольдса ($Re = 10^4\text{-}2\times10^6$).

Для технологических приложений важно понять физику работы наносекундных плазменных актуаторов (см. раздел «Наносекундные поверхностные барьерные разряды») и отличия этого типа разряда от других SDBD разрядов с точки зрения их эффективности [33, 35, 213-217].

С этой точки зрения имеется несколько важных этапов. В [208] экспериментально показано, что импульсный наносекундный высоковольтный разряд, позволяя управлять отрывом потока в широком диапазоне скоростей набегающего потока, не приводит к ускорению газа. В работе [218] численно анализировалась возможность использования AC-SDBD актуатора, генерирующего поверхностную струю навстречу основному потоку. Согласно расчетам, такой актуатор вызывал переход пограничного слоя в турбулентный режим, создавая более «полный» профиль скорости. Эта особенность была использована для задержки отрыва потока на профиле NACA-0015 при больших углах атаки с использованием импульсного актуатора со встречным потоком. Таким образом, в [218] показано, что ускорение газа вдоль потока не является необходимым для управления пограничным слоем. С учетом того, что AC-SDBD приводит к существенному нагреву газа в пограничном слое из-за рекомбинации плазмы и релаксации внутренних степеней свободы газа, механизм действия плазменных актуаторов на основе медленно меняющихся напряжений все еще нуждается в детальном обосновании. Возможно, именно не слишком эффективный (по сравнению с наносекундными разрядами) нагрев газа в пограничном слое является определяющим механизмом влияния AC-SDBD актуаторов на поток, а незначительное ускорение приповерхностного слоя из-за электростатических эффектов является побочным и малозначимым эффектом.

В [213] экспериментально продемонстрирован механизм влияния импульсного наносекундного высоковольтного разряда на отрыв пограничного слоя. Было показано, что быстрая термализация неравновесной плазмы (в масштабе времени сотен наносекунд) производит в разрядной зоне слой нагретого газа высокого давления, после чего следует образование сильной ударной волны. Было высказано предположение о том, что распространение области нагретого газа вызывает сильные возмущения потока и способствует его присоединению благодаря образованию крупных вихрей в слое, разделяющем свободный поток и отрывной пузырь [213]. Позже измерения распределения скоростей потока [219] показали, что плазма наносекундного SDBD выступает в качестве импульсного источника возмущения большой амплитуды, которые управляют устойчивостью потока и приводят к образованию когерентных вихрей на границе отрывной зоны на закритических углах атаки. Эти когерентные структуры взаимодействуют с основным потоком, приводя к перемешиванию газа основного потока и отрывного пузыря. Численное моделирование развития ns-SDBD также показывает быстрое формирование плазменного слоя, нагрев газа и генерацию ударных волн [32, 220].

В [215] подробно исследованы процесс взаимодействия плазменного слоя наносекундного импульсного разряда с потоком и формирования возмущений и вихрей. В эксперименте использовалась модель профиля NACA 63-618 с хордой 20 см и размахом 40 см. Исследовалась работа нескольких различных актуаторов, в том числе одиночных, двойных и тройных. Скорость потока составляла 30 м/с. Некоторые из полученных результатов показаны на рисунке 52. На рисунке ясно видны созданная актуаторами ударная волна и структура больших вихрей, развивающаяся через 40 мкс после разряда [215]. Было отмечено, что после 2-3 разрядов картина течения полностью изменялась: поток присоединялся, а зона отрыва смещалась вниз по потоку. Оказалось, что размещение второго актуатора в точке, где происходил отрыв течения после установки первого актуатора, приводил к дальнейшему сдвигу точки отрыва вниз по потоку. С использованием трех актуаторов это позволило добиться присоединения потока по всему профилю вплоть до угла атаки $\alpha = 32^0$. Суммарная потребляемая мощность была менее 1 Вт (25 мВт/см) для профиля с размерами 40×20 см$^2$ при скорости потока 30 м/с.

Типичное время реакции системы составляло 10-15 мс, и было близко к времени распространения вихря вдоль поверхности профиля, что соответствует безразмерной частоте возбуждающей силы $F_c^+ \sim 1$ (рисунок 52). На этом рисунке видно, как возмущения, порождаемые импульсным актуатором, инициируют развитие неустойчивости в слое смешения. Эта неустойчивость распространяется вдоль слоя; дополнительное перемешивание приносит дополнительный импульс от основного потока в отрывную область и присоединяет поток. Следует отметить, что энергия разряда здесь играет второстепенную роль: два различных режима (режим повторяющихся импульсов и цуговый режим) приведенные на рисунке в нечетных и четных колонках, соответственно, демонстрируют почти одинаковую динамику присоединения потока, хотя энергия разряда во втором случае в 10 раз больше. Аналогичный вывод был сделан в работе [221], где исследовалась динамика развития возмущения в зависимости от энергии, вложенной в плазму. Такой эффект возникает потому, что для управления потоком используется развитие неустойчивости, конечная амплитуда которой слабо зависит от амплитуды начального возмущения.

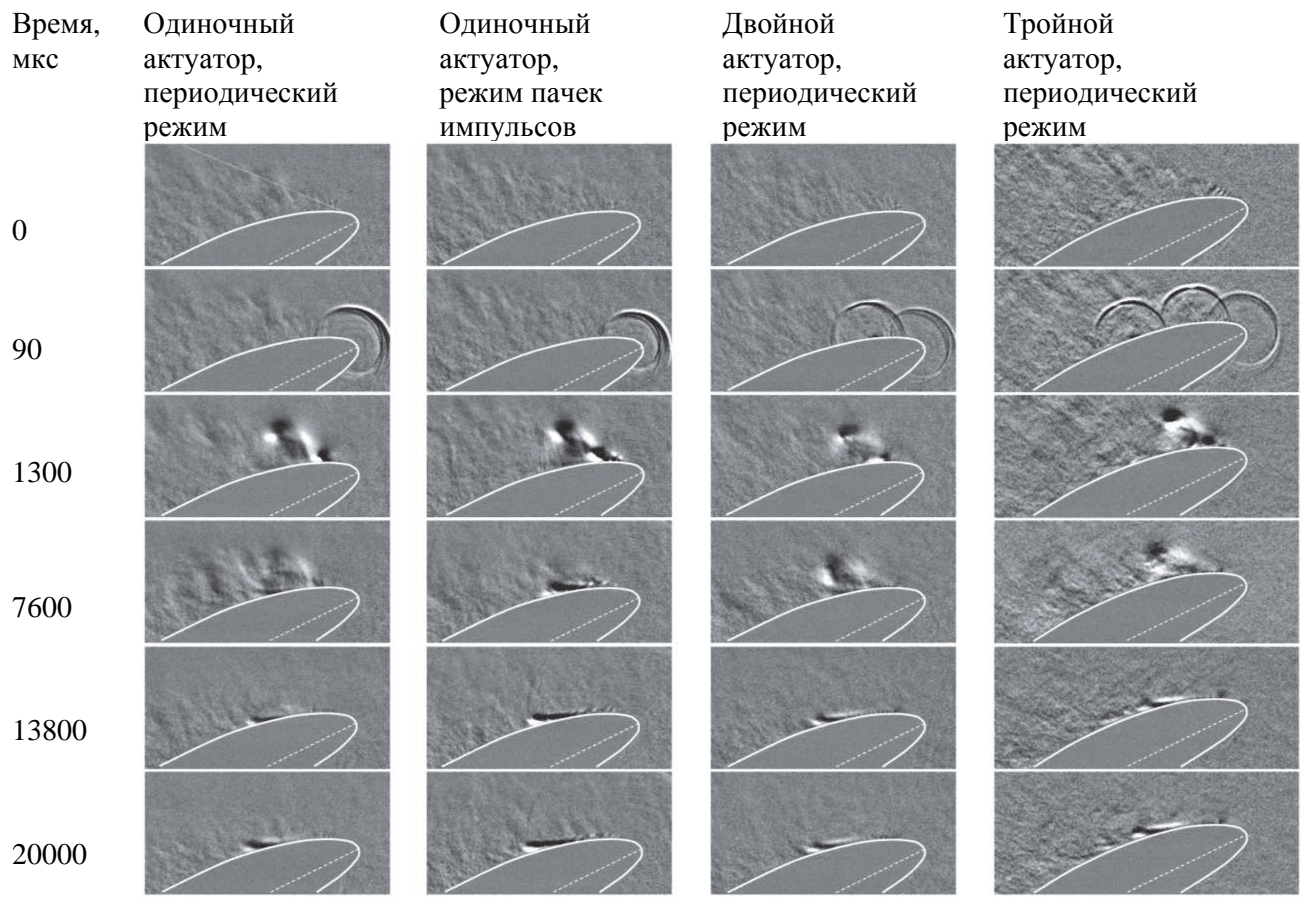

Рисунок 52. Динамика присоединения пограничного слоя. $U_\infty = 30$ м/с, $\alpha = 26^0$, профиль NACA 63-618 длиной 20 см и размахом 40 см. Энергия разряда 5 мДж в импульсе, частота разряда 200 Гц в импульсном и цуговом режимах, в одном цуге 10 импульсов [215].

Нужно отметить, что развитие неустойчивости в слое смешения вызывается не ударной волной, возникающей немедленно после термализации плазмы, и покидающей область взаимодействия на временах менее 100 мкс (вторая строчка на рис.52), а значительно позднее, когда газ низкой плотности из области разряда достигает начала отрывной зоны (1 миллисекунда и далее, строчка 3 на рис. 52). В работе [215] такой подход был использован для моделирования развития неустойчивости при воздействии ns-SDBD актуатора на отрывной поток за ступенькой. При этом классическая неустойчивость Кельвина-Гельмгольца, развивающаяся на границе основного потока и отрывного течения из-за присутствующего в этой области градиента скорости, провоцируется попаданием в область градиента скорости именно пятна газа низкой плотности, а не начальной ударной волны. Ударная волна не влияет на развитие неустойчивости в этих условиях, поскольку разница плотностей газа в отрывной зоне и основном потоке при дозвуковых скоростях течения мала, и неустойчивость Рихтмайера-Мешкова, возникающая между двумя контактирующими сплошными средами различной плотности, когда поверхность раздела испытывает импульс ускорения, например при прохождении ударной волны, не развивается. Такой же вывод сделан в недавней работе [222], где анализировалась роль различных механизмов на развитие возмущения.

В работе [223] исследовалось воздействие импульсного оптического разряда около передней кромки профиля на отрыв пограничного слоя при нулевом угле атаки. Лазерная искра создавалась поперек передней кромки профиля и потока, формируя область сильно-нагретого газа непосредственно перед моделью. Было показано, что ударная волна, образующаяся в результате расширения нагретой области, не влияет на отрывной пузырь. Воздействие на отрывную область было объяснено сильной турбулентностью в области газа низкой плотности, созданной лазерной искрой.

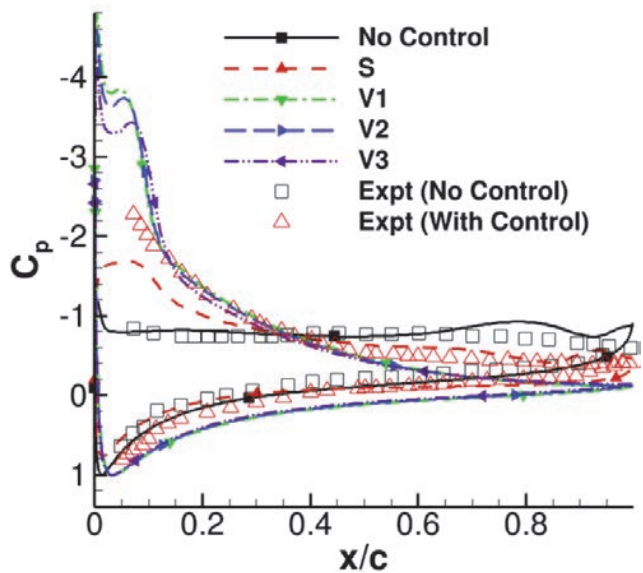

Рисунок 53. Распределение коэффициента давления по поверхности профиля. Верхняя группа кривых – данные для верхней поверхности, нижняя – для нижней. Эксперимент [219] и моделирование без плазменного актуатора и при его включении [227] при $Re = 2.5 \times 10^5$.

В работе [224], где исследовалось влияние ns-SDBD на отрыв потока, были получены практически идентичные картины взаимодействия возмущения с отрывным течением. Авторами [224] высказывается предположение, что взаимодействие актуатора с отрывным течением может быть обусловлено, кроме механизмов быстрого нагрева (тушение электронно-возбужденных молекул и рекомбинация заряженных частиц [141, 225, 226]), механизмами относительно медленной релаксации колебательных степеней свободы. Очевидно, что энергия, запасенная в колебательных степенях свободы, будет термализована. Однако, при температурах близких к комнатным, релаксация основного резервуара колебательной энергии – возбужденных состояний молекулярного азота – происходит очень медленно. На типичных временах взаимодействия (менее 1 миллисекунды на рис. 52), эта энергия остается замороженной в колебательных степенях свободы и не может участвовать в процессе. Кроме того, типичные величины приведенного электрического поля, реализующиеся в ns-SDBD, приводят к преимущественному энерговкладу в электронные степени свободы и ионизацию, что также резко ограничивает роль колебательных степеней свободы в данном процессе.

Косвенным подтверждением отсутствия заметного влияния «медленного» нагрева на эффективность работы наносекундных актуаторов являются результаты численного моделирования воздействия таких актуаторов на отрывные течения [227, 228]. В этих работах воздействие плазменного наносекундного актуатора моделировалось как распределенное в пространстве мгновенное энерговыделение. Хорошее совпадение результатов моделирования с экспериментальными данными по управлению отрывом потока говорит об отсутствии заметного влияния медленной релаксации энергии колебательных степеней свободы на процесс.

Так, в работе [227] получено хорошее совпадение с результатами эксперимента [219] по распределению давления по поверхности профиля (рис.53) для разных вариантов расчета с мгновенным объемным выделением энергии (V1-V3 на рис.53). Отметим, что чисто поверхностный нагрев (вариант S) дает заметно заниженную эффективность контроля отрыва потока. Этот вывод подтверждает данные [210] (рис.48) и анализ [216], где нагрев стенки и генерация ударной волны рассматривались как каналы потери энергии, снижающие эффективность актуатора.

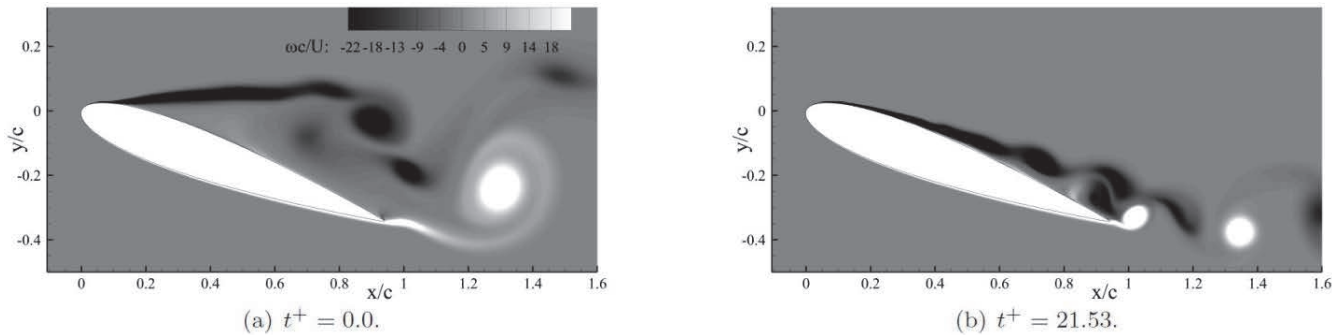

Рисунок 54. Контуры завихренности в разное время после однократного срабатывания ns-SDBD актуатора при $Re = 1.15\times10^6$. [228].

В работе [228] аналогичные результаты были получены для потока с числом Рейнольдса $Re = 1.15\times10^6$ (рис. 54). Длина хорды профиля равнялась $c = 0.2032$ м, скорость свободного потока $U_\infty = 93$ м/с и угол атаки $\alpha = 20^0$. Пиковое напряжение высоковольтного импульса $V_{peak} = 8.4$ кВ. Частота повторения импульсов составляла 915 Гц для этого режима, что приводит к значению безразмерной частоты $F_c^+ = f_c/U_\infty \simeq 2$. Предполагалось, что в быстрый нагрев газа в разряде переходит 35% энергии импульса (0.092 мДж/см), и время выделения энергии $\tau = 100$ нс. Длина и толщина плазменного слоя принималась равными 5 мм и 0.4 мм, соответственно. Рассчитанная в этой работе динамика развития вихря и присоединения потока качественно воспроизводит результат эксперимента [215] (рис. 52). Таким образом, механизм быстрого нагрева газа при распаде плазмы ns-SDBD позволяет полностью объяснить развитие возмущений в слое смешения на границе отрывной зоны, перенос импульса к поверхности профиля и присоединение потока.

### 10.3. Влияние положения актуатора на эффективность управления потоком

Необходимо признать, что систематических исследований влияния положения актуатора на эффективность подавления отрыва потока практически не проводились. В ранних работах по управлению потоком с помощью AC-SDBD в качестве основного механизма такого управления постулировалось увеличение скорости потока в пограничном слое. Такой подход привел к заключению, что актуатор может быть эффективен как при расположении перед зоной отрыва, так и непосредственно в области отрывного вихря. С другой стороны, очевидно, что ns-SDBD актуатор, который не может создавать пристеночную струю, в отрывной зоне будет неэффективен. В работе [207] для наносекундных актуаторов было использовано расположение, типичное для AC-режима – актуаторы располагались в несколько рядов на различных расстояниях от передней кромки. Однако уже в работе [208] были исследованы две конфигурации электродов – с расположением параллельно и перпендикулярно передней кромке (рис. 55). Было показано, что во втором случае эффективность актуатора заметно выше, чем в первом, и сделан вывод, что для режимов течения с фиксированным углом атаки профиля значительно более эффективен актуатор, находящийся перед точкой отрыва потока.

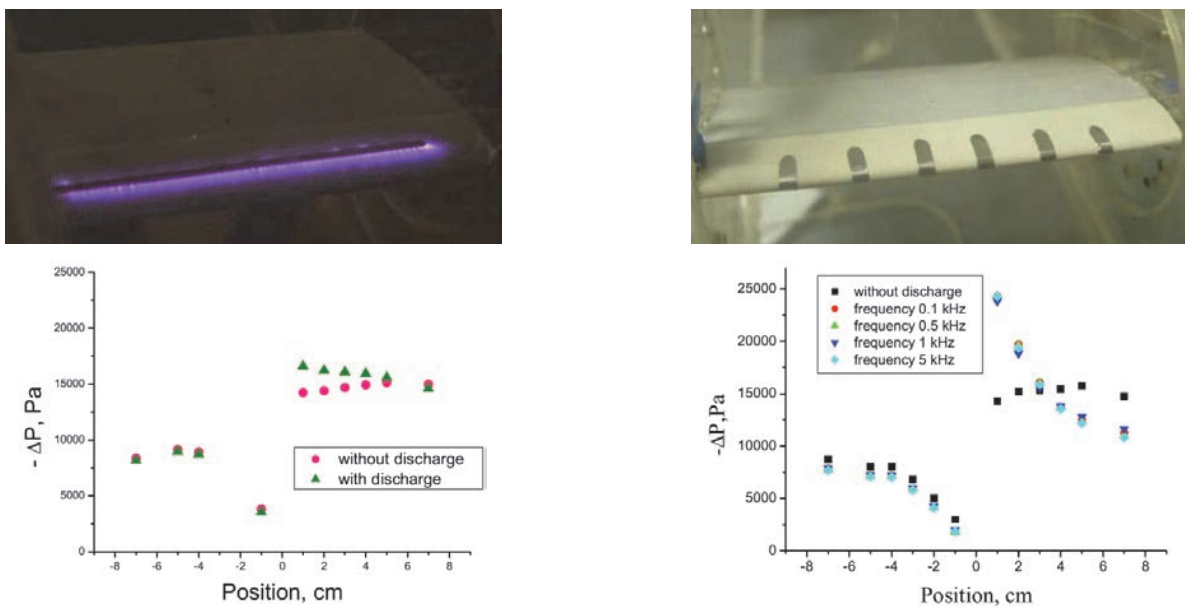

Рисунок 55. Различные конфигурации электродов для ns-SDBD актуатора (вверху) и распределение статического давления по поверхности профиля (внизу). Угол атаки $\alpha = 17^0$; $U_\infty = 110$ м/с; длительность импульса 25 нс; напряжение 25 кВ. Слева: электроды расположены параллельно передней кромке; справа: электроды расположены перпендикулярно передней кромке [208].

Характерным примером зависимости эффективности актуатора от его положения на профиле крыла является сравнение результатов работы [212] и [229], где исследовалось управление трансзвуковым обтеканием с помощью ns-SDBD актуаторов. В работе [212] актуатор располагался на передней кромке профиля, что обеспечивало сильный отклик потока на воздействие (рисунок 51,а). В более поздних работах [229, 230] актуатор был расположен в центре профиля, практически в области висячего скачка уплотнения. В отличие от [212], в работах [229, 230] влияния актуатора на распределение давления по поверхности профиля в трансзвуковых режимах практически не обнаружено. Таким образом, для эффективной работы ns-SDBD актуатора необходимо его размещение вблизи или непосредственно на передней кромке профиля. Отметим обратную – по сравнению со стандартной – конфигурацию электродной системы. Изолированный электрод вынесен вперед по потоку по отношению к внешнему, что приводит к развитию разряда от передней кромки внешнего электрода вверх по потоку.

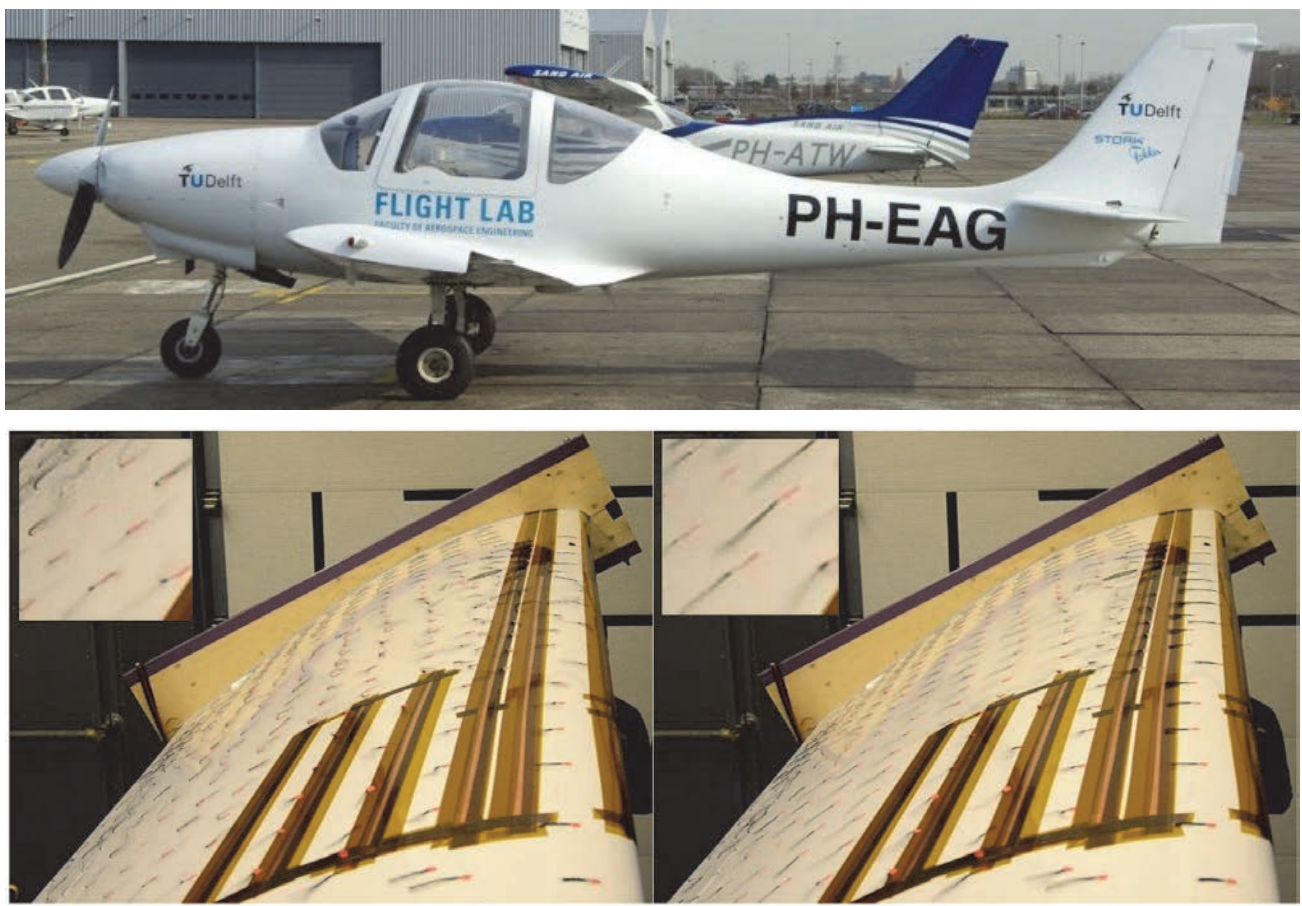

Рисунок 56. Вверху: самолет EAGLET EE-10. Внизу: визуализация отрыва потока методом нитей для профиля NACA 63-415 при скорости 30 м/с, угле атаки $27^0$ и $Re_c = 2.4 \times 10^6$ [231]. Слева: актуаторы выключены. Справа: актуаторы включены. Вставки показывают увеличенную картину визуализации пристеночного течения с помощью нитей. При выключенном актуаторе (слева) пристеночный поток направлен против, при включенном (справа) – по направлению основного течения.

Размещение актуатора на передней кромке, однако, вызывает значительные сложности, когда размер крыла становится большим. Проблема масштабирования заключается в том, что возмущение, создаваемое актуатором, быстро затухает с увеличением расстояния от электродов из-за диффузии и теплопроводности. Поскольку точка отрыва потока может перемещаться по хорде в зависимости от выбранного профиля крыла, угла атаки и скорости, возникает соблазн использовать несколько последовательных систем электродов для гарантированного попадания возмущений в требуемый диапазон расстояний от точки отрыва. В работе [231] было экспериментально продемонстрировано влияние наносекундных актуаторов на крупномасштабное присоединение потока к крылу самолета EAGLET EE-10. Одна консоль крыла этого самолета (максимальная толщина в корне 25 см, длина более 300 см, средняя хорда 120 см, профиль NACA 63-415) была закреплена на статическом стенде в аэродинамической трубе TU Delft (рис. 56).

Крыло находилось на существенно закритическом угле атаки $\alpha = 27^0$. Число Рейнольдса, вычисленное по хорде профиля, равно $Re = 2.4 \times 10^6$. Актуаторы были расположены на профиле в шести положениях: на передней кромке, на расстоянии 15, 20, 30, 40 и 50 см вниз по потоку. При этом первые три актуатора были выполнены на всю длину консоли, а три последующие – на половину (рис. 56).

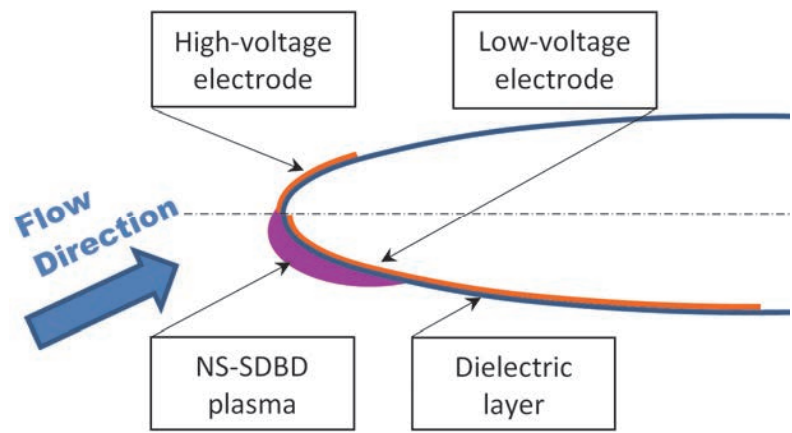

Рисунок 57. Схема расположения ns-SDBD актуатора на профиле. [232]

Хорошо видно, что при выключенных актуаторах поток отрывается от крыла практически сразу за передней кромкой (рис. 56, слева). Включение актуаторов приводило к присоединению потока. При этом наличие трех дополнительных актуаторов ниже по потоку улучшало качество обтекания (рис. 56, справа). Таким образом, при высоких числах Рейнольдса и больших длинах хорды использование нескольких актуаторов друг за другом позволяет улучшить управление потоком. В то же время положение первого актуатора должно быть вблизи передней кромки крыла.

В работе [232] было предложено размещение этого первого актуатора еще выше по потоку – таким образом, чтобы кромка высоковольтного электрода совпадала с передней кромкой профиля, а весь разряд развивался бы на его нижней поверхности (рис. 57). В этом случае при положительных углах атаки точка торможения потока смещается на нижнюю поверхность профиля и оказывается выше по потоку, чем зона максимального энерговклада разряда, которая расположена вблизи передней кромки электрода. Это гарантирует снос возмущения, создаваемого актуатором, к точке отрыва потока на любых положительных углах атаки. При таком положении актуатора вообще не наблюдается снижение подъемной силы крыла на больших углах атаки, а максимальное превышение подъемной силы по сравнению с экспериментами с выключенным актуатором достигает величины $C_L(\text{ON})/C_L(\text{OFF})$ ~ 2.14 (рис. 58).

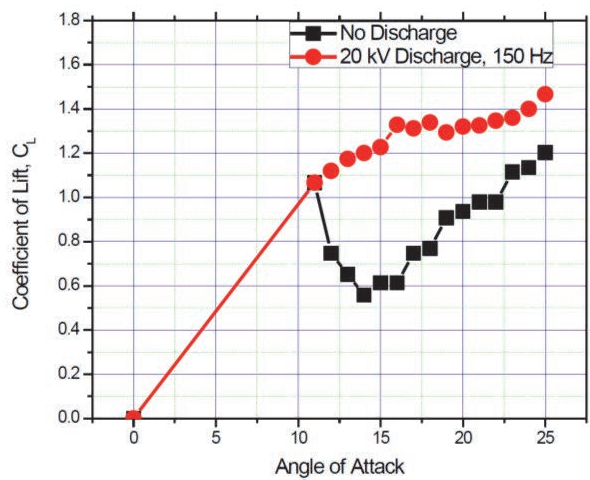

Рисунок 58. Коэффициент подъемной силы в зависимости от угла атаки [232]. Мощность плазменного актуатора 0.08 Вт/см, частота импульсов $f$ = 150 Гц, энергия на один импульс 0.6 мДж/см, скорость потока 31 м/с. Профиль NACA-0015, длина хорды 15 см, $Re = 3.06 \times 10^5$, $F^+_c = 0.73$.

В отличие от работ [232, 213], в работах [233, 234] актуатор использовался в стандартной геометрии: открытый электрод располагался выше по потоку, закрытый – ниже. Кромка открытого электрода, с которой стартовал разряд, была не на оси профиля, а смещена на 0.5% хорды от передней кромки. Поскольку в [233, 234] также использовался профиль NACA-0015 при практически одинаковых числах Рейнольдса ($Re = 3.06 \times 10^5$ для [232] и $Re = 2.68 \times 10^5$ для [234]), появляется возможность прямого сравнения эффективности этих двух геометрий ns-SDBD актуаторов.

Приведенные на рис. 59 результаты измерений коэффициента подъёмной силы в [234] показывают заметный спад на углах атаки больше $\alpha = 22^0$, в то время как для конфигурации [232] наблюдается устойчивый рост $C_L$ по крайней мере до угла атаки $\alpha = 25^0$. Максимальное увеличение подъемной силы в [234] при включении актуатора достигает только $C_L(\text{ON})/C_L(\text{OFF})$ ~ 1.61 (в [232] $C_L(\text{ON})/C_L(\text{OFF})$ ~ 2.14). Также нужно отметить, что для достижения заметно более слабого эффекта в [234] требуется более высокая частота работы актуатора ($F^+_c = 2.96$ [234] и $F^+_c = 0.73$ в [232]). Таким образом, положение и геометрия актуатора оказывают заметное влияние на эффективность его работы. По-видимому, можно утверждать, что оптимальное положение зоны максимального энерговыделения ns-SDBD находится вблизи центральной точки профиля на его передней кромке. Перемещение наносекундного актуатора на верхнюю поверхность профиля снижает его эффективность. Единственный вариант, когда использование таких актуаторов становится целесообразным, это случай, когда такие актуаторы используются в комбинации с актуатором на передней кромке для усиления эффективности последнего на больших профилях и/или в случае динамического отрыва потока, который будет обсуждаться в следующей части.

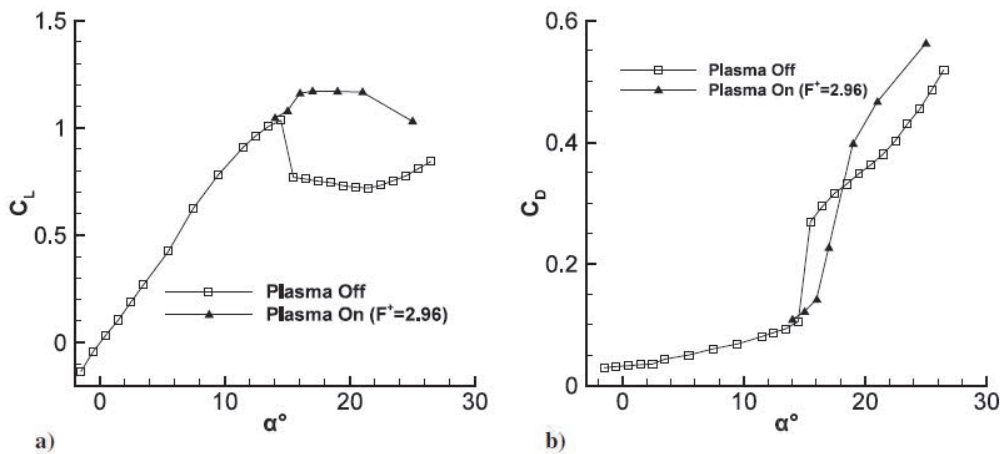

Рисунок 59. Влияние ns-SDBD на коэффициенты подъема и сопротивления в зависимости от угла атаки α. a) $C_L$; b) $C_D$ при $Re = 2.68 \times 10^5$, $F^+_c = 2.96$ [234].

Отметим, что максимум эффективности ns-SDBD актуатора вблизи $F^+_c = 1.0$ отмечен также в работе [235]. В то же время в работе [236] при исследовании управления отрывным течением на профиле Boeing Vertol VR7 актуатор был расположен при x/c = 0.04. При сравнимых числах Рейнольдса $Re = (0.217\text{-}0.307) \times 10^6$ даже для приведенной частоты $F^+_c = 4.21$ наблюдалось только частичное присоединение потока. Отличие от результатов работ [232, 235] было объяснено значительной разницей в характере аэродинамического профиля, особенно вблизи передней кромки. Такое объяснение отчасти верно, однако основное отличие заключается в сдвиге актуатора в работе [236] ниже по потоку, где зона энерговыделения частично попадает за линию отрыва потока и работа актуатора становится менее эффективной.

## 11. Управление динамическим отрывом потока

Задача управления динамическим отрывом потока важна потому, что на практике любое отрывное течение является нестационарным. Это относится и к самолетам, которые выходят на закритические углы атаки только на очень короткое время, сравнимое со временем перестройки течения; и к вертолетам, где на больших скоростях полета становится важна проблема отрыва потока с лопасти, движущейся назад; и к ветроэлектростанциям, где резкий порыв ветра может сорвать поток с лопастей до того, как ротор поменяет направление в пространстве и скорость вращения. Несмотря на это, задача контроля динамического обтекания с помощью импульсных разрядов рассматривалась только в нескольких работах из-за ее сложности.

По-видимому, одной из первых работ, где была предпринята попытка управления динамическим отрывом потока с помощью ns-SDBD, была работа [237]. Использовался профиль NACA-0015, установленный в трансзвуковой аэродинамической трубе. Угол атаки изменялся по синусоидальному закону. Приведенная частота актуатора находилась в диапазоне значений $F_c^+ = 0.78\text{-}6.06$ для $M = 0.2$ и от 0.39 до 2.03 для $M = 0.4$. Несмотря на то, что в работе был сделан оптимистический вывод о наличии возможности управления динамическим отрывом в таких условиях, больших изменений для коэффициента подъемной силы и ее распределения по профилю обнаружено не было (рис. 60). Возможно, слабое влияние воздействия актуатора на течение были связаны как с высоким уровнем начальных возмущений в аэродинамической трубе [237] (отрыв потока на неподвижном профиле не фиксировался до угла атаки $\alpha = 16^0$), так и с неудачным расположением актуатора на 4% хорды профиля – далеко от передней кромки.

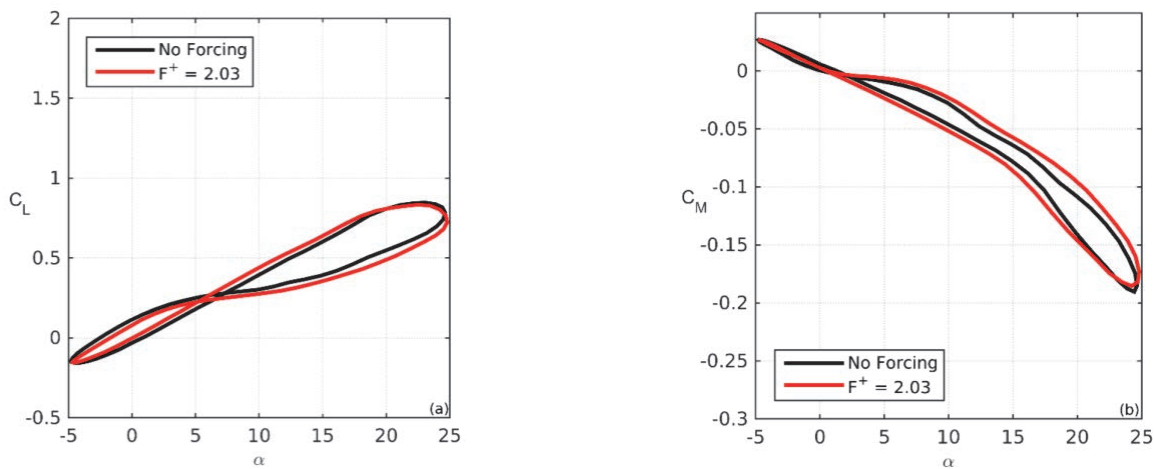

Рисунок 60. Зависимость коэффициентов подъемной силы ($C_L$) и момента силы ($C_M$) от угла атаки [237] при $M = 0.4$, $k = 0.05$, $F^+ = 2.03$, $Re = 2.2\times10^6$. Кривые соответствуют измерениям при выключенном и включенном актуаторе.

## 11.1. Динамика перестройки течения

Возможность управления динамическим отрывом потока зависит от времени присоединения потока $\tau_{\text{att}}$ после включения актуатора. Очевидно, что при выполнении условия $\tau_{\text{att}} \ll 1/f$, где $f$ – частота изменения угла атаки, задача управления динамическим отрывом не будет значительно отличаться от задачи управления статическим отрывом. Таким образом, ключевым становится вопрос о скорости реакции системы на воздействие, создаваемое актуатором.

Этот вопрос был исследован в работе [238], где были проведены прямые измерения сил и моментов, воздействующих на модельный профиль NACA-0015 при динамическом включении и выключении актуатора, расположенного на его передней кромке. Как показали измерения (рис. 61), статический отрыв потока на профиле NACA-0015 происходит на угле атаки $\alpha > 12^0$, что хорошо совпадает с известными данными. Подъёмная сила профиля при этом снижается практически в 1.8 раза при одновременном росте силы сопротивления. Как видно из данных [238], включение актуатора даже при небольшой приведенной частоте $F_c^+ = 0.725$ приводит к присоединению потока и росту подъёмной силы до величины, превышающей подъёмную силу на критическом угле атаки на 20% (рис. 61).

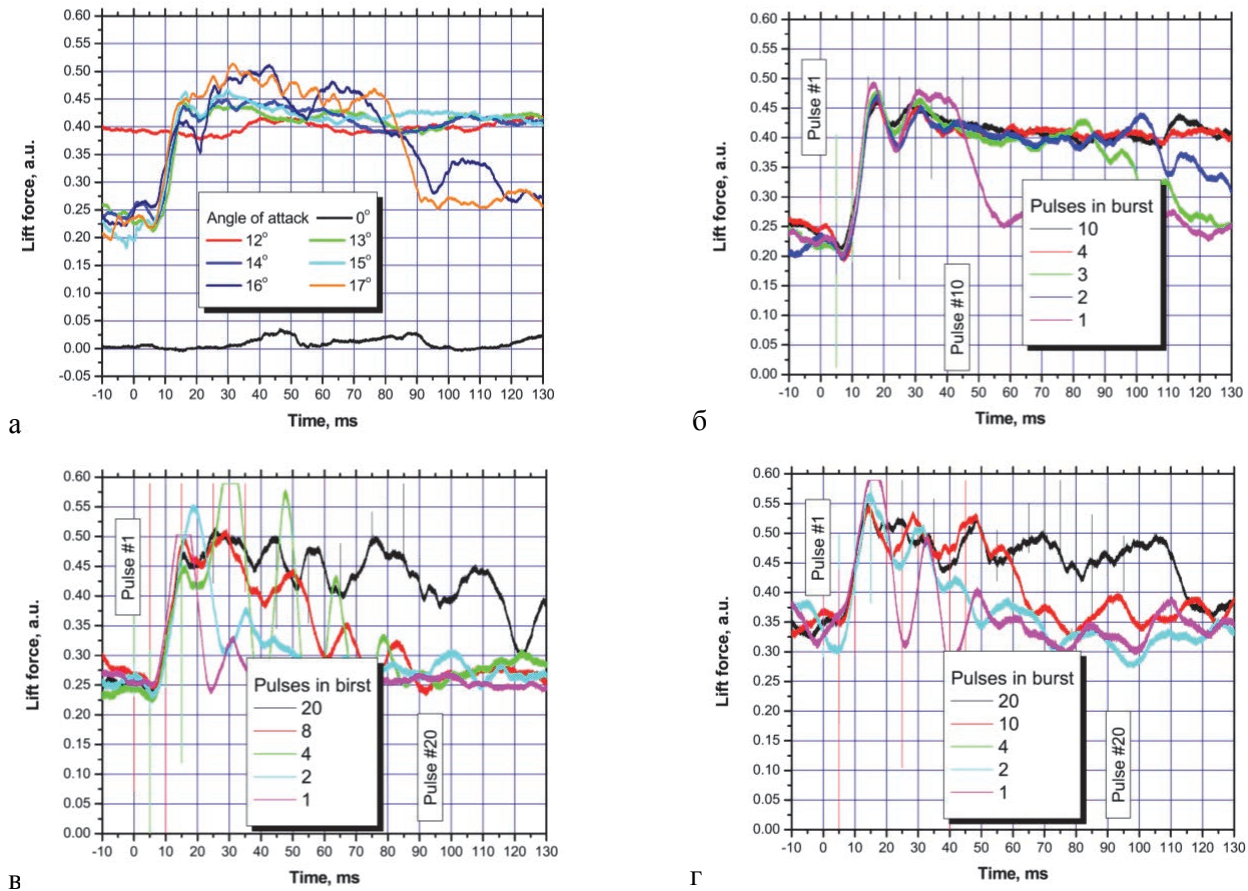

Рисунок 61. Динамика присоединения потока и изменения подъемной силы [238]. а) подъемная сила для разных углов атаки. Первый импульс актуатора происходит при $t = 0$ с. Средняя мощность разряда $P = 2$ Вт, частота следования импульсов $f = 150$ Гц ($F_c^+ = 0.725$), энергия в импульсе $Q = 15$ мДж. б, в, г) – подъемная сила в зависимости от количества импульсов в цуге. $U_\infty = 31$ м/с. $f = 200$ Гц ($F^+ = 0.97$). б) $\alpha = 16°$; в) $\alpha = 20°$; г) $\alpha = 26°$. Профиль NACA-0015, $c = 15$ см.

Еще одним важным результатом работы [238] является независимость механизма перестройки потока от наличия или отсутствия последующих импульсов в цуге. Как правило, уже первый импульс актуатора вызывает перестройку течения, присоединение потока и рост подъемной силы. В дальнейшем при относительно небольших углах атаки наблюдается сохранение данной конфигурации потока довольно длительное время (рис. 61). Только при максимальных исследованных в [238] углах атаки ($\alpha = 26^0$) поток отрывался снова практически в момент прекращения воздействия.

Показательным фактором является скорость изменения подъёмной силы профиля. В работе [238] было установлено, что время перестройки течения не превышало нескольких миллисекунд после включения актуатора (рис. 62). Интересно отметить, что время перестройки заметно сокращается при увеличении угла атаки. По-видимому, такое сокращение времени реакции системы в этом диапазоне параметров связано с усилением тангенциального разрыва на границе между отрывным пузырем и основным потоком, что приводит к более быстрому развитию возмущений в нем, более интенсивному вихреобразованию в этом слое и разрушению отрывного вихря. Вычисленное по этим данным отношение времени перестройки течения к характерному времени изменения угла атаки профиля ($\tau_{att} f$) показывает, что в широком диапазоне нестационарных течений при приведенной частоте $k < 0.2$ выполняется условие $\tau_{att} f \ll 1$, которое гарантирует, что поток будет успевать присоединяться к профилю несмотря на изменение его угла атаки во времени (рис. 62).

### 11.2. Управление динамическим отрывом прямого потока

Практически первой демонстрацией контроля динамического отрыва потока при обтекании аэродинамического профиля были работы [239, 240]. В этих работах был использован как одиночный, так и многоэлектродный актуаторы. Одиночный актуатор располагался так, чтобы кромка высоковольтного электрода совпадала с центральной линией профиля на передней кромке (рис. 57). Было продемонстрировано увеличение подъемной силы профиля для существенно закритических углов атаки для приведенных частот $k = \pi f c / U_\infty = 0$-$0.05$ в диапазоне скоростей $U_\infty = 30$-$45$ м/с ($Re = 3$-$4.5 \times 10^5$).

Было показано, что использование плазменного актуатора может предотвратить резкое уменьшение подъемной силы и увеличение силы сопротивления, даже когда максимальный угол атаки лопасти увеличивается до $\alpha = 32^0$. Включение актуатора приводило к мгновенному увеличениию подъемной силы на величину до 55%. Это позволяет значительно расширить диапазон допустимых условий полета и улучшить управляемость вертолётов в критических режимах.

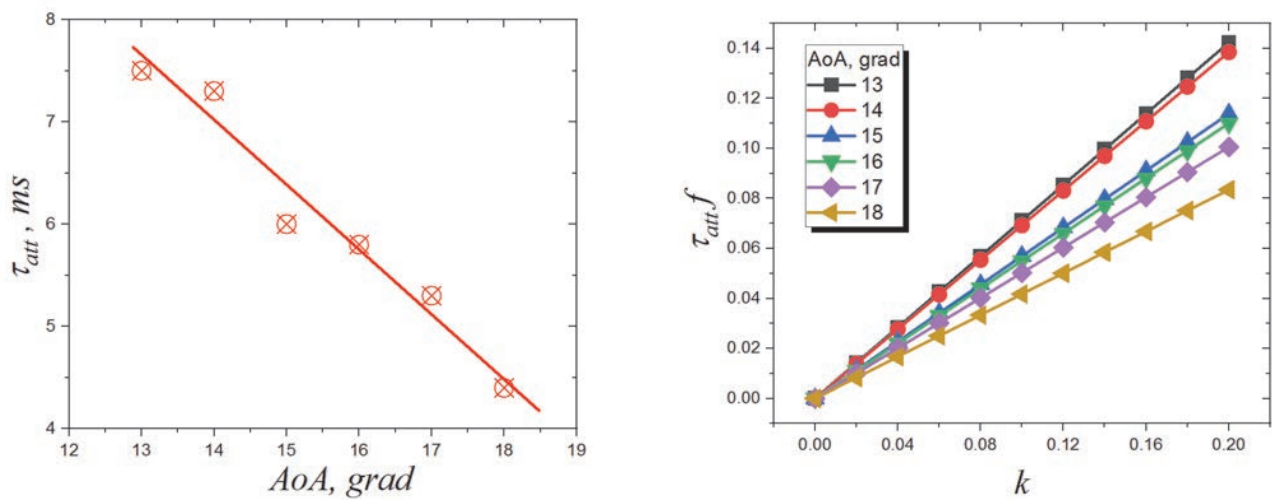

Рисунок 62. Время присоединения потока в зависимости от угла атаки (слева). Приведенное время присоединения потока в зависимости от приведенной частоты изменения угла атаки $k = \pi f c / U_\infty$. По данным [238]. Мощность разряда $P = 3$ Вт, $f = 225$ Гц ($F^+ = 0.75$), $Q = 15$ мДж/импульс. $U_\infty = 44.7$ м/с, профиль NACA-0015.

Рисунок 63 показывает динамику изменения подъемной силы профиля при изменении угла атаки от $\alpha_{min} = 0^0$ до максимального угла $\alpha_{max} = 26^0$. Видно, что из-за динамического характера взаимодействия снижение подъемной силы начинается значительно позже, чем достигается критический угол атаки для данного профиля $\alpha_{crit} = 12^0$. При достижении угла $\alpha = 16^0$ начинается отрыв потока (задержка от точки прохождения критического угла порядка 30 миллисекунд).

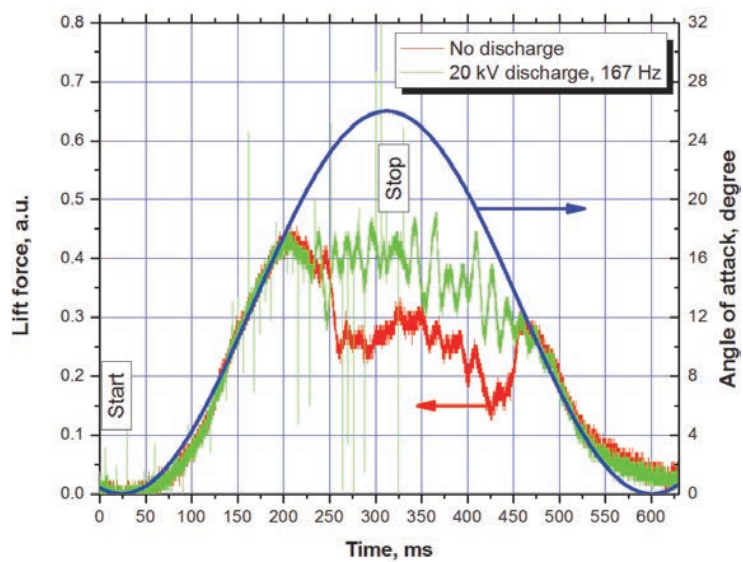

Рисунок 63. Динамика подъемной силы в течение цикла изменения угла атаки [239]. $\alpha_{max} = 26°$, синусоидальный профиль с периодом $t = 575$ мс ($k = 0.026$). $U_\infty = 31$ м/с ($Re = 3\times10^5$). Частота включения актуатора $f = 167$ Гц ($F^+ = 0.81$). Отметки «Start» и «Stop» указывают на область работы актуатора.

Распространение точки отрыва вверх по потоку приводит к полному отрыву течения и резкому падению подъёмной силы. До этого момента наличие актуатора никак не сказывается на развитии течения, поскольку для его работы необходим сформировавшийся отрывной пузырь со сдвиговым слоем, где возможно развитие возмущений, создаваемых актуатором. Развитие такого пузыря занимает приблизительно такое же время, как и задержка отрыва потока – 30 миллисекунд. После этого развитие течения становится принципиально разным для случая без воздействия и с воздействием актуатора. В случае без разряда развитие отрыва потока приводит к резкому катастрофическому падению подъёмной силы (на 40% от максимальной величины до отрыва потока). В случае наличия возбуждения актуатором падение подъемной силы практически отсутствует и остается на уровне, близком к значению, достигнутому в момент отрыва потока. В эксперименте, показанном на рис. 63, актуатор был принудительно выключен после достижения профилем максимального угла атаки ($t = 330$ мс), чтобы проследить роль актуатора на амплитуду и частоту изменений подъемной силы в отрывном режиме. Из данных рис. 63 видно, что частота колебаний величины подъемной силы несколько уменьшилась (70 Гц с включенным актуатором и 60 Гц с выключенным), в то время как амплитуда колебаний практически удвоилась. При этом начался переход к отрывному течению, который не успел завершиться до момента снижения величины угла атаки до подкритических значений (рис. 63). После уменьшения угла атаки до $\alpha < 12^0$ происходит естественное приприсоединение потока, и значение подъемной силы становится таким же, каким было на соответствующих углах атаки на восходящей ветке цикла.

Для оценки эффективности подавления динамического отрыва в работе [239] было измерено интегральное за цикл изменение подъемной силы актуатора в зависимости от частоты воздействия (рис. 64). Максимум эффективности приходится на диапазон $F_c^+ = 0.5\text{-}1.0$, что хорошо согласуется с данными по управлению статическим отрывом потока и оценкам времени перестройки течения (см. рис. 62).

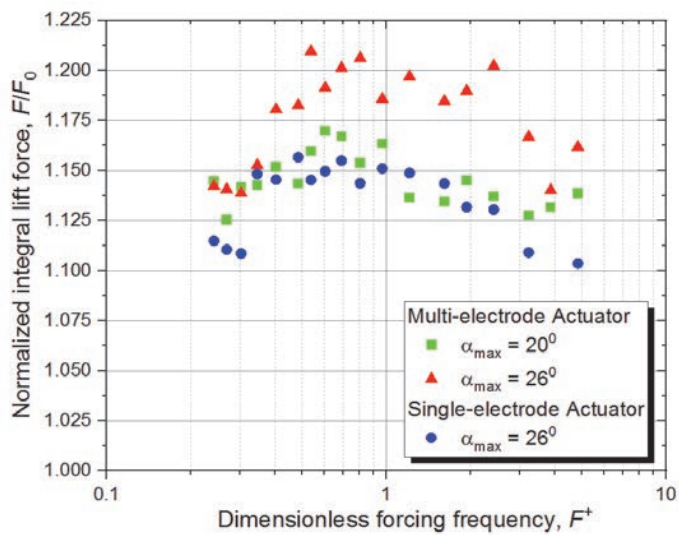

Рисунок 64. Интеграл подъемной силы за цикл изменения угла атаки в зависимости от приведенной частоты работы актуатора. $F/F_0$ – отношение интегральной за цикл подъемной силы при включенном актуаторе к интегральной подъемной силе без актуатора. $\alpha_{max} = 20^0$ и 26°, синусоидальный профиль изменения угла атаки с периодом $t = 575$ мс ($k = 0.026$). $U_\infty = 31$ м/с ($Re = 3\times10^5$). По данным [239]

В дополнение к данным по единичному актуатору, расположенному на передней кромке профиля, в работе [239] были проведены эксперименты с многоэлектродной конфигурацией актуатора по типу рис. 56. В многоэлектродной конфигурации был использован такой же электрод на передней кромке аэродинамического профиля плюс четыре дополнительных пары электродов, равномерно распределенных на верхней поверхности профиля параллельно передней кромке. Расстояние между электродами составляло 30 мм. Предполагалось, что такие дополнительные электроды могут оказывать влияние на динамику развития отрывного течения по мере продвижения точки отрыва от задней кромки крыла вперед. Из данных [239] видно, что интегральное увеличение подъёмной силы достигает 15% в случае одиночного электрода на передней кромке, и увеличивается до 20% в случае многоэлектродной системы. Таким образом, было показано, что максимальная эффективность воздействия на поток приходится на актуатор, расположенный на передней кромке. Актуаторы, расположенные ниже по потоку, вносят лишь небольшие добавки к подъёмной силе за счет взаимодействия с зоной отрыва потока в моменты, когда она находится ниже по потоку относительно данной пары электродов.

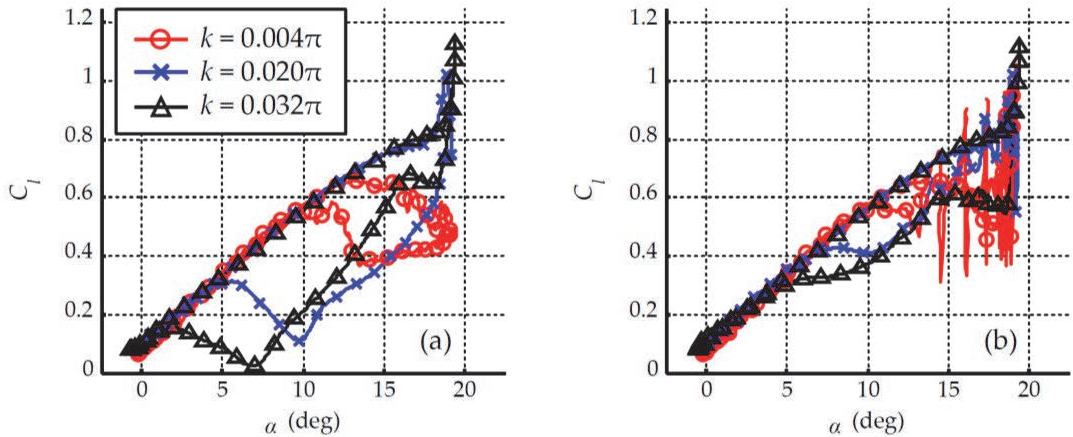

Рисунок 65. Динамика коэффициента подъемной силы для различных приведенных частот колебания угла атаки профиля [235]. Здесь a) и b) соответствуют режимам без и с возбуждением ns-SDBD актуатора соответственно.

В недавних работах [235, 241] авторы скопировали расположение и геометрию ns-SDBD актуаторов, предложенных в [238-240]. При этом выводы работ [239, 240] о высокой эффективности такого размещения актуаторов были полностью подтверждены. В работе [235] с использованием несимметричного профиля относительной толщины 12% в потоке при скорости $U_\infty = 50$ м/с исследовалось влияние воздействия актуатора на отрывное течение, возникающее при периодическом изменении угла атаки профиля в диапазоне $\alpha = 10^0\text{-}20^0$ при приведенной частоте колебаний $k = 0.012$–0.1. Было показано, что при частоте воздействия $F_c^+ = 0.13\text{-}0.96$ ns-SDBD актуатор может уменьшить гистерезис величин аэродинамических коэффициентов, и во всех случаях достигается эффект управления потоком (рис. 65).

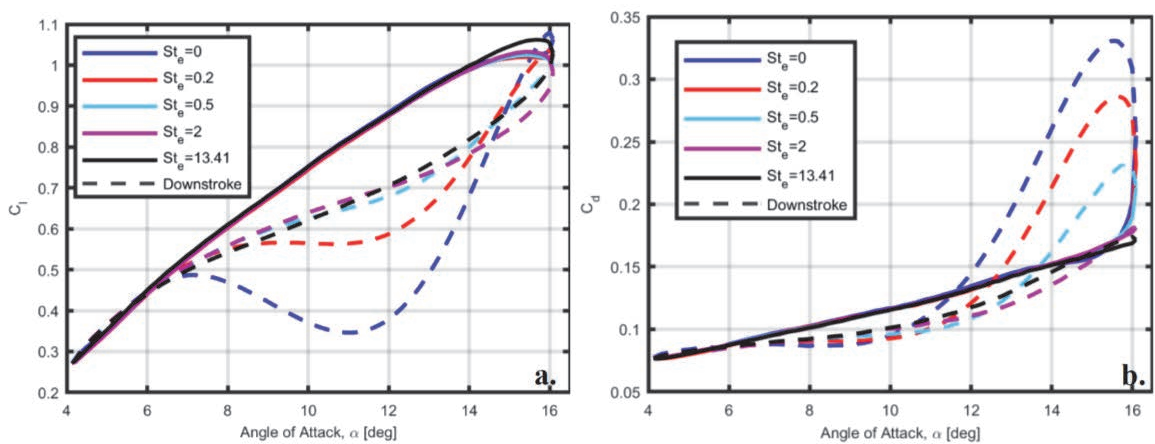

Рисунок 66. Режим слабого отрыва [241]. (а) коэффициент подъемной силы, (б) коэффициент сопротивления для $k = 0.075$, $Re = 3\times10^5$ и различных частот возбуждения $St_e$. Угол атаки изменялся по закону $\alpha(t) = 10^0 + 6^0 \sin(\omega t)$. Случай $St_e = 0$ соответствует выключенному актуатору.

В работе [241] рассматривается использование плазменных ns-SDBD актуаторов для динамического контроля потока при относительно небольших углах атаки $\alpha = 10^0$-$16^0$ для исследования режима слабого отрыва. Приведенная частота колебаний ($k$) и число Струхаля возбуждения ($St_e$) варьировались для числа Рейнольдса $Re = 3 \times 10^5$ в диапазоне $k = 0.05$-$0.1$ и $St_e = 0$-$13.4$. Во всех случаях наблюдалось значительное увеличение отношения подъемной силы к силе сопротивления. Было также отмечено, что во всех случаях с активным воздействием наблюдалось более раннее присоединение потока и уменьшение гистерезиса подъемной силы (рис.66).

### 11.3. Управление динамическим отрывом обратного течения

При больших скоростях полета вертолета, кроме проблемы динамического отрыва потока на отступающей лопасти из-за ее большого угла атаки, возникает еще одна проблема. Это проблема обратного обтекания прикорневых частей отступающей лопасти. Из-за большого угла атаки этой лопасти при ее движении острой кромкой вперед относительно набегающего воздуха практически неизбежно возникает интенсивный отрыв потока. Поскольку скорость движения прикорневых частей лопасти невелика, они не оказывают существенного влияния на подъемную силу. С другой стороны, отрыв потока с острой кромки лопасти приводит к возникновению сильных вибраций и шумов. По этой причине возможность управления таким отрывом становится очень актуальной.

Впервые возможность управления отрывом потока на отступающей лопасти была продемонстрирована в работах [239, 240]. Профиль NACA-0015 был развернут в канале аэродинамической трубы таким образом, что поток двигался от острого заднего края профиля к передней части (рис. 67). При этом был использован тот же принцип, что и в случае с нормальным направлением потока (рис. 57): кромка высоковольтного электрода была расположена точно на оси профиля на задней кромке, а разряд развивался на той стороне профиля, на которую набегает поток.

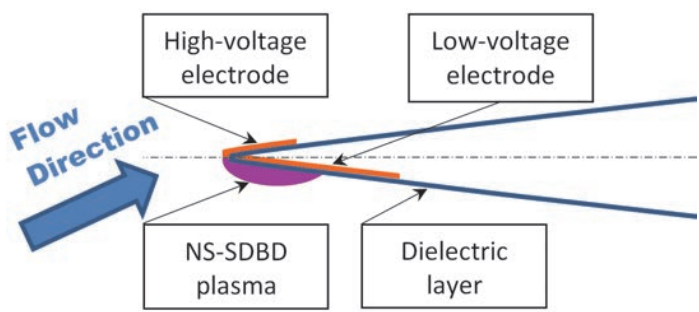

Рисунок 67. Геометрия расположения актуатора в случае обратного течения.

Чтобы продемонстрировать статические характеристики отрыва обратного потока, на рис.68,а показан режим медленного изменения угла атаки. Актуатор с энергией в импульсе около 0.6 мДж на сантиметр размаха крыла непрерывно работал в первой половине цикла, и был выключен во второй. Хорошо видно, что в случае отсутствия возбуждения картина отрыва потока имеет симметричный вид, что свидетельствует о квазистационарном характере отрыва. При этом отрыв потока происходит уже при угле атаки α = 9⁰ из-за наличия острой передней кромки. Дальнейшее увеличение угла атаки приводит к усилению отрывной зоны; при этом до α = 15⁰ происходит уменьшение подъемной силы. Уменьшение угла атаки от максимального значения приводит к зеркально-симметричной картине восстановления безотрывного обтекания при α < 9⁰ (рис.68,а).

На рисунках 68,б-г показано влияние воздействия актуатора на динамический отрыв при приведенной частоте $k$ = 0.026 и приведенной частоте актуатора $F_c^+$ = 0.6 для различных углов атаки. Видно, что для $α_{max}$ = 11.5⁰ действие актуатора полностью ликвидирует отрыв потока. При больших углах атаки $α_{max}$ = 17⁰ и $α_{max}$ = 23⁰ отрыв потока происходит значительно позже, чем в варианте с выключенным актуатором. Подъемная сила монотонно увеличивается с увеличением угла атаки до тех пор, пока не произойдет отрыв потока (α ~ 10⁰). Присоединение потока наступает на фазе обратного движения при меньших углах (α ~ 6-8⁰) из-за нестационарности процесса. При больших углах атаки (рисунки 68c,d) происходит частичное присоединение с увеличением подъемной силы практически в 2 раза.

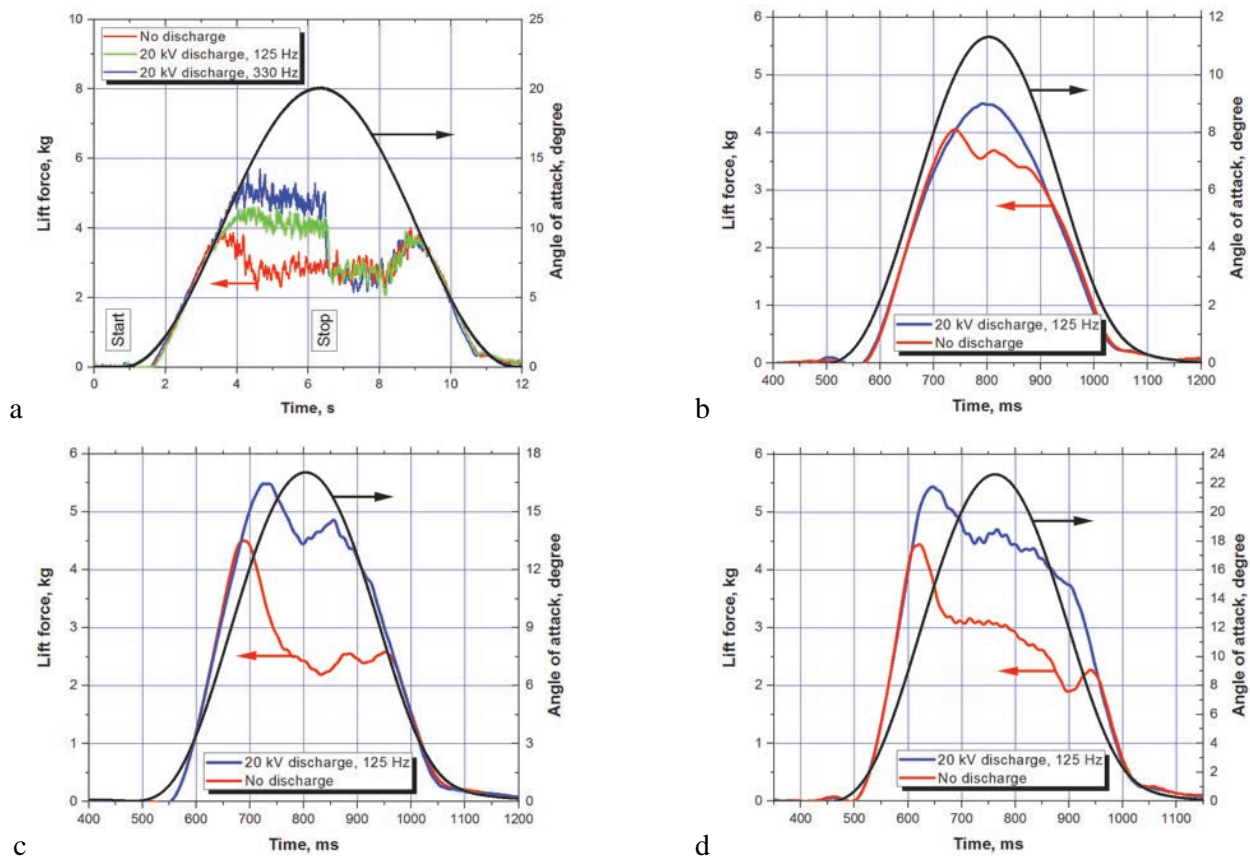

Рисунок 68. (а) Динамика подъемной силы в течение цикла [240]. Обратный поток, $U_∞$ = -31 м/с. а) медленное изменение угла атаки – управление статическим отрывом потока с острой кромки. Плазменный актуатор был включен в течение первой половины цикла; (б) –(г) быстрое изменение угла атаки – управление динамическим отрывом потока с острой кромки. (б) $α_{max}$ = 11.5⁰; (в) $α_{max}$ = 17⁰; (г) $α_{max}$ = 23⁰. Плазменный актуатор включен в течение всего цикла. Энергия импульса Q = 0.6 мДж/см.

Интегральная по периоду подъемная сила, как видно на рисунке 69, монотонно увеличивается при включенном актуаторе до максимальных углов $\alpha_{max} = 32°$ при относительном увеличении подъёмной силы до 55%, в то время как изменения силы сопротивлении остаются ниже 10%.

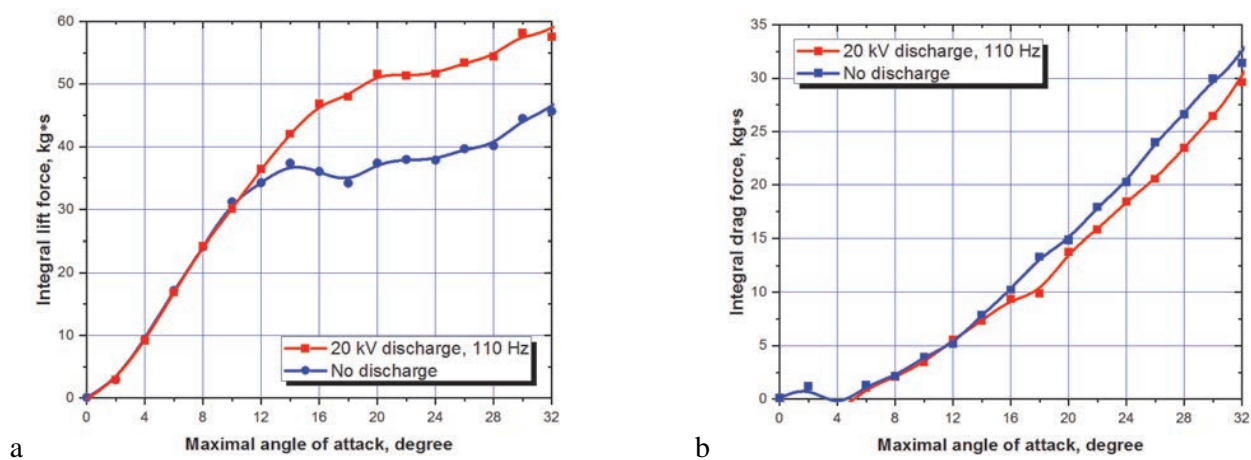

Рисунок 69. Интеграл подъемной силы (a) и силы сопротивления (b) за период изменения угла атаки для конфигурации с обратным потоком [240]. $U_\infty = 31$ м/с, $k = 0.026$, $F_c^+ = 0.53$.

Таким образом, установлено, что ns-SDBD актуатор эффективен как при управлении динамическим отрывом потока с затупленной передней кромки с большим радиусом кривизны, так и при управлении отрывом обратных потоков, когда отрыв происходит с заостренных кромок. Условиями эффективной работы актуатора являются: а) его расположение непосредственно на передней кромке профиля (рис. 57, 67); б) достаточная энергия импульса разряда, которая, по-видимому, зависит от числа Рейнольдса потока и для умеренных величин $Re = 3 \times 10^5$ близка к $Q \sim 1$ мДж/см; и в) работа актуатора в эффективном диапазоне приведенных частот $F_c^+ = F_f c/U\infty \sim 0.5$-$1.0$.

### 11.4. Управление отрывом 3D течения

Для изучения эффективности работы плазменных актуаторов в трехмерном потоке в работе [242] была проведена специальная серия экспериментов с моделью вертолета (рис.70). Для эксперимента использовалась стандартная радиоуправляемая модель вертолета (ALIGN TREX 800E). Модель была оснащена электродвигателем мощностью 4.5 кВт. Модель была дополнительно оборудована системой, которая позволяла подавать высоковольтные наносекундные импульсы от генератора FID 20-10 на вращающиеся лопасти. Импульсы имели амплитуду 20 кВ и частоту от 200 Гц до 2 кГц. Асимметричные плазменные актуаторы (схема расположения как на рис. 57) были установлены на передней кромке лопастей. Электроды располагались по всей длине передней кромки. Размах лопастей ротора составлял 1860 мм, а длина хорды лопасти была 60 мм. Во всех случаях лопасти вращались с частотой ~18 Гц.

Рисунок 71,а показывает зависимость подъёмной силы вращающегося ротора при фиксированном угле атаки от частоты работы актуатора. Почти линейное увеличение подъемной силы для диапазона частот актуатора 300-1500 Гц отражает тот факт, что разные части лопастей имеют разную линейную скорость, и оптимальная частота срабатывания увеличивается с увеличением расстояния от оси ротора. Внешние части лопастей создают большую подъемную силу, и увеличение частоты работы актуатора приводит к увеличению суммарной подъёмной силы, поскольку оптимальная точка ($F_c^+ \sim 1$) смещается при увеличении частоты работы актуатора к концам лопастей.

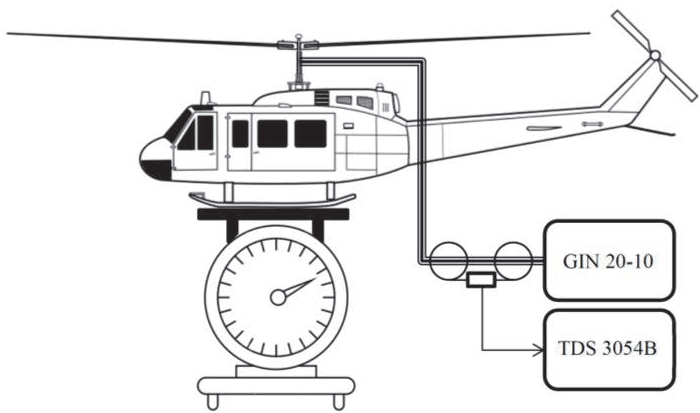

Рисунок 70. Модель вертолета на весах. [242]

На рис. 71,б показаны результаты измерений подъемной силы в статическом режиме при разных углах атаки в трех случаях: с выключенными актуаторами, при частоте разряда $f$ = 1 кГц ($F_c^+$ = 0.57), и при частоте разряда $f$ = 2 кГц ($F_c^+$ = 1.14). Без актуаторов максимум подъемной силы достигается при α = 12$^0$. Дальнейшее увеличение угла атаки приводит к отрыву потока на лопастях и уменьшению подъемной силы (рис. 71,б). Когда актуатор был включен, наблюдалось значительное увеличение подъемной силы при больших углах атаки (α = 20$^0$) по сравнению с максимальной величиной подъемной силы, полученной с выключенными актуаторами – до 7% для частоты актуатора $F_c^+$ = 0.57 и до 20% при $F_c^+$ = 1.14.

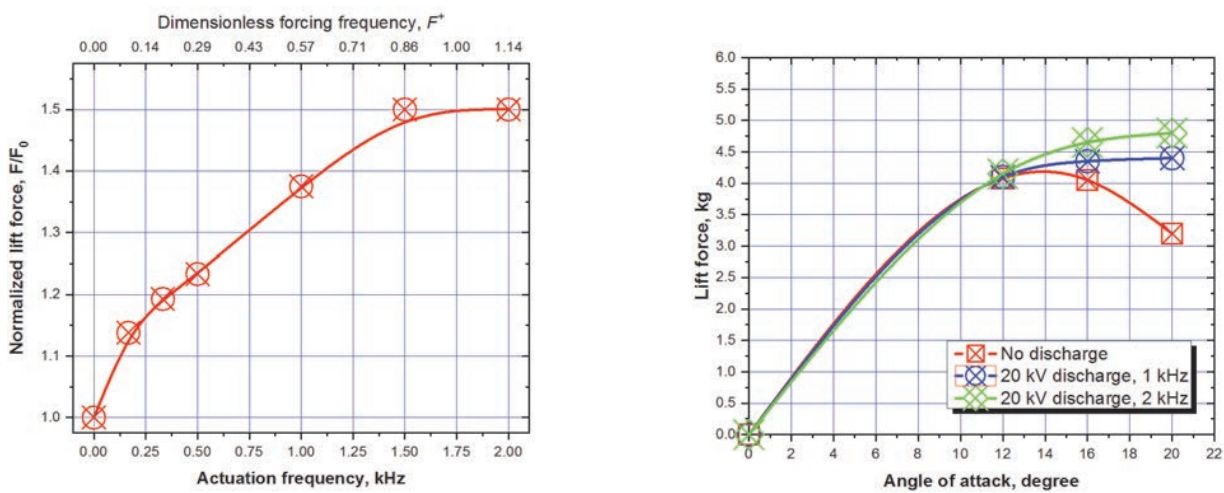

Рисунок 71. (а) Зависимость подъемной силы от частоты работы актуатора [242]. Угол атаки α = 20$^0$. (б) Зависимость подъемной силы от угла атаки. $F_c^+$ = 0.57 (1 кГц); $F_c^+$ = 1.14 (2 кГц).

При мощности разряда ~38 Вт ($F_c^+ = 1.14$, $f = 2$ кГц) было достигнуто увеличение интегральной подъемной силы более чем на 20% при постоянной мощности двигателя 1500 Вт по сравнению с оптимальной конфигурацией лопастей без актуаторов. Эти результаты показывают высокую эффективность использования актуаторов для трехмерных течений и открывают возможность их применения для реальных вертолетных систем.

## 12. Анти-обледенительные плазменные системы

Практическое применение плазменных систем для управления движением летательных аппаратов в атмосфере иногда сталкивается с возражением, что электроразрядные системы могут работать нестабильно в условиях дождя или снега. Действительно, образование сплошного толстого слоя воды или льда на поверхности электродов может привести к тому, что разряд не будет развиваться, и актуатор прекратит работать. Однако образование такой достаточно толстой пленки воды на поверхности крыла, движущегося с большой скоростью, невозможно. Вода будет сноситься потоком воздуха, а отдельные капли и струи не скажутся на эффективности работы актуатора.

Более сложным является вопрос образования льда на поверхности крыла. В этом случае слой льда может накапливаться и блокировать формирование плазменного слоя. Обычно с обледенением крыла и других элементов конструкции самолета борются с помощью нагревательных элементов, встроенных в конструкцию. Отметим, что наличие плазменного актуатора не снижает функциональность таких устройств. Если штатная система антиобледенения самолета или вертолета может справиться с образованием льда, она автоматически очистит и электродную систему актуатора. Однако плазменные актуаторы являются источниками тепла сами по себе. Интересно проверить, могут ли такие актуаторы одновременно работать и как антиобледенительная система.

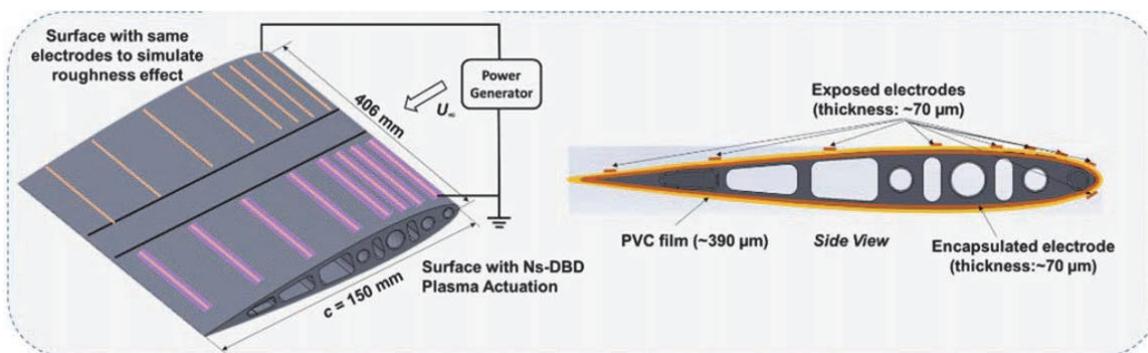

Рисунок 72. Схемы модели и аэродинамического профиля, использованные в эксперименте по контролю над обледенением с помощью ns-SDBD [243].

По сравнению с обычной антиобледенительной системой, основанной на нагреве поверхности профиля, система антиобледенения, построенная на основе плазменных актуаторов, может иметь значительные преимущества. Во-первых, плазменные актуаторы изначально нагревают не элементы конструкции крыла, а воздух в пограничном слое; во-вторых, создаваемые такими актуаторами сильные возмущения (ударные волны) могут приводить к механическому, а не тепловому, удалению появляющихся ледяных элементов с поверхности; в третьих, такие актуаторы не требуют для своего размещения места внутри крыла и имеют малый вес.

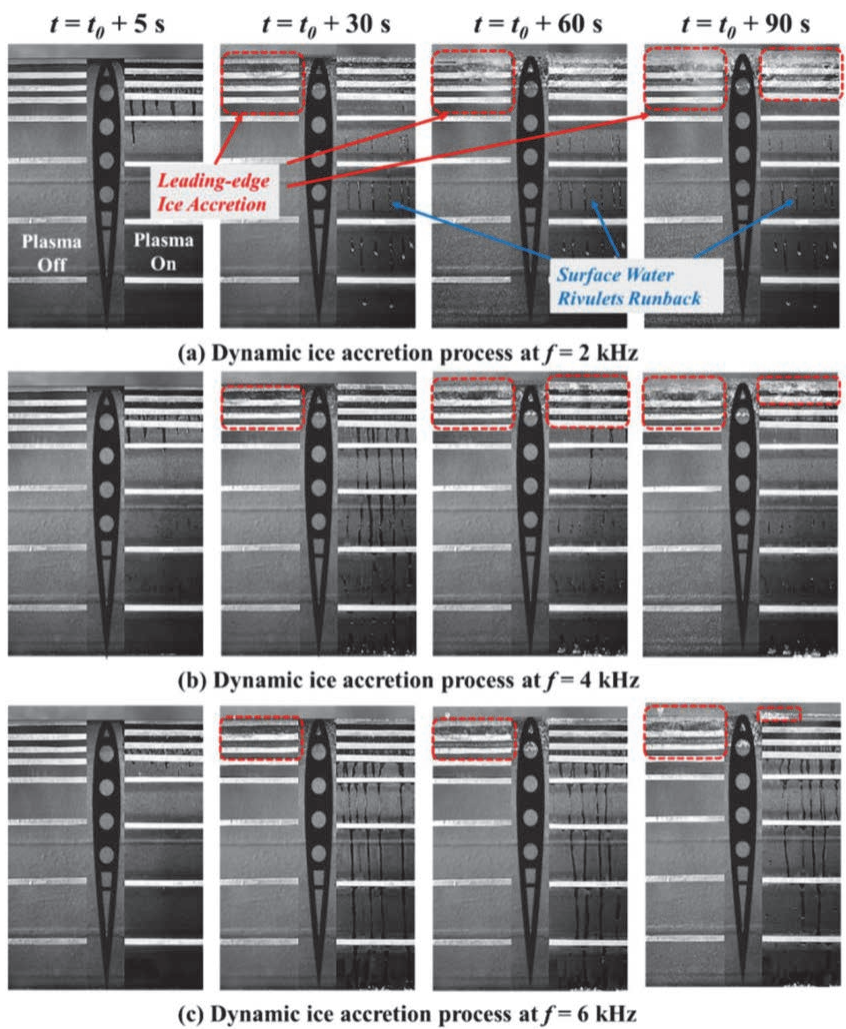

Рисунок 73. Эволюция во времени динамических процессов аккреции льда на поверхности аэродинамического профиля в условиях обледенения при $U_\infty = 40$ м/с, расходе жидкой воды 1.0 г/м$^3$ и $T_\infty = -10^0$С [244]. ns-SDBD актуатор работал на частоте (а) $f = 2$ кГц, (б) $f = 4$ кГц и (с) $f = 6$ кГц.

В работах [243, 244] проведено исследование тепловых характеристик плазмы ns-SDBD на поверхности аэродинамического профиля и были получены оценки антиобледенительных характеристик таких плазменных актуаторов. Был проведен ряд экспериментов для оценки влияния различных параметров окружающей среды на эффективность нагрева газа под действием ns-SDBD. С помощью системы высокоскоростной визуализации и инфракрасной визуализации с синхронизацией по времени и пространственному разрешению были изучены переходные тепловые характеристики плазмы над поверхностью крыла и антиобледенительные характеристики ns-SDBD. Такие характеристики были оценены при различных условиях обледенения – изморозь, смешанные условия и ледяная глазурь. Исследовано влияние скорости набегающего воздушного потока, температуры воздуха и угла атаки модели на скорость обледенения. Было обнаружено, что антиобледенительные характеристики плазменных актуаторов значительно улучшаются при увеличении частоты следования разрядов, что является очевидным следствием увеличения средней мощности, выделяемой разрядом в поступательные степени свободы газа.

На рисунке 72 показана схема модели, использованная в эксперименте по контролю над обледенением с помощью ns-SDBD. Модель была разделена на две симметричные половины, на которых были смонтированы идентичные актуаторы. Актуаторы на одной из половин модели были подключены к источнику наносекундных высоковольтных импульсов, вторая половина модели служила как контрольный образец. Были проведены измерения распределения температуры по поверхности крыла в условиях обледенения при работе плазменного актуатора на различных частотах и показано, что при высокой частоте работы актуатора выделяемой им энергии достаточно не только для нагрева приповерхностного слоя воздуха, но и для нагрева самой поверхности на несколько градусов.

Импульсный нагрев воздуха в пограничном слое около носка профиля приводит к формированию ударных волн, которые способствуют механическому удалению льда с поверхности профиля. Кроме того, формирующийся около поверхности теплый слой не дает частичкам льда и каплям воды перекристаллизоваться на поверхности крыла ниже по потоку (рис. 73). Хорошо видно, что при высокой частоте следования импульсов актуатора практически не происходит аккреции льда на поверхности. В работах [245, 246] результаты [243, 244] были подтверждены с использованием продольного расположения электродов наносекундного актуатора, аналогичного использованному в работе [212] для управления отрывом пограничного слоя в трансзвуковых режимах (рис. 74).

Результаты, полученные в [243-246], показывают, в первую очередь, значительную устойчивость работы ns-SDBD плазменных актуаторов при наличии в набегающем потоке большого количества воды, частиц льда и снега, а также перспективы разработки новых стратегий антиобледенительной защиты на основе плазмы, специально разработанных для снижения обледенения воздушных судов в полете.

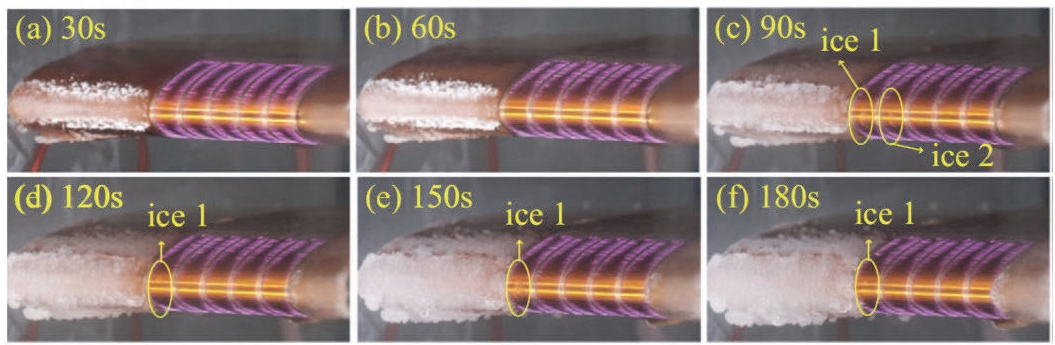

Рисунок 74. Снимки на протяжении 3-х минут от начала динамического антиобледенительного процесса с помощью плазменного наносекундного актуатора в условиях: частота разряда $f = 6$ кГц, напряжение $V = 7.7$ кВ, длина электрода 0.484 м, расход жидкой воды $Q = 0.5$ г/м$^3$, средний диаметр капель $d = 25$ мкм, $T = -15\text{C}^0$, $U_\infty = 65$ м/с [245].

## 13. Выводы

В данном обзоре продемонстрирован огромный потенциал управления скоростными потоками газа с помощью сверхбыстрого локального нагрева среды при создании и распаде сильнонеравновесной импульсной плазмы. Было показано, что такой локальный нагрев позволяет управлять конфигурацией ударных волн при сверхзвуковом обтекании, взаимодействием ударной волны с пограничным слоем, эффективен как на малых дозвуковых, так и на сверхзвуковых скоростях. Плазменные методы управления позволяют контролировать как статический, так и динамический отрыв потока на больших углах атаки, в том числе в режиме обратного течения, характерного для отступающей вертолетной лопасти. Значительный интерес представляют последние работы по управлению турбулентным сопротивлением и ламинарно-турбулентным переходом.

Главными факторами для получения высокой эффективности управления потоком с помощью плазменных актуаторов является их размещение по отношению к ключевым точкам профиля объекта и потока. Так, для управления движением сверхзвукового объекта критичным является вытянутый вдоль траектории движения профиль создаваемого возмущения и его значительная амплитуда, которая зависит от числа Маха и угла наклона лидирующей ударной волны. Эффективное управление отрывом пограничного слоя с помощью импульсного энерговыделения возможно в случае размещения актуаторов непосредственно на передней кромке профиля объекта перед точкой отрыва потока. Управление динамикой слоя смешения наиболее эффективно при размещении актуаторов на границе или в самом слое.

Другой важный аспект плазменного управления потоками – частота работы актуатора. Как показывают многочисленные эксперименты, эффективное управление потоком в значительной степени зависит от частоты воздействия. В случае отрывных течений наиболее эффективны частоты, соответствующие числу Струхаля порядка единицы. Напротив, для антиобледенительных систем такая низкая частота является малоэффективной из-за низкой средней мощности, и требуемый диапазон частот в этом случае гораздо выше. Управление сверхзвуковым движением приводит к заметному перераспределению нагрузок на летящий объект пока возмущение, созданное разрядом, движется вдоль его поверхности. Таким образом, для эффективного воздействия на конфигурацию ударных волн и создаваемых ими нагрузок, диапазон чисел Струхаля также должен быть порядка единицы.

Важным моментом является энергия актуатора в импульсе. Нужно отметить, что все рассмотренные случаи не предполагают изменение глобальных параметров потока (например, изменение скорости потока как целого или его нагрев) с помощью плазмы. Напротив, энерговыделение разряда всегда используется точечно, только для создания либо неоднородности поля параметров течения, либо как инициатор развития неустойчивости основного потока или слоя. При таком подходе не требуется значительной энергии даже для управления сверхзвуковым обтеканием объектов с высоким аэродинамическим качеством, возможна эффективная борьба с образованием льда на аэродинамических поверхностях, и при незначительной мощности управляющего воздействия становится возможным контроль статического и динамического отрыва потока при больших скоростях и существенно закритических углах атаки.

Из такого анализа становится очевидным, что эффективное применение плазменных актуаторов определяется возможностью управления скоростью, положением и интенсивностью энерговыделения в плазме. В обзоре рассмотрены основные механизмы быстрой генерации сильнонеравновесной плазмы в скоростных потоках газа. Как правило, к таким механизмам относятся импульсные разряды при высоком и сверхвысоком перенапряжении. Значительное превышение электрического поля над пороговым значением позволяет быстро нагреть электронный ансамбль. При неизменном направлении электрического поля (например, в наносекундных импульсных разрядах) это приводит к быстрому достижению электронами порога ионизации и эффективному формированию сильнонеравновесной плазмы. В случае переменного поля – СВЧ или лазерного – время свободного пробега электрона и время изменения направления поля в зависимости от условий могут стать сопоставимыми. В этих условиях для достижения электронами энергии ионизации и развития лавин может потребоваться большое количество циклов изменения поля, а сами значения электрических полей должны быть значительно выше, чем в случае импульсного наносекундного разряда. Нужно отметить, что для любого типа разряда ключевым фактором его применимости для управления потоком за счет импульсного локализованного выделения энергии является возможность создания относительно плотной плазмы при типичном удельном энерговкладе 0.05-0.5 эВ на молекулу в электронные степени свободы газа и ионизацию за время, меньшее времени газодинамического расширения плазменной области.

Требование преимущественного направления энерговклада разряда на возбуждение высокоэнергичных состояний диктуется необходимостью быстрой конверсии возбуждения внутренних и химических степеней свободы газа в тепловую энергию, которая может быть использована для управления потоком. Как показали эксперименты, наибольшая эффективность и скорость такой релаксации энергии в последовательности событий «наложение электрического поля – нагрев электронов – возбуждение и ионизация молекул – термализация энергии» достигается именно при высоких и сверхвысоких полях, когда преимущественным типом возбуждения газа является создание высоковозбужденных состояний молекул и ионизация. В этом случае быстрые тушение сильновозбужденных частиц и рекомбинация плазмы приводит к образованию поступательно-горячих атомов и молекул, которые за несколько столкновений обмениваются своей энергией с остальным газом. Эта ситуация кардинально отличается от возбуждения, например, колебательных степеней свободы молекул при умеренных электрических полях, когда релаксация возбуждения и увеличение температуры газа могут занимать длительное с точки зрения газодинамики время.

В заключение упомянем о нерешенных проблемах плазменного управления потоками. В первую очередь, конечно, это погодные условия. Несмотря на наличие прямых экспериментальных демонстраций возможности успешной работы плазменных актуаторов даже в условиях сильного обледенения, остаются вопросы возможного снижения эффективности работы таких систем при наличии сильного дождя, снега или запыленности воздуха. Другой важной стороной проблемы, ограничивающей возможности применения плазменных актуаторов, являются создаваемые ими электромагнитные возмущения. Как уже отмечалось, наибольшей эффективностью обладают импульсные системы, генерирующие сильновозбужденную плазму за единицы-десятки наносекунд. Поскольку эффективность излучения электромагнитного сигнала любым проводником, как правило, резко возрастает с увеличением частоты, то повышение эффективности работы плазменных систем неизбежно означает увеличение электромагнитных помех, создаваемых ими. Кроме того, практически не исследовались еще вопросы устойчивости таких систем при воздействии на них атмосферного электричества. Если в случае оптического пробоя, инициируемого лазерным импульсом, всегда есть возможность достаточно надежно заэкранировать лазерную систему от тока молнии, то электроды SDBD актуаторов и генераторы импульсных напряжений защитить заметно сложнее. К той же группе нерешенных вопросов относятся массо-габаритные и энергетические характеристики плазменных актуаторов. Несмотря на их относительно малое энергопотребление, необходимый диапазон мощности источников питания может достигать, в зависимости от размера объекта, нескольких киловатт. Для лазерных систем из-за невысокого коэффициента преобразования электроэнергии в энергию луча и затем разряда необходимые мощности на борту летательного аппарата могут быть еще больше. В заключение упомянем чисто технологическую проблему. Наилучшими характеристиками по износостойкости, химической стойкости, электроизоляционным параметрам и их стабильности обладают электротехнические керамики. Однако, такие материалы тяжело поддаются механической обработке и являются хрупкими. Применяемые в лабораторных условиях диэлектрические полимерные пленки (Каптон®, поливинилхлорид, Фторопласт®) обеспечивают уникальную электрическую прочность слоя, высокую химическую и физическую стабильность, но обладают по сравнению с керамикой заметно более низкой износостойкостью, что накладывает определенные ограничения на их использование в условиях сильнозапыленных потоков.

Решение данных вопросов откроет совершенно новые возможности для сверхбыстрого управления потоками и высокоскоростными объектами с помощью сильнонеравновесной импульсной разрядной плазмы.

## 14. Благодарности

## 15. Список литературы